\documentclass[letterpaper]{JHEP3}
\usepackage{epsfig}

\def\ba{\begin{eqnarray}}
\def\ea{\end{eqnarray}}
\def\be{\begin{equation}}
\def\ee{\end{equation}}
\def\nn{\nonumber}
\def\exd{{\rm d}}
\def\pd{\partial}
\makeatletter
\def\x@arrow{\DOTSB\Relbar}
\def\xlongequalsignfill@{\arrowfill@\x@arrow\Relbar\x@arrow}
\newcommand{\xlongequal}[2]{%
    \ext@arrow 0099\xlongequalsignfill@{#1}{#2}}
\makeatother

\newcommand{\roughly}[1]{\mathrel{\raise.3ex\hbox{$#1$\kern-0.85em
\lower1ex\hbox{$\sim$}}}}

\newcommand{\lsim}{\roughly<}

\def\endignore{}
\def\ignore #1\endignore{} 

\def\be{\begin{equation}}
\def\beq\begin{equation}
\def\ee{\end{equation}}
\def\bea{\begin{eqnarray}}
\def\eea{\end{eqnarray}}

\def\eqa{\begin{eqnarray}}
\def\eeqa{\end{eqnarray}}
\def\eq{\begin{equation}}
\def\eeq{\end{equation}}

\def\nn{\nonumber}

\def\pref#1{(\ref{#1})}

\def\ol#1{\overline{#1}}

\def\dint{{\hat d}}
\def\exd{{\rm d}}
\def\nn{\nonumber}
\def\pref#1{(\ref{#1})}
\def\be{\begin{equation}}
\def\ee{\end{equation}}
\def\erf{{\rm erf}}

\def\beq{\begin{equation}}
\def\eeq{\end{equation}}
\def\beqa{\begin{eqnarray}}
\def\eeqa{\end{eqnarray}}

\def\cA{{\cal A}}
\def\cB{{\cal B}}
\def\cC{{\cal C}}

\def\cF{{\cal F}}
\def\cG{{\cal G}}
\def\cH{{\cal H}}

\def\cL{{\cal L}}

\def\cN{{\cal N}}
\def\cO{{\cal O}}

\def\cR{{\cal R}}
\def\cS{{\cal S}}

\def\cV{{\cal V}}

\def\ssA{{\scriptscriptstyle A}}
\def\ssB{{\scriptscriptstyle B}}
\def\ssC{{\scriptscriptstyle C}}
\def\ssD{{\scriptscriptstyle D}}
\def\ssE{{\scriptscriptstyle E}}
\def\ssF{{\scriptscriptstyle F}}

\def\ssK{{\scriptscriptstyle K}}
\def\ssL{{\scriptscriptstyle L}}
\def\ssM{{\scriptscriptstyle M}}
\def\ssN{{\scriptscriptstyle N}}
\def\ssP{{\scriptscriptstyle P}}
\def\ssQ{{\scriptscriptstyle Q}}
\def\ssR{{\scriptscriptstyle R}}

\def\ssV{{\scriptscriptstyle V}}

\def\Vone{\cV_{\rm 1-loop}}

\newcommand{\bmat}{\left(\begin{array}}
\newcommand{\emat}{\end{array}\right)}

\def\-{\hphantom{-}}

\def\s2{\frac{1}{2}}

\def\vev#1{\langle#1\rangle}
\def\tr{{\rm tr \,}}
\def\Tr{{\rm Tr \,}}

\def\IF{\relax{\rm I\kern-.18em F}}
\def\II{\relax{\rm I\kern-.18em I}}
\def\IP{\relax{\rm I\kern-.18em P}}
\def\IC{\relax{\rm I\kern-.48em C}}
\def\IR{\relax{\rm I\kern-.18em R}}
\def\IK{\relax{\rm I\kern-.20em K}}
\def\IM{\relax{\rm I\kern-.25em M}}

\def\nott#1{\setbox0=\hbox{$#1$}                
   \dimen0=\wd0                                 
   \setbox1=\hbox{/} \dimen1=\wd1               
   \ifdim\dimen0>\dimen1                        
      \rlap{\hbox to \dimen0{\hfil/\hfil}}      
      #1                                        
   \else                                        
      \rlap{\hbox to \dimen1{\hfil$#1$\hfil}}   
      /                                         
   \fi}                                         %

\def\y2{Y_{\ssM\ssN} Y^{\ssM\ssN}}
\def\Riem2{R_{\ssA\ssB\ssM\ssN} R^{\ssA\ssB\ssM\ssN}}
\def\Ricci2{R_{\ssM\ssN} R^{\ssM\ssN}}

\def\f2{F^{a}_{\ssM\ssN} F^{\ssM\ssN}_a}
\def\Y{Y_{\ssM\ssN}}

\def\Dsl{\,\raise.15ex\hbox{/}\mkern-13.5mu D}
\def\ksl{\,\raise.15ex\hbox{/}\mkern-10.5mu k}
\def \one{\relax{\rm 1\kern-.26em I}}

\def\exd{{\rm d}}

\def\V{\mathcal{V}}

\def\nn{\nonumber}

\def\({\left(}
\def\){\right)}

\preprint{FTUAM-12-114, IFT-UAM/CSIC-12-112}

\title{
Running with Rugby Balls:\\
Bulk Renormalization of Codimension-2 Branes}

\author{

M.~Williams,${}^{1}$
C.P.~Burgess,${}^{1,2}$
L.~van Nierop${}^1$ and
A.~Salvio${}^{3}$\\
$^1$ Department of Physics \& Astronomy, McMaster University,
 Hamilton ON, Canada\\
$^2$   Perimeter Institute for Theoretical Physics,
 Waterloo ON, Canada\\
$^3$ Scuola Normale Superiore and INFN, Piazza dei Cavalieri 7, 56126 Pisa, Italy
\\
\hspace{0.25cm}Departamento de F\'isica Te\'orica,  Universidad Aut\'onoma de Madrid and \\
\hspace{0.25cm}Instituto de F\'isica Te\'orica IFT-UAM/CSIC, Cantoblanco, 28049 Madrid, Spain
}

\date{}

\abstract { We compute how one-loop bulk effects renormalize both bulk and brane effective interactions for geometries sourced by codimension-two branes. We do so by explicitly integrating out spin-zero, -half and -one particles in 6-dimensional Einstein-Maxwell-Scalar theories compactified to 4 dimensions on a flux-stabilized 2D geometry. (Our methods apply equally well for $D$ dimensions compactified to $D-2$ dimensions, although our explicit formulae do not capture all divergences when $D>6$.) The renormalization of bulk interactions are independent of the boundary conditions assumed at the brane locations, and reproduce standard heat-kernel calculations. Boundary conditions at any particular brane do affect how bulk loops renormalize this brane's effective action, but not the renormalization of other distant branes. Although we explicitly compute our loops using a rugby ball geometry, because we follow only UV effects our results apply more generally to any geometry containing codimension-two sources
with conical singularities. Our results have a variety of uses, including calculating the UV sensitivity of one-loop vacuum energy seen by observers localized on the brane. We show how these one-loop effects combine in a surprising way      with bulk back-reaction to give the complete low-energy effective cosmological constant, and comment on the relevance of this calculation to proposed applications of codimension-two 6D models to solutions of the hierarchy and cosmological constant problems.

}

\begin{document}
\section{Introduction}
\label{sec:introduction}

Does the vacuum have energy? If so, does it gravitate? Much of what we do not understand about quantum field theory is contained in these deceptively simple questions because calculations robustly indicate the vacuum should have lots of zero-point energy, yet cosmological observations indicate that this energy gravitates very little.

This disagreement is particularly sharp if there are only four dimensions because then the Lorentz invariance of the vacuum makes its energy equivalent to a cosmological constant, which acts as a homogeneous and isotropic obstruction to having the comparatively flat universe in which we appear to live. By contrast localized energy sources need not curve all directions equally; for example General Relativity predicts the world-sheet geometry of a cosmic string to be flat regardless of the value of its tension, whose main effect is to curve the dimensions transverse to the string world sheet (and in particular to produce a conical singularity at the string position) \cite{Vil}.

This observation suggests exploring whether extra dimensions can help understand how the vacuum energy gravitates, such as if we were to live on a four-dimensional analog of a cosmic string within a spacetime having a few relatively large dimensions. Fewer dimensions are better in this context since bulk fields fall off less quickly with distance, and so allow all branes to compete with the bulk regardless of how far apart they are in the extra dimensions.\footnote{More than two extra dimensions might exist, but would not be relevant at low energies if they were much smaller than the ones of interest here.} This has sparked the construction of several such brane-world systems, within one \cite{5Dbackreaction} or two \cite{6Dprelim, conical, Towards, GGP, OtherConical, laterconical} extra dimensions. The explicit solutions identified in this way have the property that their on-brane geometry is flat despite having large on-brane tensions.

Of course this in itself does not provide a solution to the cosmological constant problem, which also requires an understanding of why these geometries should be robust against quantum corrections. In particular, although solutions with flat on-brane geometries exist, so too do solutions with curved on-brane geometries. What is required is an understanding of how the on-brane curvature depends on physical choices for the bulk and the branes, and whether these choices remain stable against renormalization as high-energy modes are integrated out.

The most progress understanding these issues has been made for 6D models, for which fairly general yet explicit calculations can be made. Although no special magic is found for non-supersymmetric models \cite{localizedflux}, theories with a bulk described by 6D supergravity have very attractive features \cite{Towards, SLEDrefs, TNCC}. In particular, although explicit solutions with on-brane de Sitter geometries are known\footnote{These solutions are interesting in their own right as a counter-example to the many no-go theorems for de Sitter solutions to higher-dimensional supergravity \cite{6DdSnogo}.} \cite{6DdS}, a sufficient condition for the absence of on-brane curvature is the absence of a coupling between the branes and a particular bulk field: the scalar dilaton that is related to the graviton by 6D supersymmetry \cite{OtherConical, localizedflux}. This is attractive since it is the kind of condition that is stable against arbitrary loops involving only on-brane particles \cite{TNCC, uvcaps}.

Such a condition would {\em not} be stable against bulk loops, however, since the brane must couple to the metric and the metric couples to the dilaton. So bulk loops must provide an important part of any naturalness story, particularly  tracking how loops of heavy bulk states contribute to the low-energy effective vacuum energy. A missing step in this story is an explicit calculation of the UV sensitivity of Casimir energy calculations on the 6D geometries of interest. (See, however, \cite{UVsensitivity, GHBQ} for an assessment of bulk UV sensitivity for Ricci-flat geometries -- including also the gravity sector, but in the absence of branes.)

This paper and its companion \cite{Companion} close part of this `bulk gap', by computing explicitly the UV-sensitive part of the corrections to both brane and bulk interactions obtained by integrating out low-spin (spins zero, half and one) bulk fields in an extra-dimensional spacetime sourced by two codimension-two branes. This paper presents general results for arbitrary low-spin fields, while the companion specializes the results to the case where the bulk matter comes from a 6D supergravity. Our restriction to low-spin fields is a temporary one due to the technical complications of diagonalizing the full gravity-sector spectrum in the geometries of interest, and we intend to report on calculations using the full spectrum at a later date.\footnote{See also \cite{OtherRugbyCasimir} for a partial calculation of the Casimir energy of the gravity sector and
\cite{OtherotherRugbyCasimir} for the bosonic part of
the spectrum on a rugby ball.}

The background geometry of the extra dimensions with which we work is a rugby ball \cite{conical, Towards, moreconical}. This is described in \S\ref{sec:bulksugra} below, and is basically a flux-stabilized sphere with conical singularities at both of its poles corresponding to the back-reaction of codimension-two branes located there. In a nutshell, \S\ref{sec:genloops} argues that the Casimir energy obtained by integrating out a bulk field of 6D mass $m$ has the generic form
\be
 V(\alpha, m, r) = \frac{\cF ( m r , \alpha)}{(4\pi r^2)^2}  \,,
\ee
where $r$ is the rugby ball's radius and $\alpha$ is related to its defect angle, $\delta$, (see below for more precise definitions) by $\alpha = 1 - \delta/2\pi$. In the limit of large $mr$ the dimensionless function, $\cF$, becomes
\be \label{eq:introVform}
 \cF(mr, \alpha)
 \simeq \cF_0(\alpha) + \left[ \frac{(m r)^6}{6} \, s_{-1}(\alpha)  - \frac{(m r)^4}{2} \, s_0(\alpha) + (mr)^2 s_1(\alpha)  - s_2(\alpha) + \cdots \right] \ln ( m r )  \,,
\ee
where $\cF_0(\alpha)$ is $m$-independent and the $s_k(\alpha)$'s are calculated explicitly for spins zero, half and one in \S\ref{sec:lowspin}. These constants contain the dependence on $\alpha$ in this limit, and so also encode the dependence of the answer on the boundary conditions of the bulk fields near the branes situated at the poles. Our results reduce in special cases to results in the literature for spheres \cite{CandelasWeinberg,SphereCasimir, KandM}.

The logarithm appearing in eq.~\pref{eq:introVform} is slightly more complicated than what normally arises for loop calculations with tori \cite{GHBQ, OtherTori}, and \S\ref{sec:genloops} shows this ultimately can be traced to the existence of a number of effective interactions involving the curvature (or the background flux), that happen to vanish when evaluated for tori. These effective interactions arise both in the bulk and on the branes, and are renormalized by quantum loops of bulk fields. The resulting running produces the logarithmic coefficient, and because it can be traced to the renormalization of UV divergences this logarithmic running (and the power-law dependence on $m$ that pre-multiplies it) captures the dominant sensitivity to very heavy bulk loops that we seek.

Furthermore, in the special case that the renormalized interactions are in the bulk lagrangian, the coefficients $s_i$ are known for arbitrary geometries using very general Gilkey-de Witt methods \cite{GilkeyDeWitt, GdWrev}, a result we summarize in Appendix \ref{app:gilkeydewitt}. We check that our results reduce to these general results in the appropriate limit: $\alpha \to 1$. A well-known property of the Gilkey-de Witt coefficients is that their contributions to bulk counterterms never depend on boundary conditions. Physically this is because they capture the effects of very short-wavelength modes, and because these see only local properties of the fields they don't `know' about conditions imposed at the boundaries. More precisely, the only UV divergences that directly involve the boundary conditions are those that renormalize the brane action at which the boundary conditions are applied.

For the rugby ball this implies the renormalization of bulk interactions is identical to that for the sphere, and once these universal bulk counter-terms are subtracted we can separately identify how the brane action is renormalized by bulk loops. This exposes how these renormalizations depend on boundary conditions, and how they contribute to the coefficients $s_i$. Again, because short-wavelength modes cannot know about conditions at distant boundaries, our results for the renormalization of brane-bulk interactions are not specific to rugby balls, and apply equally well to {\em any} codimension-2 brane situated within a 6D geometry, and give rise there to a conical singularity. Explicit results for the $s_k$'s generated by loops of low-spin bulk fields are given in \S\ref{sec:lowspin}, where we also see that our results agree (when appropriate) with known divergence calculations for spacetimes with conical singularities \cite{conicaldivergences}.

Finally, \S\ref{sec:4DVE} identifies a subtlety that brane back-reaction introduces when using standard Casimir energy calculations to identify the effective 4D cosmological constant as seen by a low-energy observer on the brane. The quantity $V$ computed above is the standard fare of Casimir energy calculations: it is the (negative of the) loop-corrected 1PI effective lagrangian density evaluated at the background classical geometry (in this case a rugby-ball). And in the absence of branes this quantity is also the effective cosmological constant seen by a 4D observer, at least for systems where the zeroth-order 4D geometry is flat. This is because the 1PI action of the low-energy 4D effective theory has the generic form, $\cL_{\rm eff} / \sqrt{-g} = - \Lambda + \hbox{curvature terms}$, which becomes $- \Lambda$ if evaluated at the classical 4D background (which was assumed to be flat). As \S\ref{sec:4DVE} shows, what changes in this argument once brane back-reaction is included is that it is no longer
sufficient to
evaluate the 1PI action at the uncorrected classical spacetime, even at first order in the loop corrections. This is most clear for systems where back-reaction cancels a classical brane tension, since in this case renormalizations of the tension should also be cancelled in the same way.

In summary, what we present here as new is: an explicit calculation of the divergent part of the Casimir energy obtained from loops of low-spin bulk fields in a 6D geometry compactified on a flux-stabilized 2D rugby ball. We explicitly show how these divergences are renormalized into bulk and brane counter-terms, and compute how the corresponding effective interactions run as a result. Finally, we show how these renormalizations can be matched to the low-energy 4D effective theory, to see how they feed through to the cosmological constant as seen by a low-energy 4D observer.

While we think these calculations can have a variety of applications to loop effects in extra-dimensional spacetimes, our main application is to use them to examine how supersymmetry ameliorates the UV sensitivity of the vacuum energy, as described in a companion paper \cite{Companion}. Brief comments on the relevance to using codimension-2 branes to address the cosmological constant problem are summarized in \S\ref{sec:concl}.

\section{Bulk field theory and background solution}
\label{sec:bulksugra}

We begin by summarizing the field content and dynamics of the bulk field theory of interest: $D$-dimensional matter (with spins zero, half and one) coupled to gravity. We then describe two-dimensional compactifications of this system in the presence of $d = D-2$ dimensional brane sources. We do not work within the probe limit, and so explicitly include the back-reaction of these sources on the bulk geometry. In our explicit one-loop calculations we specialize to the case $D = 6$ and $d=4$.

\subsection*{Field content and action}
\label{S:6DSS}

The fields of interest consist of a metric $g_{\ssM\ssN}$, plus a collection of scalar fields $\phi^i$, gauge potentials $A^a_\ssM$, and spin-half fermions $\psi^r$. We imagine the scalars and spin-half fields to transform under the gauge group, respectively represented on these by hermitian generators ${(t_a)^i}_j$ and ${(T_a)^r}_s$.

The bosonic part of the classical lagrangian for these fields
is, in the Einstein frame:\footnote{Our metric is `mostly plus' and we follow Weinberg's curvature conventions \cite{GandC}, which differ from those of MTW \cite{MTW} only by an overall sign in the definition of the Riemann tensor.}
\be
\label{E:Baction}
    \frac{{\cal L}_\ssB}{\sqrt{- g}} = -\, \frac{1}{2\kappa^2}  R - \frac12 \, \cG_{ij}(\phi)
    D_{\ssM} \phi^i \, D^\ssM \phi^j  - \frac14 \cH_{ab}(\phi) \; F^a_{\ssM\ssN} F^{b\ssM\ssN} -  U(\phi) \,,
\ee
where $D_\ssM$ denote gauge-covariant derivatives for the scalars, $F_{\ssM\ssN}^a$ is the field strength for the gauge potentials, and the functions $\cG_{ij}(\phi)$, $\cH_{ab}(\phi)$ and $U(\phi)$ are to be specified.

\subsection*{Rugby-ball compactifications}
\label{ssec:RugbyBall}

In general some scalars carry gauge charge and so having these be nonzero in the background would give some gauge bosons masses. Since in what follows our main interest is in background configurations for the massless gauge fields (though we do include massive gauge fluctuations about these backgrounds), when solving the classical field equations we assume that all nonzero background scalars do not carry the charges of the nonzero background gauge fields. In this case, the equations of motion which follow from the lagrangian, eq.~\pref{E:Baction}, are:
\eqa \label{E:Beom}
 && \cG_{ij}(\phi) \, \Box \, \phi^j - \frac14 \, \cH_{ab,\,i} (\phi) \; F^a_{\ssM\ssN} F^{b\ssM\ssN} - U_{,\,i}(\phi) = 0 \,,\nn\\
 &&D_\ssM \Bigl( \cH_{ab}(\phi) \, F^{b\ssM\ssN} \Bigr) = 0 \,, \\
 &&R_{\ssM\ssN} + \kappa^2 \cG_{ij}(\phi) D_\ssM\phi^i \, D_\ssN \phi^j + \kappa^2 \cH_{ab}(\phi) \, F^a_{\ssM\ssP} {F^b_\ssN}^\ssP \nn\\
 && \qquad\qquad\qquad
 + \frac{2\kappa^2}{2-D} \,  \left[-  U(\phi) + \frac14 \, \cH_{ab}(\phi) F^a_{\ssP\ssQ} F^{b\ssP\ssQ} \right]\, g_{\ssM\ssN} = 0 \,, \nn
\eeqa
where
\be
 \Box \phi^j := g^{\ssM \ssN} \left[ \partial_\ssM \partial_\ssN \phi^j - \Gamma^\ssP_{\ssM \ssN} \partial_\ssP \phi^j + \gamma^j_{kl}(\phi) \, \partial_\ssM \phi^k \, \partial_\ssN \phi^l \right] \,,
\ee
and $\Gamma^\ssP_{\ssM\ssN}$ and $\gamma^j_{kl}(\phi)$ respectively denote the Christoffel symbols constructed from the spacetime metric, $g_{\ssM \ssN}$, and the target-space metric, $\cG_{ij}(\phi)$.

The simplest compactifications \cite{SS} are found using the Freund-Rubin {\em ansatz} \cite{FR} for which $\phi^i$ is a constant and
\eq \label{E:FRansatz}
    {g}_{\ssM\ssN} = \pmatrix{
    {g}_{\mu\nu}(x) & 0 \cr 0 & {g}_{mn}(y) \cr}
    \qquad \hbox{and} \qquad
    {F}_{\ssM\ssN} = \pmatrix{0 & 0 \cr 0 &
    f\, {\epsilon}_{mn}(y) \cr}  ,
\eeq
where ${g}_{\mu\nu}$ is a maximally-symmetric Lorentzian metric ({\it i.e.} de Sitter, anti-de Sitter or flat space), and ${g}_{mn}$ is the metric on the two-sphere, $S_2$, whose volume form is ${\epsilon}_{mn}$. The quantity $f$ appearing in the background gauge field --- which could be any one of the gauge fields present in the theory --- is a constant. All other fields vanish.

The gauge potential, ${A}_m$, that gives rise to such a field strength, $F_{mn}$, is the potential of a magnetic monopole and so satisfies a flux-quantization condition in the presence of charged matter. We denote the background gauge coupling constant by
\be
 \frac{1}{\tilde g^2} := \cH \,,
\ee
with $\cH$ the component of $\cH_{ab}$ corresponding to the nonzero background flux $F_{mn}$. With this definition, requiring gauge transformations be single-valued for charged matter fields with charge $q \tilde g$ for integer $q$ gives the quantization condition
\be \label{eq:fluxqtzn}
2\pi N = q \int_{S_2} F  = 4 \pi r^2 qf \,,
\ee
where $N = 0, \pm 1, ...$ is an arbitrary integer and $r$ is the sphere's radius, in our conventions satisfying $R_{mn} = - g_{mn}/r^2$. Quantization requires the normalization constant, $f$, to satisfy
\eq \label{E:fquant}
    f = {N \over 2  q {r}^2} \,, \quad (\textrm{without brane sources})
\eeq
for all matter fields in the theory. If all charged fields have the same charge it is conventional to define $\tilde g$ so that $q = 1$. When more than one nonzero charge is present we choose $q = 1 $ for the smallest nonzero charge (say) and then for any second charged field with $q \ne 1$ eq.~\pref{E:fquant} requires there to exist another integer $N$ such that $N/q = N_1$ is also an integer, and so all charges are rational multiples of the smallest one.

With the above ansatz --- and using the identities $F_{mp}{F_n}^p = f^2 g_{mn}$ and $F_{mn} F^{mn} = 2 f^2$ --- the field equations boil down to the following three conditions:
\bea \label{E:redeqs}
 R_{\mu\nu} &=& \frac{\kappa^2}{D-2} \left( \frac{ f^2}{\tilde g^2} - 2 U \right) g_{\mu\nu} = \frac{\kappa^2}{D-2} \left( \frac{ N^2}{4 \,q^2\tilde g^2 r^4} - 2 U \right) g_{\mu\nu} \nn\\
 \frac{1}{r^2} &=& \frac{\kappa^2}{D-2} \left[2U + (D-3)\, \frac{ f^2}{\tilde g^2} \right] = \frac{\kappa^2}{D-2} \left[2U + (D-3)\, \frac{ N^2}{4 \,q^2\tilde g^2 r^4} \right]\\
 \hbox{and}&& \quad \left[ 2 U - \frac{ f^2}{\tilde g^2}\right]_{,\,i} = \left[  2 U - \frac{ N^2}{4 \,q^2 \tilde g^2 r^4} \right]_{,\,i}  = 0 \,. \nn
\eea
With $f$ fixed by flux quantization, eq.~\pref{E:fquant}, these equations can be solved for $\phi^i$ --- and so also $\tilde g(\phi)$ and $U(\phi)$ --- as well as $r$ and the curvature in the $D-2$ directions spanned by the $(\mu\nu)$ coordinates.

\subsubsection*{Brane sources}

The solutions as outlined so far describe an extra-dimensional 2-sphere supported by flux, with metric
\be
 \exd s^2 = r^2 \Bigl( \exd \theta^2 + \sin^2 \theta \, \exd \varphi^2 \Bigr) \,,
\ee
without the need for brane sources \cite{SS}. However a class of solutions with brane sources can be included very simply \cite{Towards}, just by allowing the angular coordinate to be periodic with period $\varphi \simeq \varphi + 2 \pi \alpha$ with $\alpha$ not necessarily equal to unity. Geometrically, this corresponds to removing a wedge from the sphere along two lines of longitude and identifying points on opposite sides of the wedge \cite{conical, Towards}. This introduces a conical singularity at both the north and south poles, with defect angle $\delta = 2\pi (1 - \alpha)$, a geometry called the {\em rugby ball}.\footnote{`Rugby ball' is used rather than `football' to avoid a cultural ambiguity in what the shape of a football is.}

Physically, such a construction corresponds to the introduction of two identical brane sources, one situated at each of the singularities, with Einstein's equations relating the defect angle to the properties of the branes. Concretely, take the action of the brane to be\footnote{A more covariant way of writing the term linear in $F_{mn}$ is as the integral of the Hodge dual, ${}^\star F$, over the $d$-dimensional brane world-sheet \cite{localizedflux}.}
\bea \label{eq:genbraneaction}
 S_b &=& - \int \exd^dx \, \sqrt{-g} \; L_b \nn\\
 \hbox{with} \quad L_b &=& T_b - \frac{ \cA_b}{2 \tilde g^2} \, \epsilon^{mn} F_{mn}
 + \cdots \,,
\eea
%
where the ellipses denote other terms involving two or more derivatives, and the coefficients $T_b$, $\cA_b$, $\cB_b$, $\cC_b$ and so on could depend on the extra-dimensional scalars $\phi^i$. However, the existence of a rugby ball solution does require the derivative with respect to $\phi$ of the total brane Lagrangian to vanish at the background, because the near-brane boundary condition for the bulk scalars requires \cite{uvcaps}
\be
 \lim_{\rho \to 0} \Bigl( \cG_{ij} \, \rho \, \pd_\rho\phi^j \Bigr) = \frac{\kappa^2}{2\pi} \left( \frac{\delta S_b}{\delta\phi^{\,i}} \right) \,,
\ee
where $\rho$ denotes proper distance from the brane.

For conical singularities, the near-brane boundary conditions for the metric imply \cite{Vil, localizedflux, uvcaps} the defect angle at the brane's position is given by
\be
 \delta_b = \kappa^2 L_b \,.\label{deltab-Lb}
\ee
A rugby-ball solution requires identical branes at each pole,\footnote{See \cite{GGP, OtherConical, laterconical} for solutions with conical singularities that can differ at the two poles.} for which
\be
 1 - \alpha = \frac{\kappa^2 L_\pm}{2\pi} = 4 G_6 L_\pm \,,
\ee
where $L_\pm$ is the lagrangian for either of the source branes.

The presence of the brane sources complicates the flux quantization condition in two important ways. The first complication arises because the resulting defect angle changes the volume of the sphere, which appears in the flux-quantization condition when integrating over the bulk magnetic field,
\be
 \int_{S_2(\alpha)} F = 4\pi\alpha \, r^2 f \,.
\ee
The second complication arises because the branes themselves can carry a localized flux, given by
\be
 \Phi_{1b} = \frac{\cA_b}{2\pi} \,.
\ee
For two identical branes the total flux localized in this way is $\Phi_1 := \sum_b \Phi_{1b}$, in terms of which the flux-quantization condition becomes
\be \label{eq:BLFquantization}
 2 \pi N_1 = \frac{2\pi N}{q} = \sum_b \cA_b + \int_{S_2(\alpha)} F
 = 2\pi \, \Phi_1 + 4 \pi \alpha\, r^2 f \,.
\ee
The condition on the normalization constant $f$ is then
\be
f= \frac{\cN}{2 q r^2} \,,\quad (\textrm{with brane sources})
\ee
where
\be
\cN := \omega(N-\Phi)
\ee
and (for later convenience) $\omega := 1/\alpha$, $\Phi := q\Phi_1$ (and also $\Phi_b := q\Phi_{1b}$).

\subsubsection*{Control of approximations}

Since our entire discussion takes place within the semi-classical approximation we must demand all fields vary slowly enough to trust the low-energy effective-field-theory approximation \cite{GREFTrev, JFDEFT} for whatever (possibly a string theory) ultimately provides its ultraviolet completion.

In practice, without knowing the details of this UV completion, we ask fields to vary slowly relative to the length scale, $\ell$, set by the gravitational coupling: $\kappa^2 = \ell^{D-2}$. Since (barring unnatural cancellations) eqs.~\pref{E:redeqs} imply $1/r^2 \sim \kappa^2 V \sim \kappa^2 N^2/\tilde g^2 r^4$, we also see that $r^2 \sim \kappa^2 N^2/\tilde g^2 \sim 1/\kappa^2 V$ and so $r \gg \ell$ also implies $\tilde g^2 \ll N^2 \ell^{D-4}$ and $V \ll \ell^{-D}$.

Finally, once brane sources are included we must also demand them not to curve excessively the background geometry, and for branes with tension $T$ this requires $\kappa^2 T \ll 1$. For rugby-ball geometries this ensures the defect angle satisfies $\delta \ll 2\pi$.

\section{General features of bulk loops}
\label{sec:genloops}

We now turn to the size of one-loop quantum fluctuations about the rugby-ball background just discussed. In particular, we calculate the UV-divergent part of loops computed for the various fields $\phi^i$, $\psi^r$ and $A_\ssM^a$ expanded about this background. Notice for these purposes that we need not restrict ourselves to the loop contributions of fields that are nonzero in the background.

The main assumption we use to compute a field's one-loop contribution to divergences is to suppose that its kinetic operator can be written in the form\footnote{For one-loop purposes this can also be done for fermions by working with the square of $\Dsl$.} $\Delta = - \Box + X + m^2$, for some choice of local quantity $X$ (perhaps a curvature or background flux) and a squared mass, $m^2$. This is sufficiently general to include most of the spin-zero, -half and -one particles of interest in later sections. As we shall argue in more detail below, because we restrict to the UV sensitive part of the calculation our results apply more generally than just to rugby-ball geometries. On the other hand, our calculation is insensitive to effective interactions that cannot be distinguished using only a rugby-ball geometry, such as differences between $R^2$ and $R_{\ssM\ssN} R^{\ssM\ssN}$ interactions (which are indistinguishable for spheres), or those interactions like $\nabla_\ssM R \nabla^\ssM R$ involving gradients
of the curvature (which vanish for spheres).

The other main restriction to our calculations that emerges in subsequent sections is the need to avoid fields whose fluctuations mix nontrivially with those of the metric. These must be avoided because for them it is not straightforward to show that $\Delta$ takes the desired form. In particular this precludes our computing the effects of those fields that are nonzero in the classical background.

Finally, for simplicity our final expressions also specialize to $D=6$ and $d=4$. Although our methods work equally well for any $D$ and $d = D-2$, our explicit evaluations only capture {\em all} one-loop divergences for $D \le 6$. For $D>6$ they give only a subset (the most divergent) of UV divergences.

\subsection*{One-loop calculations}

Writing the generator of 1PI correlators as $\Gamma = S + \Sigma$, then the UV-sensitive part of one loop quantum corrections, $\Sigma$, can be calculated using heat-kernel methods. For a real $D$-dimensional field of mass $m$ we have the 1-loop quantum action\footnote{We return below to how the kinetic operator for higher-spin fields can be put into the form $-\Box + X + m^2$.}
\eq{
    i\Sigma = -i \int \exd^d x \, \Vone
     = -(-1)^\ssF\, \frac{1}{2} \, \hbox{Tr}\;
    \hbox{Log} \,\left( \frac{ -\Box_\ssD + X + m^2}{\mu^2} \right) \,,
    \label{eqn: sigma} }
\eeq
%
where $(-1)^\ssF=+1$  for bosons and $-1$ for fermions, $X$ is a local combination of background fields (such as curvatures and background fluxes) and $\Vone$ denotes the effective 1-loop scalar potential (or Casimir energy density) in $d = D-2$ dimensions. We assume that for the compactification of interest the higher-dimensional d'Alembertian splits into the sum of $d$- and two-dimensional pieces: $\Box_\ssD = \Box_d + \Box_2$.

Anticipating the need to dimensionally regularize ultraviolet divergences we write the spacetime dimension as $d = \dint - 2 \, \varepsilon$, where $\dint$ is an integer and $\varepsilon \to 0$ at the end of the calculation, so that
\eq
    \Vone = (-1)^\ssF\frac12 \, \mu^{2\varepsilon} \sum_{jn} \int \frac{\exd^d k_\ssE}{(2\pi)^d} \, \ln \left( \frac{k_\ssE^2 + m^2 + m_{jn}^2}{\mu^2}  \right)\,,
\eeq
where $m_{jn}^2$ denote the eigenvalues of $-\Box_2 + X$ in the compactified space, and we Wick rotate to Euclidean signature using $\exd^d k = i \, \exd^d k_\ssE$. Using the identity $\ln X = - \int_0^\infty (\exd s/s) \exp(-s X)$ (which is valid up to an infinite constant that is independent of $X$) we then have
\eq
    \Vone =  -(-1)^\ssF\,\frac12 \, \mu^{2\varepsilon} \sum_{jn} \int \frac{\exd^d k_\ssE}{(2\pi)^d} \, \int_0^\infty \frac{\exd s}{s} \, \exp \left[ - s \left(k_\ssE^2 + m^2 + m_{jn}^2  \right) \right]  \,,
\eeq
where $\mu^2$ in the exponential is absorbed into a rescaling of $s$. Performing the $\exd^d k_\ssE$ integral using
\eq
    \int \frac{\exd^d k_\ssE}{(2\pi)^d} \, e^{-sk_\ssE^2} = \left( \frac{1}{4 \pi s} \right)^{d/2} \,,
\eeq
gives
\eqa \label{eq:Vdfirst}
    \Vone &=&  -(-1)^\ssF\,\frac{\mu^{2\varepsilon}}{2(4 \pi)^{d/2}} \sum_{jn} \int_0^\infty \frac{\exd s}{s^{1 + d/2}} \, \exp \left[ - s \left( m^2 + m_{jn}^2  \right) \right] \nn\\
    &=& -\,\frac{\mu^{2\varepsilon}}{2(4 \pi r^2)^{d/2}}  \int_0^\infty \frac{\exd t}{t^{1 + d/2}} \, e^{- t (m r)^2} \, S(t) \,,
\eeqa
where $t = s/r^2$ is dimensionless,
\be
 S(t) := (-1)^\ssF\sum_{jn} \exp \left[ - t \lambda_{jn}  \right] \,,
\ee
and the dimensionless quantities $\lambda_{jn}$ are defined by
\be
 m_{jn}^2 := \frac{\lambda_{jn}}{r^2} \,.
\ee
In the examples of interest $1/r$ is the generic Kaluza-Klein scale for the compactification.

In the appendices, we show that the function $S(t)$ has the following small-$t$ limit:
\be
 S(t) =  \frac{s_{-1}}{t} +
 \frac{s_{-1/2}}{\sqrt t} + s_0 + s_{1/2} \sqrt{t} + s_1 \, t + s_{3/2} t^{3/2} + s_2 \, t^2 + \cO(t^{5/2}) \,,
\ee
where the coefficients $s_i$ depend on the spectra, $\lambda_{jn}$, and so also on the spin of the particle involved, and in principle also on the boundary conditions used near any branes situated within the background geometry. Much of what follows is devoted to computing these coefficients explicitly for the fields and boundary conditions of interest.

Using this small-$t$ expansion in eq.~\pref{eq:Vdfirst} gives
\bea
 \Vone &=& -\frac{\mu^{2\varepsilon}}{2(4 \pi r^2)^{d/2}} \int_0^\infty \exd t \, e^{- t (m r)^2} \, \bigg[ \frac{s_{-1}}{t^{2+d/2}} +
 \frac{s_{-1/2}}{t^{3/2+ d/2}} + \frac{s_0}{t^{1+d/2}} + \frac{s_{1/2}}{t^{1/2+d/2}} \nn\\
 && \qquad\qquad\qquad\qquad\qquad\qquad\qquad   + \frac{s_1}{t^{d/2}} + \frac{s_{3/2}}{t^{-1/2+d/2}} + \frac{s_2}{t^{-1+d/2}} + \cO(t^{-2+d/2}) \bigg] \nn\\
 &=& -\frac{m^d \mu^{2\varepsilon}}{2(4 \pi)^{d/2}}  \sum_{i} s_i \, (mr)^{-2i} \, \Gamma(i-d/2)  \,,
\eea
which shows that the result diverges when $d \to \dint$ for all terms, $s_i$, for which $i - \dint/2$ is a non-positive integer.

At this point we specialize to $\dint = 4$ so that all divergences are captured by $s_i$ with $i \le 2$. For this case we have
\be
 \Vone = \frac{m^d \mu^{2\varepsilon}}{(4 \pi)^{d/2}} \left[ \frac{s_{-1}}{6} (m r)^2 - \frac{s_0}{2} + \frac{s_1}{(mr)^2} - \frac{s_2}{(mr)^4} \right] \Gamma(-d/2) + (\hbox{finite as $d\to4$}) \,.
\ee
Using $x^{4-d} \Gamma(-d/2) = ({4-d})^{-1} + \ln x + \hbox{finite}$, the divergent part of $\Vone$ emerges as
\be \label{eq:Vinfinity}
 \Vone = \frac{\cC}{(4\pi r^2)^2} \left[ \frac1{4-d} + \ln\left(\frac\mu{m}\right)\right] + \cV_f \,,
\ee
where $\cV_f$ is finite and $\mu$-independent in the limit $d \to 4$, and
\be \label{eq:Cdef}
 \cC := \frac{s_{-1}}{6} (m r)^6 - \frac{s_0}{2}(m r)^4 + s_1 (mr)^2 - s_2 \,.
\ee

What is important in what follows is that the coefficient $\cC$ depends on the independent external variables that control the properties of the background geometry, like $r$, $N$, $\Phi_b$ and $\alpha$. Subsequent sections use this dependence to extract more information about the effective interactions that renormalize these divergences.

\subsection*{Renormalization}

Ultraviolet divergences are renormalized, as usual, into counter-terms in the action, and in the setup of interest here this action has both bulk and brane contributions. The goal of the next subsections is to separate each of these types from one another.

\subsubsection*{Bulk counterterms}

The crucial feature of bulk counterterms that allows them to be separated from brane counterterms is their insensitivity to the boundary conditions satisfied by the bulk fields in the vicinity of the branes. This is most easily seen if they are computed using Gilkey-de Witt heat-kernel techniques \cite{GilkeyDeWitt, GdWrev} --- see Appendix \ref{app:gilkeydewitt} --- since this calculation is explicitly boundary-condition independent (for bulk counterterms). Physically this arises because divergences are sensitive only to modes with wavelengths much shorter than the physical size of the spacetime, since this both ensures their effects are captured by local interactions and that they are too short to build correlations between points in the bulk and distant boundaries.

Because they do not depend on the boundary information, bulk counterterms can be computed using only the bulk geometry, without making reference to brane properties. In particular, they may be obtained by specializing the coefficients $s_i$ to the special case where the background geometry has no defect angle: $\alpha = 1$.

The interactions to be renormalized are found by writing the most general local bulk lagrangian consistent with the given field content and symmetries, organized into a derivative expansion: $\cL_\ssB = \cL_{\ssB 0} + \cL_{\ssB 2} + \cL_{\ssB 4} \cdots$. Restricting to terms that are nonzero in the background, this expansion gives
\bea
 \cL_{\ssB 0} &=& - \sqrt{-g} \;  U  \nn\\
 \cL_{\ssB 2} &=&  - \sqrt{-g} \left[ \frac{1}{2 \kappa^2} \, R + \frac{1}{4\tilde g^2} \, F_{\ssM \ssN} F^{\ssM \ssN} \right] \\
 \cL_{\ssB 4} &=& - \sqrt{-g} \left[ \frac{ \kappa \zeta_{\ssA\ssR}}{8 \tilde g^2} \,  R \, F_{\ssM \ssN} F^{\ssM \ssN} + \frac{\zeta_{\ssR^2}}{\kappa} \, \ol R^2 \right]  \\
 \cL_{\ssB 6} &=& - \sqrt{-g} \bigg[ \zeta_{\ssR^3}\, \ol R^3 + \cdots \bigg]
\eea
and so on. Here we define
\be \label{Rsqol}
 \ol R^2 = a_\ssR\, R^2 + 2 b_\ssR\, R_{\ssM \ssN} R^{\ssM \ssN} + c_\ssR\, R_{\ssM \ssN \ssP \ssQ} R^{\ssM \ssN \ssP \ssQ} 
 \,,
\ee
with $a_\ssR + b_\ssR + c_\ssR = 1$ so that $\ol R^2 = R^2$ when specialized to a spherical geometry (for which $R_{mnpq} R^{mnpq} = 2 R_{mn} R^{mn} = R^2 = 4/r^4$). A similar, but more elaborate, definition is used for $\ol R^3$. Calculations on a sphere can only track the overall renormalization $\zeta_{\ssR^2}$, $\zeta_{\ssR^3}$ and not how the separate parameters such as $a_\ssR$, $b_\ssR$ and $c_\ssR$ renormalize, but --- as summarized in Appendix \ref{app:gilkeydewitt} --- for the bulk contributions these separate renormalizations are known explicitly for general geometries from earlier work \cite{GdWrev}. A similar story also holds for terms that involve gradients of the background scalar fields, which vanish for the rugby-ball configurations. Since this includes in particular kinetic terms like $(\partial \phi)^2$, it represents an obstruction to computing wave-function renormalizations for the scalars. This limits the generality of our later formulae for the renormalizations of scalar couplings.

Evaluating the bulk action at the background rugby-ball solution and integrating over the compact directions gives
\bea \label{VbulkAtEom}
 \cV_\ssB = - \int \exd^2 x \, \cL_\ssB &=& \left( 4 \pi \alpha \, r^2 \right) \left\{ U - \frac{1}{\kappa^2 r^2} + \frac{f^2}{2 \tilde g^2} \left[ 1 - \frac{\kappa \zeta_{\ssA\ssR} }{r^2} \right]  + \frac{4 \zeta_{\ssR^2}}{\kappa \,r^4}  - \frac{8\zeta_{\ssR^3}}{r^6} + \cdots \right\} \\
 &=& \left( 4 \pi \alpha \, r^2 \right) \left\{ U - \frac{1}{\kappa^2 r^2} + \frac{\cN^2}{8 \,q^2 \tilde g^2 r^4} \left[ 1 - \frac{\kappa \zeta_{\ssA\ssR} }{r^2} \right]  + \frac{4 \zeta_{\ssR^2}}{\kappa \,r^4} - \frac{8\zeta_{\ssR^3}}{r^6} +\cdots \right\} \,, \nn
\eea
showing that $U$, $1/\kappa^2$, $\zeta_{\ssR^2}$, and $\zeta_{\ssR^3}$ terms can be read off respectively from the $r^2$, $r^0$, $r^{-2}$, and $r^{-4}$ terms in $\cV_\ssB$, while the $1/\tilde g^2$ and $\zeta_{\ssA\ssR}$ terms are identified as the coefficients of $\cN^2/r^2$ and $\cN^2/r^4$, respectively. In particular, the power of $\cN$ acts as a proxy for the power of $f$, and so does not appear at all if the particle in the loop does not carry the charge gauged by the background gauge field.

The divergences in $\Vone$ are absorbed by splitting all couplings -- {\em i.e.} $U$, $1/\kappa^2$ {\em etc}. -- into a renormalized and counter-term part, with the infinities of $\Vone$ that arise in the bulk canceling the divergences in the bulk counter-terms as $d \to 4$. Once this is done the $\mu$-dependence of the renormalized quantities also must cancel the explicit $\mu$-dependence in the corresponding finite parts of $\Vone$. And the bulk part of the divergences in $s_i$ can be identified because they do not depend on the brane boundary conditions, and so are the same as they would have been for a calculation on a sphere. Explicitly, if we denote by $s^{\rm sph}_i$ what the coefficients $s_i$ would have been if evaluated on a sphere ({\em i.e.} with $\alpha = 1$ and $\Phi_b = 0$), then the full result for $s_i$ can be written
\be \label{eq:deltasidef}
 s_i := \alpha \, s_i^{\rm sph} + \delta s^{\rm tot}_i \,,
\ee
which can be regarded as the definition of $\delta s^{\rm tot}_i$. (The factor of $\alpha$ pre-multiplying $s_i^{\rm sph}$ arises from the integration over the extra dimensions, as in eq.~\pref{VbulkAtEom}.) This split is useful because only the divergences in the first term, $\alpha s_i^{\rm sph}$, can be absorbed into renormalizations of the bulk interactions of $\cV_\ssB$, while those of $\delta s_i^{\rm tot}$ must be absorbed into brane interactions.

With this definition the running of the renormalized bulk couplings satisfy
\bea
 \mu \, \frac{\partial U}{\partial \mu} = -\frac{m^6}{6 (4\pi)^3} \; s_{-1}^{\rm sph,\, 0} \,,&&\quad
 \mu \, \frac{\partial}{\partial \mu} \left( \frac{1}{\kappa^2} \right) = -\frac{ m^4}{2 (4\pi)^3} \; s_{0}^{\rm sph,\,0} \,,\\
 \mu\,\frac{\partial }{\partial \mu}\left(\frac{\zeta_{\ssR^2}}\kappa\right) = -\frac{m^2}{4(4\pi)^3} \; s_1^{\rm sph,\,0} \,,&&\quad
\quad  \mu\,\frac{\partial \zeta_{\ssR^3}}{\partial\mu} = -\frac{1}{8(4\pi)^3} \;s_2^{\rm sph,\,0} \,.
\eea
and so on. Here the quantities\footnote{Notationally, for $s_i^{{\rm sph},\,k}$ the ``sph'' emphasizes that these quantities are evaluated on the sphere, and the superscript `$k$' denotes terms involving $k$ powers of the gauge-field normalization, $\cN$.} $s_{-1}^{\rm sph,\,0}$, $s_0^{\rm sph,\,0}$, $s_1^{\rm sph,\,0}$ and $s_2^{\rm sph,\,0}$ are computed below by explicitly summing over the KK spectrum on a sphere, giving results that agree with those obtained in Appendix \ref{app:gilkeydewitt} using general heat-kernel methods. We note at this point that since we derive the running of the couplings by assuming they cancel only the explicit $\ln(\mu/m)$ dependence of $\Vone$, our explicit formulae exclude the case where additional $\mu$-dependence enters through the appearance of renormalized couplings and fields pre-multiplying the pole in $1/(d-4)$ (though the formulae are easily generalized to include this more general case).

The renormalization of the gauge-field terms, $1/\tilde g^2$ and $\zeta_{\ssA\ssR}$, is similarly done by keeping track of those divergences involving $f$. This can be done, for example by comparing loops of particles that couple directly to the background flux with those that do not. The result is
\be
 \mu \, \frac{\partial}{\partial \mu} \left(\frac{1}{\tilde g^2}\right) = -\frac{2\, m^2}{(4\pi)^3 r^4 f^2} \; s_{1}^{\rm sph,\,2} = -\frac{8\, q^2 m^2}{(4\pi)^3 \cN^2} \; s_{1}^{\rm sph,\,2} \,,
\ee
and
\be
 \quad \mu \, \frac{\partial}{\partial \mu} \left(\frac{\kappa \zeta_{\ssA\ssR}}{\tilde g^2}\right) = -\frac{2}{(4\pi)^3 r^4 f^2} \; s_{2}^{\rm sph,\,2} = -\frac{8 \, q^2}{(4\pi)^3 \cN^2} \; s_{2}^{\rm sph,\,2}
\ee
where the particle in the loop has charge $q \tilde g$, and $s_i^{\rm sph, \,2}$ represents that part of $s_i^{\rm sph}$ that is proportional to $\cN^2$.

\subsubsection*{Brane counterterms}

A similar reasoning can be applied to brane-localized interactions, which (unlike the bulk counterterms) can depend on the boundary conditions used near the brane but should be independent of those boundary conditions imposed on distant branes. (For earlier treatments of divergences in the presence of conical singularities, see \cite{conicaldivergences}.)

To identify the brane contributions we first subtract the contributions of the universal bulk counterterms found above, using eq.~\pref{eq:deltasidef}, and use $\delta s^{\rm tot}_i = \sum_b \delta s_{i(b)}$ to extract how each individual brane-localized interaction renormalizes. This can be done as before, by distinguishing those contributions that come from couplings to the background gauge field from those that do not.

An additional complication in the case of brane counterterms is the necessity to disentangle which contributions come from which branes. This complication arises because although the KK spectra encountered below depend explicitly the flux, $\Phi_b$, localized on each brane, they only depend on the brane tensions through the common defect angle $\alpha = 1/\omega = 1 - \delta/2\pi$. We write
\be \label{deltastot}
 \delta s^{\rm tot}_i = \delta s^{\rm same}_i + \sum_b \delta s^{\rm diff}_{i(b)} \,,
\ee
where $\delta s^{\rm same}_i$ receives equal contributions from each brane and where the $\delta s^{\rm diff}_{i(b)}$ depends on the explicit $\Phi_b$ that in general differ on each brane. Since we are interested in tracking the renormalization of each brane separately, the quantity of interest is
\be \label{deltasdef}
 \delta s_{i(b)} := \frac{\delta s^{\rm same}_i}2 + \delta s^{\rm diff}_{i(b)} \,,
\ee
since this is the one for which we can write $\delta s^{\rm tot}_i = \sum_b \delta s_{i(b)}$. To avoid any extra notational clutter, we choose to drop the subscript $(b)$ when writing $\delta s_{i(b)}$ in what follows.

Writing the most general local brane lagrangian in a derivative expansion:  $\cL_b = \cL_{b0} + \cL_{b1} + \cL_{b2} + \cL_{b3} + \cdots$, and dropping terms that vanish when evaluated at the rugby ball background, we have
\be
 \cL_{b 0} =  - \sqrt{-\gamma} \;  T_b
\ee
\be
 \cL_{b 1} =  \sqrt{-\gamma} \left[ \frac{ \cA_b }{2\tilde g^2} \, \epsilon^{mn} F_{mn} \right]
\ee
\be
 \cL_{b 2} =  - \sqrt{-\gamma} \left[ \frac{
 \zeta_{\ssR \,b} }{\kappa} \, R + \frac{\kappa \zeta_{\ssA b}}{4 \tilde g^2} \, F_{\ssM \ssN} F^{\ssM \ssN} \right]
\ee
\be
 \cL_{b 3} = \sqrt{- \gamma} \left[ \frac{\kappa \zeta_{\tilde{\ssA} \ssR \, b}}{2 \tilde g^2} \,  R \, \epsilon^{mn} F_{mn} \right] \,,
\ee
\be
 \cL_{b 4} = - \sqrt{-\gamma} \left[ \zeta_{\ssR^2 b} \, \ol R^2 + \frac{\kappa^2 \, \zeta_{\ssA \ssR \,b}}{8 \tilde g^2} \, R \, F_{\ssM \ssN} F^{\ssM \ssN} \right]  \,,
\ee
and so on, where $\gamma_{\mu\nu} := g_{\ssM \ssN} \partial_\mu x^\ssM \partial_\nu x^\ssN$ is the induced metric on the brane and the $\epsilon^{mn} F_{mn}$ terms arise covariantly as the integral of the Hodge dual, ${}^\star F$, over the brane world-volume.

Evaluating these at the background solution gives a contribution from each brane of size
\bea
 \cV_b &=&  T_b - \frac{\cA_b f}{\tilde g^2} - \frac{2 \zeta_{\ssR \, b}}{\kappa \,r^2} + \frac{\kappa \zeta_{\ssA b} f^2}{2 \tilde g^2} + \frac{2 \, \kappa \zeta_{\tilde{\ssA}\ssR \, b} f}{\tilde g^2 r^2} + \frac{4 \zeta_{\ssR^2 b}}{r^4} - \frac{\kappa^2\zeta_{\ssA \ssR \, b} f^2}{2 \, \tilde g^2 r^2} + \cdots \nn\\
 &=&  T_b - \frac{\cA_b \cN}{2 \, q \tilde g^2 \,r^2} - \frac{2 \zeta_{\ssR \, b}}{\kappa \,r^2} + \frac{\kappa \zeta_{\ssA b} \cN^2}{8 \, q^2 \tilde g^2 r^4} + \frac{ \kappa \zeta_{\tilde{\ssA}\ssR \, b} \cN}{q \tilde g^2 r^4} + \frac{4 \zeta_{\ssR^2 b}}{r^4} - \frac{\kappa^2 \zeta_{\ssA \ssR \, b} \cN^2}{8 \,  q^2\tilde g^2 r^6} + \cdots \,,
\eea
to the 1PI potential, with the sum over branes giving $\cV^{\rm tot} := \sum_b \cV_b$. Requiring the counter-term parts of the sum of the two brane actions to cancel the remaining divergences implies the renormalized quantities satisfy
\bea
 \mu \, \frac{\partial \, T_b}{\partial \mu} = \frac{m^4}{2(4\pi)^2} \; \delta s^0_{0}\,,\quad&& \mu\,\frac{\partial}{\partial \mu} \left( \frac{\cA_b}{\tilde g^2}\right) =  \frac{2\,q m^2}{(4\pi)^2 \cN} \; \delta s^{1}_1 \,, \nn\\
 \mu\,\frac{\partial}{\partial \mu} \left( \frac{\zeta_{\ssR b}}{\kappa}\right) =  \frac{m^2}{2(4\pi)^2} \; \delta s^0_1 \,,\quad&& \mu\,\frac{\partial}{\partial \mu} \left( \frac{\kappa \zeta_{\tilde{\ssA} \ssR \, b}}{\tilde g^2}\right) = \frac{q}{(4\pi)^2 \cN} \; \delta s^{1}_{2} \,, \\
 \mu\,\frac{\partial \zeta_{\ssR^2 b}}{\partial \mu} = \frac1{4(4\pi)^2}\; \delta s_2^0 \,,\quad&& \mu\,\frac{\partial}{\partial \mu} \left( \frac{\kappa \zeta_{\ssA b}}{\tilde g^2}\right) =  \frac{8\, q^2}{(4\pi)^2 \cN^2} \; \delta s^{2}_{2} \,,\nn
\eea
where both branes are assumed to be identical, $\delta s_i$ is the quantity defined by eqs.~\pref{eq:deltasidef}, \pref{deltastot} and \pref{deltasdef}, for which the superscripts `0', `1' and `2' respectively correspond (as above) to the terms independent of, linear in, or quadratic in the background gauge field (or its proxy, $\cN$).

What remains is to compute the coefficients $s_{-1}$ through $s_2$ for the fields of interest. This is the aim of the next sections.

\subsection*{KK mode sums}

This section now sketches how the coefficients $s_i$ are computed, by performing the sum defining $S(t)$ using a hypothetical eigenvalue spectrum, $m_{jn}$, that is general enough to include most of the special cases of practical interest.

We can obtain our later results by performing the sum over the mode labels, $n$ and $j$, in either of two different ways. The most reliable (and more cumbersome) method first performs a Poisson resummation, which has the advantage of casting the sums in a way that converges more quickly for small $t$. The second (and easier) method avoids the complications of Poisson resummation, instead using zeta-function regularization to regularize the part of the mode sum that is non-singular as $t \to 0$. We present both methods because although the zeta-function technique is much simpler to use, its validity ultimately relies on the more complicated calculation based on Poisson resummation. More details on the equivalence of these two techniques can be found in Appendix \ref{app:usefulsums}.

To explain these two techniques, consider the following expression for the KK spectrum, $\lambda_{jn}$, that is general enough to include many of the cases met in later sections,
\be \label{eq:GenLambda}
 \lambda_{jn} = \left(j + \frac{\omega}{2} \left| n + b_+ \right| + \frac{\omega}{2} \left| n - b_- \right| + a \right)^2 - \tau \,.
\ee
Here $n$ is an arbitrary integer and $j = 0,1,2,\dots$ is a whole number. The quantities $a$, $b_\pm$ and $\tau$ are real parameters that differ for different spin fields in the loop and for different background flux quantum number, $N$. For example, in the case of a complex scalar field that is charged under the background flux we show in the Appendix \ref{app:SpecNSum} that the appropriate choices are $a = \frac12$, $b_+ = |N|$, $b_- = 0$ and $\tau = \frac14$, where $N$ is the background flux quantum.\footnote{These expressions work in a particular patch for the background monopole gauge potential, with $b_\pm$ interchanged in the opposite patch.}

Our interest is in tracking how the sum, $S(t)$, defined using this spectrum depends on the geometrical quantities $N$ and $r$, as well as the rugby-ball defect angle that is encoded in the quantity
\be
 \omega = \frac{1}{\alpha} \quad \hbox{where} \quad \alpha = 1 - \frac{\delta}{2\pi} \,,
\ee
and $\delta$ is the rugby-ball defect angle described in \S\ref{sec:bulksugra}. In particular, the limiting case $\omega = 1$ corresponds to the sphere, for which a variety of results are known \cite{CandelasWeinberg,SphereCasimir, KandM, Lutken:1987pc} for the Casimir energy. For instance, in this limit and with no gauge flux ($N = 0$) the scalar spectrum becomes $\lambda_{jn} = \ell (\ell + 1)$, where $\ell = j + |n|$. In this case the sums can be re-ordered to give the usual form for scalars on a sphere:
\be
 \sum_{n = - \infty}^\infty \sum_{j = 0}^\infty = \sum_{n = -\infty}^\infty \sum_{\ell = |n|}^\infty
 =  \sum_{\ell = 0}^\infty \sum_{n = -\ell}^\ell  \,.
\ee

\subsubsection*{Poisson resummation technique}
\label{Poisson}

We first sketch the calculation that best controls the convergence of the sums at intermediate steps. For clarity of explanation we do not work with the most general form for the spectrum, but specialize to the following special case
\be
 S(\omega,t) = e^{\tau t} \sum_{j=0}^\infty \sum_{n=-\infty}^\infty \exp \left[-t(j+\omega |n| +a)^2\right] \,.
\ee
The difficulty with this sum is that it converges poorly in the regime of interest: where $t$ is small.

To remedy this we use the Poisson resummation formula, which relates the sum over a function $f(x)$ to the sum over its Fourier transform $\cF(q)$. That is, if
\be
 \cF(q) =  \int_{-\infty}^\infty \exd x \; f(x) \, e^{-i q x} \,,
\ee
then
\be
 \sum_{n=-\infty}^\infty f(n) = \sum_{k=-\infty}^\infty \cF(2\pi k) \,.
\ee
To apply this in the case of interest define
\be
 f_j(x) := \exp \left[-t(j+\omega |x| +a)^2\right]
\ee
and Poisson resum the $n$-sum, giving $S(\omega, t)$ as
\be
 S(\omega, t) = e^{\tau t} \sum_{j=0}^\infty \sum_{k=-\infty}^\infty \cF_j(2\pi k)
\ee
where
\be \label{cF}
 \cF_j(q) = \frac{1}{2\omega} \sqrt{\frac{\pi}{t}}\, e^{-\bar q^2} \left[ e^{2i\bar j \bar q} \bigg( 1 - \erf(\bar j +i \bar q) \bigg) + c.c. \right] \,,
\ee
and
\be
\bar j:= \sqrt{t} (j+a) \,,\quad \bar q := \frac{q}{2\,\omega\sqrt{t}} \,.
\ee

Now comes the important observation, that is justified in some detail in Appendix \ref{app:honestcalc}. Because of the factor $e^{-\bar q^2}$ and the inverse power of $\omega \sqrt t$ appearing in $\bar q$, all of the terms with $k \ne 0$ in the sum are `regular', in the sense that their sum vanishes in the limit that either $\omega$ or $t$ vanishes. Only the $k = 0$ term contributes to the singular part of the small-$t$ limit of $S(\omega, t)$. That is,
\be \label{eq:singregsplit}
 S(\omega, t) = S^{\mathrm{sing}}(\omega, t) + S^{\mathrm{ reg}}(\omega, t) \,,
\ee
with
\be \label{eq:SsingDef}
 S^{\mathrm{sing}}(\omega, t) := e^{\tau t} \sum_{j = 0}^\infty \cF_j(0) =   \frac{e^{\tau t}}{2\,\omega} \sqrt{\frac{\pi}{t}} \sum_{j = 0}^\infty \left[ \bigg( 1 - \erf(\bar j) \bigg) + c.c. \right] \,,
\ee
and
\be
 S^{\mathrm{reg}}(\omega, t) := 2\, e^{\tau t} \sum_{k=1}^\infty \sum_{j = 0}^\infty \cF_j(2 \pi k) \,.
\ee
This last equality uses $\cF_j(-q) = \cF_j(q)$, which follows from $f_j(-x) = f_j(x)$.

Appendix \ref{app:honestcalc} evaluates the remaining sums explicitly in the small-$t$ limit, giving results that agree with the somewhat simpler techniques we now describe.

\subsubsection*{A simpler zeta-function method}

A somewhat simpler way to compute the small-$t$ limit of $S(\omega, t)$ is to start with eqs.~\pref{eq:singregsplit} and eq.~\pref{eq:SsingDef}, but not to perform the Poisson resummation for $S^{\mathrm{reg}}(\omega, t)$. The regular part is instead computed by zeta-function regularizing the initial sum over $n$ and $j$.

To see how this works use the Euler-Maclaurin formula, which states that for any analytic function $f(x)$,
\be \label{EulerMac}
 \sum_{j=0}^\infty f_n(j) = \int_0^\infty \! \exd x\, f_n(x) - \sum_{i=1}^\infty \frac{B_i}{i!} f_n^{(i-1)}(0) \,,
\ee
where $B_i$ denote the Bernoulli numbers (of which the relevant ones are $B_1 = - \frac12$, $B_2 = \frac16$, $B_3 = 0$, $B_4 = - \frac{1}{30}$, and $B_5=0$) and where $f^{(i-1)}(0)$ denotes the $(i-1)$-th derivative of $f(x)$ with respect to $x$, evaluated at $x=0$.

Applying this formula to the $j$-sums in $S(\omega, t)$, requires using the function
\be
 f_n(x) = \exp\left[ -t (x+a_n)^2 \right]
 \quad \hbox{with} \quad a_n = \omega n + a \,,
\ee
and so
\be
 \cS_n(\omega, t) := \sum_{j=0}^\infty f_n(j) = \frac{1}{2}\sqrt{\frac{\pi}{t}}  \left( 1- \erf(a_n \sqrt{t})\right) - \sum_{i=1}^\infty \frac{B_i}{i!} f^{(i-1)}(0) \,,
\ee
where the first few $f_n^{(i-1)}(0)$ terms are
\bea
 f_n^{(0)}(0) &=& e^{-t a_n^2} \,,\quad f_n^{(1)}(0) = -2ta_n \, e^{-t a_n^2} \,,\quad f_n^{(2)}(0) = (-2t + 4 t^2 a_n^2) \, e^{-t a_n^2} \nn\\
 f_n^{(3)}(0) &=& (12 \, t^2 a_n -8 \,t^3 a_n^3 ) \, e^{-t a_n^2} \,,\quad f_n^{(4)}(0) = (12\, t^2-48 \, t^3 a_n^2+16 \, t^4a_n^4) \, e^{-t a_n^2}
\eea
and $f_n^{(i)}(0) \sim \cO(t^3)$ for all $i \ge 5$.

Using this in
\be
 S^{\mathrm{reg}}(\omega,t) = e^{\tau t} \left[ \cS_0(\omega, t) + 2 \sum_{n=1}^\infty \cS_n(\omega, t) \right] \,,
\ee
and Taylor expanding the error function gives a series expression for $S^{\mathrm{reg}}(\omega, t)$ that involves divergent sums of the form $\sum_{n=1}^\infty n^k$ with $k$ a non-negative integer. Remarkably, defining these as $\zeta_\ssR(-k)$, where $\zeta_\ssR(s)$ is Riemann's zeta-function, gives a finite expression, which agrees with $S^{\mathrm{reg}}(\omega, t)$ as computed using Poisson resummation.

To this must be added $S^{\mathrm{sing}}(\omega, t)$, computed using eq.~\pref{eq:SsingDef}. Once this is done the resulting small-$t$ expansion for $S(\omega, t)$ can be compared with previous calculations of the small-$t$ limit using Gilkey-de Witt heat-kernel expansions, when these are known. They are known in particular for the sphere, where $\omega = 1$, and Appendix \ref{app:gilkeydewitt} shows that they agree in this limit.

\section{Results for low-spin bulk fields}
\label{sec:lowspin}

We next collect results for the coefficients $s_i$ of the small-$t$ limit for bulk fields with spins zero, half and one.

\subsection{Scalars}
\label{ssec:Scalars}

Consider first the simplest case of a single minimally coupled real scalar field, satisfying $\Box \phi = 0$, that is coupled to the background gauge field with monopole number $N$ and brane--localized fluxes $\Phi_b$. In this case the scalar spectrum (as derived in Appendix \ref{app:SpecNSum}, in the case where the north patch\footnote{As Appendix \ref{app:SpecNSum} also shows, an equivalent result is obtained if the south patch is instead used.} of the gauge potential is used) is given by
\be \label{eq:simplescalarspec}
 \lambda^{\rm s}_{jn} = \left(j + \frac{\omega}2 |n - \Phi_+|+\frac{\omega}2 |n-N+\Phi_-|+\frac12\right)^2 - \frac{1 +\cN^2}4 \,,
\ee
where the superscript `s' is meant to emphasize that this (and later formulae in this section) applies only for scalars. The Casimir sum becomes
\be
 S_{\rm s}(\omega, t) = e^{t(1+\cN^2)/4} \sum_{j = 0}^\infty \sum_{n = -\infty}^\infty \exp\left[-t\left(j+\frac{\omega}2 |n - \Phi_+|+\frac{\omega}2 |n-N+\Phi_-|+\frac12\right)^2\right] \,.
\ee
Using the results of Appendix \ref{app:SpecNSum}, and its notation
\be 
F_b:=|\Phi_b|\left(1-|\Phi_b|\right)\,,\quad F^{(n)}:= \sum_b F_b^n \,,\quad F^{(1)}:=F \,, \quad G(x):=(1-x)(1-2x) \,,
\ee
we find the following small-$t$ coefficients:
\bea \label{eq:simplescalars1}
s^{\rm s}_{-1} &=& \frac1\omega \,,\\
s^{\rm s}_0(\omega,N,\Phi_b) &=& \frac1\omega \left[ \frac16 + \frac{\omega^2}6(1-3F) \right] \,, \\
s^{\rm s}_1(\omega,N,\Phi_b) &=& \frac1\omega\Bigg[ \frac1{180} - \frac{\cN^2}{24} + \frac{\omega^2}{18}(1 - 3 F) -\frac{\omega^3\cN}{12} \sum_b \Phi_b \, G(|\Phi_b|) + \frac{\omega^4}{180} (1 -15F^{(2)}) \Bigg] ,\qquad \\
s^{\rm s}_2(\omega,N,\Phi_b) &=& \frac1\omega \Bigg[ -\frac1{504} - \frac{11\,\cN^2}{720} + \left(\frac1{90} -\frac{\cN^2}{144} \right) (1-3F) \omega^2 -\frac{\omega^3\cN}{24} \sum_b \Phi_b \, G(|\Phi_b|)  \nn\\
&& \qquad  + \frac{\omega^4(1-\cN^2)}{360}(1 - 15F^{(2)}) - \frac{\omega^5\cN}{120}\sum_b \Phi_b \,G(|\Phi_b|)(1+3F_b) \label{eq:simplescalars2} \\
&&\qquad   + \Bigg(\frac1{1260} -\frac{F^{(2)}}{120} - \frac{F^{(3)}}{60} \Bigg)\omega^6 \Bigg]  \,.\nn
\eea
When $\omega = 1$ and $\Phi_b=0$, these become
\bea
 s_{-1}^{\rm sph} = 1 \,, \quad
 s_{0}^{\rm sph} &=& \frac13 \,, \quad
 s_{1}^{\rm sph,\,0} = \frac{1}{15} \,, \quad
 s_{1}^{\rm sph, \,2} = -\frac{\cN^2}{24} \,, \\
 \quad s_{2}^{\rm sph,\,0} &=& \frac{4}{315} \,,
  \quad \hbox{and} \quad s_{2}^{\rm sph,\, 2} = -\frac{\cN^2}{40} \nn
\eea
in agreement with the result in \cite{KandM}, as well as with the Gilkey-de Witt coefficients as computed for a 6D scalar on a sphere in Appendix \ref{app:gilkeydewitt}, using the general results found in \cite{GdWrev}. If the scalar couples to the background field with strength $q \tilde g$, its contribution to the running of the leading bulk counterterms therefore is
\bea
 \mu \, \frac{\partial U}{\partial \mu} = -\frac{m^6}{6 (4 \pi)^3} \,,\quad&&
 \mu \, \frac{\partial}{\partial \mu} \left( \frac{1}{\kappa^2} \right) = -\frac{m^4}{6 (4\pi)^3} \,,\nn\\
 \mu\,\frac{\partial }{\partial\mu}\left(\frac{\zeta_{\ssR^2}}\kappa\right) = -\frac{m^2}{60(4\pi)^3} \,,\quad && \mu \, \frac{\partial \zeta_{\ssR^3}}{\partial\mu} = -\frac1{630(4\pi)^3}  \,,\\
 \mu \, \frac{\partial}{\partial\mu}\left(\frac{1}{\tilde g^2}\right) = \frac{2\,q^2 m^2}{3(4\pi)^3} \,,\quad && \mu \, \frac{\partial}{\partial \mu} \left(\frac{\kappa \zeta_{\ssA\ssR}}{\tilde g^2}\right) = \frac{2\,q^2}{5(4\pi)^3} \,.\nn
\eea
It is straightforward to check that the above expression for the loop component of the renormalization of the gauge coupling agrees with the result obtained directly by evaluating the Feynman graphs for the vacuum polarization in 6D flat space, following standard methods \cite{WeinbergVol1}, despite its initially unfamiliar sign.

Returning to general $\omega$ and $\Phi_b$, to identify the brane-localized renormalizations we must first subtract the contribution of the bulk counterterms to obtain $\delta s_i$. As described earlier, because these counterterms do not depend on the boundary conditions at the branes for $\omega \ne 1$ their contribution to $V$ contains only the trivial proportionality to $1/\omega = \alpha$ due to the volume, $4\pi \alpha \,r^2$, that appears when $V$ is computed by integrating the counterterms over the rugby ball. This is consistent with the overall factor of $1/\omega$ that pre-multiplies all of the $s_i$ in eqs.~[\ref{eq:simplescalars1}--\ref{eq:simplescalars2}]. We then identify $\delta s_i$ as prescribed by the discussion surrounding eq.~\pref{deltasdef} and find that $\delta s_{-1} = 0$,
\bea
 \delta s_0 &=&  \frac{\omega^2 - 1}{12\, \omega} -\frac{\omega F_b}2 = \frac1{\omega} \left( \frac{\delta\omega}6 + \frac{\delta\omega^2}{12} - \frac{\omega^2 F_b}2 \right) \simeq \frac{\delta\omega}6 - \frac{|\Phi_b|}2 \,,\\
 \delta s^0_1 &=& \frac1{\omega}\left(\frac{\omega^2-1}{36} + \frac{\omega^4-1}{360}- \frac{\omega^2 F_b}6 -\frac{\omega^4 F_b^2}{12}\right)  \nn\\
 &=& \frac1\omega\left(\frac{\delta\omega}{15} + \frac{2\,\delta\omega^2}{45} + \frac{\delta\omega^3}{90} + \frac{\delta\omega^4}{360} - \frac{\omega^2 F_b}6 - \frac{\omega^4 F^2_b}{12} \right) \simeq \frac{\delta\omega}{15} - \frac{|\Phi_b|}6 \,,\\
 \delta s^{\rm 1}_1 &=& -\frac{\omega^2 \cN}{12} \Phi_b \, G(|\Phi_b|) \simeq -\frac{\cN\Phi_b}{12} \,,\\
 \delta s^{\rm 2}_1 &=& s^{\rm 2}_1 -\left(- \frac{\cN^2}{24\,\omega}\right) = 0 \,,\\
 \delta s^0_2 &=& \frac1\omega\Bigg[ \frac{\omega^2-1}{180} + \frac{\omega^4-1}{720} + \frac{\omega^6-1}{2520} -\omega^2\bigg(\frac{F_b}{30} +\frac{\omega^2F_b^2}{24} +\frac{\omega^4 F^2_b}{120} + \frac{\omega^4 F^3_b}{60}\bigg)\Bigg] \nn\\
 &=& \frac1\omega\Bigg[ \frac{2\,\delta\omega}{105} + \frac{5\,\delta\omega^2}{252} + \frac{17\,\delta\omega^3}{1260} + \frac{37\,\delta\omega^4}{5040} + \frac{\delta\omega^5}{420} + \frac{\delta\omega^6}{2520} \nn\\
 &&\qquad -\omega^2\bigg(\frac{F_b}{30} +\frac{\omega^2F_b^2}{24} +\frac{\omega^4 F^2_b}{120} + \frac{\omega^4 F^3_b}{60}\bigg)\Bigg]  \simeq \frac{2\,\delta\omega}{105} - \frac{|\Phi_b|}{30} \,,\\
 \delta s^{\rm 1}_2 &=& -\frac{\omega^2 \cN}{24}  \Phi_b \, G(|\Phi_b|) -\frac{\omega^4 \cN}{120}  \Phi_b \, G(|\Phi_b|) (1+3F_b) \simeq - \frac{\cN\Phi_b}{20} \,,\\
 \delta s^{\rm 2}_2 &=&  -\frac{\cN^2}\omega \left( \frac{\omega^2-1}{288} + \frac{\omega^4-1}{720} - \frac{\omega^2 F_b}{48} -\frac{\omega^4 F_b^2}{24} \right) \nn\\
 &=& -\frac{\cN^2}\omega \left( \frac{\delta\omega}{80} + \frac{17\,\delta\omega^2}{1440} + \frac{\delta\omega^3}{180} + \frac{\delta\omega^4}{720} - \frac{\omega^2 F_b}{48} -\frac{\omega^4 F_b^2}{24} \right) \simeq - \cN^2 \left(\frac{\delta\omega}{80} - \frac{|\Phi_b|}{48} \right) .\quad
\eea
%
Here $\delta\omega:=\omega-1$, and the last, approximate, equalities give the leading dependence in the limit where $\delta\omega \ll 1$ and $|\Phi_b| \ll 1$. The corresponding contributions to the running of the brane counterterms are
\bea
 \mu \, \frac{\partial T_b}{\partial \mu} &=& \frac{m^4}{2 (4\pi)^2 \omega} \left( \frac{\delta\omega}6 + \frac{\delta\omega^2}{12} - \frac{\omega^2 F_b}2 \right) \simeq \frac{m^4}{4(4\pi)^2} \left(\frac{\delta\omega}{3}-|\Phi_b| \right) \,,\\
   \mu\,\frac{\pd}{\pd\mu}\left(\frac{\cA_b}{\tilde g^2}\right) &=&  -\frac{q \Phi_b\,\omega^2 m^2}{6(4\pi)^2} \, G(|\Phi_b|) \simeq -\frac{q^2 m^2\, \cA_b}{3(4\pi)^3} \,,\\
  \mu \, \frac{\partial }{\partial \mu} \left(\frac{\zeta_{\ssR b}}{\kappa}\right) &=& \frac{m^2}{2(4\pi)^2\omega} \left(\frac{\delta\omega}{15} + \frac{2\,\delta\omega^2}{45} + \frac{\delta\omega^3}{90} + \frac{\delta\omega^4}{360} - \frac{\omega^2 F_b}6 - \frac{\omega^4 F^2_b}{12} \right) \nn\\
  &\simeq& \frac{m^2}{2(4\pi)^2}\left(\frac{\delta\omega}{15} - \frac{|\Phi_b|}6\right) \,,\\
    \mu\,\frac{\pd}{\pd\mu}\left(\frac{\kappa \zeta_{\tilde\ssA\ssR b}}{\tilde g^2}\right) &=& -\frac{q\Phi_b \,\omega^2}{24(4\pi)^2} \left(  G(|\Phi_b|) +\frac{\omega^2 }{5}   \, G(|\Phi_b|) (1+3F_b) \right) \simeq -\frac{q^2 \, \cA_b }{10(4\pi)^3} \,,\\
  \mu\,\frac{\partial \,\zeta_{\ssR^2 b}}{\partial \mu} &=& \frac{1}{4(4\pi)^2\omega} \Bigg[ \frac{2\,\delta\omega}{105} + \frac{5\,\delta\omega^2}{252} + \frac{17\,\delta\omega^3}{1260} + \frac{37\,\delta\omega^4}{5040} + \frac{\delta\omega^5}{420} + \frac{\delta\omega^6}{2520} \nn\\
&&\qquad\qquad\quad -\omega^2\bigg(\frac{F_b}{30} +\frac{\omega^2F_b^2}{24} +\frac{\omega^4 F^2_b}{120} + \frac{\omega^4 F^3_b}{60}\bigg)\Bigg]  \nn\\
  &\simeq& \frac{1}{4(4\pi)^2} \left(\frac{2\,\delta\omega}{105} - \frac{|\Phi_b|}{30}\right) \,,\\
  \mu \, \frac{\partial }{\partial \mu} \left(\frac{\kappa\zeta_{\ssA b}}{\tilde g^2}\right) &=& -\frac{8\,q^2}{(4\pi)^2\omega}\left( \frac{\delta\omega}{80} + \frac{17\,\delta\omega^2}{1440} + \frac{\delta\omega^3}{180} + \frac{\delta\omega^4}{720} - \frac{\omega^2 F_b}{48} -\frac{\omega^4 F_b^2}{24} \right) \nn\\
  &\simeq& -\frac{q^2}{(4\pi)^2} \left(\frac{\delta\omega}{10} - \frac{|\Phi_b|}{6} \right)  \,.
\eea

The appearance in these expressions of non-analytic terms involving $|\Phi_b|$ bears some comment. As we see in Figure \ref{cusp}, the signs of these terms are such that they represent maxima of the potential at $\Phi_b = 0$, which are cuspy in that derivatives with respect to $\Phi_b$ are discontinuous at this point. We believe this discontinuous derivative arises because of a level crossing of the ground state at these points, for the following reasons. On the one hand, the explicit calculations given above reveal $\Vone$ as a polynomial in $|\Phi_b|$, which must therefore grow without bound for large $|\Phi_b|$. On the other hand, $\Vone$ should be periodic under the replacements $\Phi_b \to \Phi_b + 1$ and $N \to N +1$, as can be seen from the invariance of the KK spectrum, eq.~\pref{eq:simplescalarspec}, under this shift. (When shifting $\Phi_+ \to \Phi_+ + 1$ both $N$ and $n$ must be shifted by unity, but the shift in $n$ is lost in $\Vone$ once the KK mode sum is performed.) This shows that the energy
is not minimized for the same value of $N$ as $\Phi_b$ is varied to sufficiently large values; instead a new vacuum with $N \to N + 1$ is energetically preferred once $\Phi_b$ become larger than unity. Cuspy maxima in the potential can arise at the points where this crossover between vacua occurs.
\FIGURE[h]{
\epsfig{file=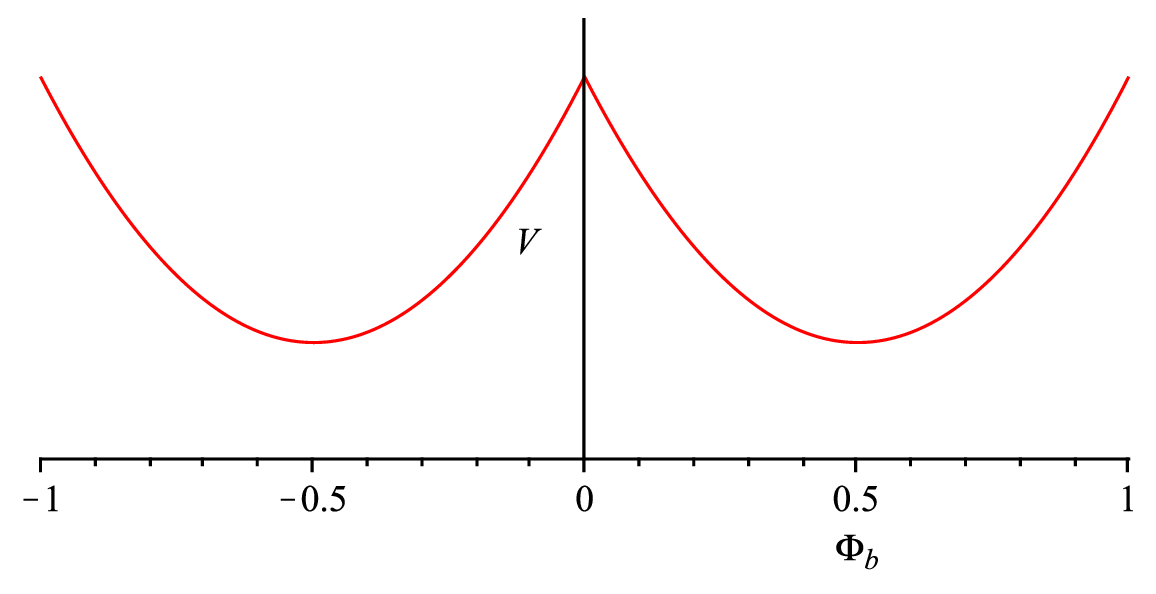,width=0.6\hsize}
\caption{Schematic plot of the potential as a function of $\Phi_b$.}
\label{cusp} }

\subsection{Spin-half fermions}
\label{ssec:fermions}

As shown in Appendix \ref{app:SpecNSum}, the KK spectrum for a fermion that is charged under the $U(1)$ whose flux supports the background rugby ball (using the north patch of the gauge potential) is given by \cite{fermionspectrumN}
\be
 \lambda^{\rm f\sigma}_{jn} = \left( j + \frac\omega2\left|n_{1/2} -\Phi_+ -\frac\sigma{2\omega}\right| + \frac\omega2 \left|n_{1/2}-N+\Phi_- +\frac\sigma{2\omega} \right| +\frac12\right)^2 - \frac{\cN^2}4
\ee
where $n_{1/2} = n-\sigma/2$ and where $\sigma \in \{\pm 1\}$ specifies the 4D helicity of each component of the spinor, of which there are 2 (4)  each in the case of a 6D Weyl (Dirac) spinor. By identifying
\be
 N_{\rm f\sigma} := N-\sigma \,,\quad \Phi_b^{\rm f\sigma} := \Phi_b - \sigma \Phi_0^{\rm f} \,, \quad \Phi_0^{\rm f} := \frac12 \left(1- \omega^{-1}\right)
 = \frac{1 - \alpha}2 = \frac{\delta}{4\pi}
\ee
(and so $\cN_{\rm f\sigma} := \omega(N_{\rm f\sigma} - \Phi_{\rm f \sigma}) = \cN - \sigma$ as well), we can relate these fermionic spectra to the scalar spectrum considered previously:
\be
\lambda^{\rm f\sigma}_{jn} (\omega, N, \Phi_b) +\frac{\cN^2}4 = \lambda^{\rm s}_{jn} (\omega, N_{\rm f\sigma}, \Phi_b^{\rm f\sigma}) + \frac{(1+\cN_{\rm f\sigma}^2)}4 \,.
\ee
%

As discussed in Appendix \ref{app:SpecNSum}, the corresponding small--$t$ coefficients take different forms depending on whether or not $|\Phi_b|\leq \Phi^{\rm f}_0$. When $|\Phi_b|\leq \Phi^{\rm f}_0$ (for both $\Phi_b$'s), the mode sum over the above spectrum yields the following small--$t$ coefficients for a 6D Weyl spinor: $s^{\rm f}_{-1} = -4/\omega$,
\bea
s^{\rm f}_0(\omega,N,\Phi_b) &=& \frac1\omega \left[ \frac13 + \left( \frac13 -2\sum_b\Phi_b^2 \right)\omega^2 \right] \,, \\
s^{\rm f}_1(\omega,N,\Phi_b) &=& \frac1\omega\Bigg[ \frac7{360} -\frac{\omega\cN\Phi}2 - \frac{\cN^2}{3} + \left(\frac1{36} -\frac{1}6\sum_b\Phi_b^2  \right)\omega^2  \nn\\
&&\qquad -\frac{\omega^3\cN}{6}\sum_b \Phi_b (1-4\Phi_b^2) + \Bigg(\frac7{360} -\frac1{6}\sum_b \Phi_b^2 (1-2\Phi_b^2) \Bigg)\omega^4 \Bigg] \,, \\
s^{\rm f}_2(\omega,N,\Phi_b) &=& \frac1\omega \Bigg[ \frac{31}{10080} -\frac{\omega\cN\Phi}{16} - \frac{31\,\cN^2}{720}  + \left(\frac7{1440} -\frac{\cN^2}{72}\bigg(1-6\sum_b\Phi_b^2 \bigg) - \frac7{240} \sum_b\Phi_b^2 \right) \omega^2   \nn\\
&& -\frac{\omega^3\cN}{24}\sum_b \Phi_b (1-4\Phi^2_b)  + \Bigg(\frac{7}{1440} -\frac{7\,\cN^2}{720} -\frac{(1-2\cN^2)}{24}\sum_b \Phi_b^2(1-2\Phi_b^2) \Bigg) \omega^4 \nn\\
&&  -\omega^5\cN\left( \sum_b \Phi_b \bigg( \frac7{240} -\frac{\Phi_b^2}6 +\frac{\Phi_b^4}5\bigg) \right)  \nn\\
&&  + \Bigg(\frac{31}{10080} -\sum_b \Phi_b^2\left(\frac7{240} -\frac{\Phi_b^2}{12} + \frac{\Phi_b^4}{15} \right) \Bigg)\omega^6 \Bigg] \,.
\eea
(Note that the above expressions would be the ones valid when considering the limit $\Phi_b \to 0$ while holding $\omega$ fixed at some value $\neq 1$.) When $|\Phi_b|\geq \Phi^{\rm f}_0$, we find that $s^{\rm f}_{-1}$ is unchanged, but that
\bea
s^{\rm f}_0(\omega,N,\Phi_b) &=& \frac1\omega \left[ -\frac23 +2\tilde F+2\omega  -\frac{2\,\omega^2}3 \right] \,, \\
s^{\rm f}_1(\omega,N,\Phi_b) &=& \frac1\omega\Bigg[ -\frac1{45}  +\frac{\rho\cN}6 - \cN\sum_b \tilde\Phi_b \bigg( \tilde F_b + \frac{\tilde \Phi_b^2}3\bigg) - \frac{\cN^2}{3} + \frac{\tilde F^{(2)}}3 \nn\\
&&\qquad+ \left(\frac19 -\frac{\tilde F}3 +\frac{\cN\tilde\Phi}3 - \frac{\rho\cN}6 \right)\omega^2 - \frac{\omega^4}{45} \Bigg] \,,\\
s^{\rm f}_2(\omega,N,\Phi_b) &=& \frac1\omega \Bigg[ -\frac1{315} +\frac{\tilde F^{(2)}}{30} + \frac{\tilde F^{(3)}}{15} + \frac{\rho\cN}{30} - \frac{\cN}3 \sum_b \tilde \Phi_b \bigg( \tilde F_b^2 + \frac{\tilde\Phi_b^2 \tilde F_b}2 +\frac{\tilde\Phi_b^4}{10} \bigg)  \nn\\
&&\qquad -\cN^2\bigg(\frac1{45}+ \frac{\tilde F^{(2)}}6\bigg) + \Bigg( \frac1{90} - \frac{\tilde F^{(2)}}6  - \frac{\cN}6 \sum_b \tilde \Phi_b \, G(|\tilde \Phi_b|) \nn\\
&&\qquad - \cN^2\bigg(\frac{1}{18} - \frac{\tilde F}6\bigg)\Bigg)\omega^2 + \Bigg( \frac{1}{90} -\frac{\tilde F}{30} -\frac{\rho\cN}{30} + \frac{\cN\tilde\Phi}{30} +\frac{\cN^2}{90} \Bigg)\omega^4 -\frac{\omega^6}{315} \Bigg]   \,\,\qquad
\eea
where
\be
 \rho_b := \mathrm{sgn}(\Phi_b) = \Phi_b/|\Phi_b| \,, \quad
 \rho := \sum_b \rho_b \quad \hbox{and} \quad
 \tilde \Phi_b := \omega(\Phi_b -\rho_b \Phi_0^{\rm f}) \,.
\ee
(We also use tilded versions of the notational contractions, such as $\tilde F_b:=|\tilde \Phi_b|(1-|\tilde \Phi_b|)$, in the same way as is done in the previous section on scalars.) As a check of these expressions, we can evaluate them when $|\Phi_b| = \Phi_0^{\rm f}$ (or simply $\tilde\Phi=0$ in the second case), and find they each give the same result.

In the limit $\omega \to 1$, $\Phi_b \to 0$ the above results agree that $s_{-1}^{\rm sph} = -4$,
\be
 s_0^{\rm sph} =  \frac23 \,, \quad
 s_1^{\rm sph,\,0} = \frac{1}{15} \,, \quad
 s_1^{\rm sph,\,2} = - \frac{\cN^2}3 \,, \quad
 s_2^{\rm sph,\,0} = \frac{1}{63}  \quad
 \hbox{and} \quad
 s_2^{\rm sph,\,2} =  - \frac{\cN^2}{15}  \,.
\ee
These also agree with the results found using Gilkey-de Witt methods in Appendix \ref{app:gilkeydewitt}. The special case where $q$ also vanishes then gives exactly one-half the result found in \cite{KandM} for a fermion on a sphere, as is appropriate due to our use here of 6D Weyl (rather than Dirac) fermions.

For a fermion with charge $q \tilde g$, the corresponding contributions to the running of the bulk couplings are
\bea
 \mu\, \frac{\partial U}{\partial\mu} = \frac{2m^6}{3(4\pi)^3} \,,\quad&& \mu\,\frac{\partial}{\partial\mu} \left(\frac{1}{\kappa^2}\right) = -\frac{m^4}{3(4\pi)^3} \,,\nn\\
 \mu\,\frac{\partial}{\partial\mu}\left(\frac{\zeta_{\ssR^2}}\kappa\right) = -\frac{m^2}{60(4\pi)^3} \,,\quad&& \mu\,\frac{\partial \zeta_{\ssR^3}}{\partial\mu} = -\frac1{504(4\pi)^3} \,,\\
 \mu\, \frac{\partial}{\partial\mu} \left(\frac{1}{\tilde g^2}\right) = \frac{8\,q^2m^2}{3(4\pi)^3} \,,\quad&& \mu\,\frac{\partial}{\partial\mu}\left(\frac{\kappa \zeta_{\ssA\ssR}}{\tilde g^2}\right) = \frac{8\,q^2}{15(4\pi)^3} \,. \nn
\eea
Subtracting the contribution of these bulk counterterms leaves the contributions to the Gilkey-de Witt coefficients that renormalize the brane action. When $|\Phi_b| \leq \Phi_0^{\rm f}$, we find that $\delta s_{-1} = 0$ and
\bea
 \delta s_0^0 &=& \frac1\omega\left(\frac{\omega^2-1}6-2\omega^2\Phi_b^2\right) = \frac{1}{\omega} \left(\frac{\delta\omega}3 + \frac{\delta\omega^2}6 -2\omega^2\Phi_b^2\right) \simeq \frac{\delta\omega}{3} - 2\Phi_b^2 \,,\\
 \delta s_1^0 &=& \frac1\omega\left( \frac{\omega^2-1}{72} + \frac{7(\omega^4-1)}{720} - \frac{\omega^2\Phi_b^2}6 - \frac{\omega^4 \Phi_b^2(1-2\Phi_b^2)}6\right)  \\
 &=& \frac1\omega\left( \frac{\delta\omega}{15} + \frac{13\,\delta\omega^2}{180} + \frac{7\,\delta\omega^3}{180}+ \frac{7\,\delta\omega^4}{720} - \frac{\omega^2\Phi_b^2}6 - \frac{\omega^4 \Phi_b^2(1-2\Phi_b^2)}6\right)
 \simeq \frac{\delta\omega}{15} - \frac{\Phi_b^2}3 \nn\,,\\
 \delta s_1^{\rm 1} &=&-\frac{\cN\Phi_b}2 -\frac{\omega^2\cN\Phi_b}6 (1-4\Phi_b^2) \simeq -\frac{2\cN\Phi_b}3 \,,\qquad \delta s_1^{\rm 2} = 0 \,,\\
 \delta s_2^0 &=& \frac1\omega \Bigg[ \frac{7(\omega^2-1)}{2880} + \frac{7(\omega^4-1)}{2880} + \frac{31(\omega^6-1)}{20160} - \frac{7\,\omega^2\Phi_b^2}{240} - \frac{\omega^4}{24} \Phi_b^2 (1-2\Phi_b^2) \nn \\
 &&\qquad-\omega^6 \Phi_b^2 \bigg(\frac7{240} -\frac{\Phi_b^2}{12} + \frac{\Phi_b^4}{15} \bigg)\Bigg] \nn\\
 &=& \frac1\omega \Bigg[ \frac{\delta\omega}{42} + \frac{101\,\delta\omega^2}{2520} + \frac{17\,\delta\omega^3}{420} + \frac{257\,\delta\omega^4}{10080} +\frac{31\delta\omega^5}{3360} + \frac{31\,\delta\omega^6}{20160}  \nn\\
 &&\qquad - \frac{7\,\omega^2\Phi_b^2}{240} - \frac{\omega^4}{24} \Phi_b^2 (1-2\Phi_b^2) -\omega^6 \Phi_b^2 \bigg(\frac7{240} -\frac{\Phi_b^2}{12} + \frac{\Phi_b^4}{15} \bigg)\Bigg] \simeq \frac{\delta\omega}{42} - \frac{\Phi_b^2}{10} \,,\\
 \delta s_2^{\rm 1} &=& -\frac{\cN\Phi_b}{16} - \frac{\omega^2\cN\Phi_b}{24}\Big(1-4\Phi_b^2\Big) - \omega^4\cN\Phi_b \left( \frac7{240} -\frac{\Phi_b^2}6 +\frac{\Phi_b^4}5\right) \simeq -\frac{2\,\cN\Phi_b}{15} \,,\\
 \delta s_2^{\rm 2}  &=& -\frac{\cN^2}\omega\left[\frac{\omega^2-1}{144} +\frac{7(\omega^4-1)}{1440} - \frac{\omega^2\Phi_b^2}{12} -\frac{\omega^4\Phi_b^2}{12} \Big( 1-2\Phi_b^2 \Big) \right] \nn\\
 &=& -\frac{\cN^2}\omega \left[\frac{\delta\omega}{30} + \frac{13\,\delta\omega^2}{360}+\frac{7\,\delta\omega^3}{360} + \frac{7\,\delta\omega^4}{1440} - \frac{\omega^2\Phi_b^2}{12} -\frac{\omega^4\Phi_b^2}{12} \Big( 1-2\Phi_b^2 \Big)\right]  \nn\\
&\simeq& -  \cN^2\left( \frac{\delta\omega}{30} - \frac{\Phi_b^2}6 \right) \,,
\eea
where, as before, $\delta \omega = \omega - 1$ and the approximate equalities give the leading result for $\delta \omega, \Phi_b \ll 1$. The brane counterterms therefore renormalize as follows:
\bea
 \mu\,\frac{\partial T_b}{\partial\mu} &=& \frac{m^4}{2(4\pi)^2\omega} \left(\frac{\delta\omega}3 + \frac{\delta\omega^2}6 -2\omega^2\Phi_b^2\right) \simeq \frac{m^4}{(4\pi)^2}\left(\frac{\delta\omega}6 - \Phi_b^2\right) \,,\\
  \mu \, \frac{\partial }{\partial \mu} \left(\frac{\cA_b}{\tilde g^2}\right) &=& -\frac{q\Phi_b \, m^2}{(4\pi)^2} \left(1 +\frac{\omega^2}3 (1-4\Phi_b^2)\right) \simeq -\frac{8\,q^2m^2}{3(4\pi)^3} \cA_b \,,\\
 \mu\,\frac\partial{\partial\mu} \left(\frac{\zeta_{\ssR b}}{\kappa}\right) &=& \frac{m^2}{2(4\pi)^2\omega} \left( \frac{\delta\omega}{15} + \frac{13\,\delta\omega^2}{180} + \frac{7\,\delta\omega^3}{180}+ \frac{7\,\delta\omega^4}{720} - \frac{\omega^2\Phi_b^2}6 - \frac{\omega^4 \Phi_b^2(1-2\Phi_b^2)}6\right) \nn\\
 &\simeq& \frac{m^2}{2(4\pi)^2} \left(\frac{\delta\omega}{15} - \frac{\Phi_b^2}3\right)  \,,\\
 \mu \, \frac{\partial }{\partial \mu} \left(\frac{\zeta_{\tilde\ssA\ssR b}}{\tilde g^2}\right) &=& -\frac{q\Phi_b }{(4\pi)^2} \left[ \frac{1}{16} + \frac{\omega^2}{24}\Big(1-4\Phi_b^2\Big) + \omega^4 \left( \frac7{240} -\frac{\Phi_b^2}6 +\frac{\Phi_b^4}5\right)  \right] \nn\\
 &\simeq& -\frac{4\,q^2}{15(4\pi)^3} \cA_b \,,\\
 \mu \, \frac{\partial \zeta_{\ssR^2 b}}{\partial\mu} &=& \frac{1}{4(4\pi)^2\omega} \Bigg[ \frac{\delta\omega}{42} + \frac{101\,\delta\omega^2}{2520} + \frac{17\,\delta\omega^3}{420} + \frac{257\,\delta\omega^4}{10080} +\frac{31\delta\omega^5}{3360} + \frac{31\,\delta\omega^6}{20160}  \nn\\
 &&\qquad - \frac{7\,\omega^2\Phi_b^2}{240} - \frac{\omega^4}{24} \Phi_b^2 (1-2\Phi_b^2) -\omega^6 \Phi_b^2 \bigg(\frac7{240} -\frac{\Phi_b^2}{12} + \frac{\Phi_b^4}{15} \bigg)\Bigg] \nn \\
 &\simeq& \frac{1}{4(4\pi)^2}\left(\frac{\delta\omega}{42} - \frac{\Phi_b^2}{10}\right) \,,\\
  \mu \, \frac{\partial }{\partial \mu} \left(\frac{\zeta_{\ssA b}}{\tilde g^2}\right) &=& -\frac{8\, q^2}{(4\pi)^2\omega} \left[\frac{\delta\omega}{30} + \frac{13\,\delta\omega^2}{360}+\frac{7\,\delta\omega^3}{360} + \frac{7\,\delta\omega^4}{1440} - \frac{\omega^2\Phi_b^2}{12} -\frac{\omega^4\Phi_b^2}{12} \Big( 1-2\Phi_b^2 \Big)\right]  \nn\\
&\simeq& -\frac{4\, q^2}{3(4\pi)^2}\left( \frac{\delta\omega}{5} - \Phi_b^2 \right)  \,.
\eea

When instead $|\Phi_b| \geq \Phi_0^{\rm f}$, we find that (as always) $\delta s_{-1} = \delta s_1^{\rm 2} = 0$ and
\bea
 \delta s_0^0 &=& \frac1\omega\left( (\omega-1) -\frac{\omega^2-1}3 +2\tilde F_b \right) = \frac{1}{\omega} \left(\frac{\delta\omega}3 - \frac{\delta\omega^2}3 +2\tilde F_b\right) \simeq \frac{\delta\omega}{3} + 2|\tilde\Phi_b| \,,\\
 \delta s_1^0 &=& \frac1\omega\left( \frac{\omega^2-1}{18} - \frac{(\omega^4-1)}{90} +\frac{\tilde F^2_b}3 -\frac{\omega^2\tilde F_b}3 \right)  \nn\\
 &=& \frac1\omega\left( \frac{\delta\omega}{15} -\frac{\delta\omega^2}{90} - \frac{2\,\delta\omega^3}{45} - \frac{\delta\omega^4}{90} +\frac{\tilde F^2_b}3 -\frac{\omega^2\tilde F_b}3 \right)
 \simeq \frac{\delta\omega}{15} - \frac{|\tilde\Phi_b|}3 \,,\\
 \delta s_1^{\rm 1} &=& \frac\cN\omega\Bigg[\frac{\rho_b(1-\omega^2)}6 - \tilde\Phi_b\bigg(\tilde F_b + \frac{\tilde\Phi_b^2}3\bigg) + \frac{\omega^2\tilde\Phi_b}3  \Bigg]\simeq -\frac{\rho_b\cN\,\delta\omega}3+\frac{\cN\tilde\Phi_b}3  \,,\\
 \delta s_2^0 &=& \frac1\omega \Bigg[ \frac{(\omega^2-1)}{180} + \frac{(\omega^4-1)}{180} - \frac{(\omega^6-1)}{630} + \frac{\tilde F_b^2}{30} + \frac{\tilde F_b^3}{15} -\frac{\omega^2 \tilde F_b^2}6 - \frac{\omega^4 \tilde F_b}{30} \Bigg] \nn\\
 &=& \frac1\omega \Bigg( \frac{\delta\omega}{42} + \frac{19\,\delta\omega^2}{1260} - \frac{\delta\omega^3}{105} - \frac{23\,\delta\omega^4}{1260} -\frac{\delta\omega^5}{105} - \frac{\delta\omega^6}{630}  \nn\\
 &&\qquad + \frac{\tilde F_b^2}{30} + \frac{\tilde F_b^3}{15} -\frac{\omega^2 \tilde F_b^2}6 - \frac{\omega^4 \tilde F_b}{30} \Bigg) \simeq \frac{\delta\omega}{42} - \frac{|\tilde\Phi_b|}{30} \,,\\
 \delta s_2^{\rm 1} &=& \frac1\omega\Bigg[ \frac{\rho_b\, \cN(1-\omega^4)}{30} - \frac{\cN\tilde\Phi_b}3  \bigg( \tilde F_b^2 + \frac{\tilde\Phi_b^2 \tilde F_b}2 + \frac{\tilde\Phi_b^4}{10} \bigg) - \frac{\omega^2\cN\tilde\Phi_b}6   \, G(|\tilde\Phi_b|) +\frac{\omega^4\cN\tilde\Phi_b}{30} \Bigg] \nn\\
 &\simeq& - \frac{2\rho_b \, \cN \,\delta\omega}{15} -\frac{2\cN\tilde\Phi_b}{15} \,,\\
 \delta s_2^{\rm 2}  &=& \frac{\cN^2}\omega\left[-\frac{\omega^2-1}{36} +\frac{(\omega^4-1)}{180} -\frac{F^2_b}6 + \frac{\omega^2 \tilde F_b}6 \right] \nn\\
 &=& \frac{\cN^2}\omega \left( -\frac{\delta\omega}{30} + \frac{\delta\omega^2}{180} +\frac{\delta\omega^3}{45}+\frac{\delta\omega^4}{180} -\frac{F^2_b}6 + \frac{\omega^2 \tilde F_b}6 \right)  \nn\\
&\simeq& \cN^2\bigg(- \frac{\delta\omega}{30} + \frac{|\tilde\Phi_b|}6 \bigg) \,.
\eea
Therefore, when $|\Phi_b| \geq \Phi_0^{\rm f}$, the brane counterterms renormalize as follows:
\bea
 \mu\,\frac{\partial T_b}{\partial\mu} &=& \frac{m^4}{2(4\pi)^2\omega} \left(\frac{\delta\omega}3 - \frac{\delta\omega^2}3 +2\tilde F_b\right) \simeq \frac{m^4}{(4\pi)^2}\left(\frac{\delta\omega}6 + |\tilde\Phi_b|\right) \,,\\
  \mu \, \frac{\partial }{\partial \mu} \left(\frac{\cA_b}{\tilde g^2}\right) &=& \frac{q \, m^2}{(4\pi)^2\omega} \Bigg[\frac{\rho_b(1-\omega^2)}6 - \tilde\Phi_b\bigg(\tilde F_b + \frac{\tilde\Phi_b^2}3\bigg) + \frac{\omega^2\tilde\Phi_b}3  \Bigg] \nn\\
  &\simeq&  \frac{q\, m^2}{(4\pi)^2} \left(-\frac{\rho_b\,\delta\omega}3+\frac{\tilde\Phi_b}3\right) \,,\\
 \mu\,\frac\partial{\partial\mu} \left(\frac{\zeta_{\ssR b}}{\kappa}\right) &=& \frac{m^2}{2(4\pi)^2\omega} \left( \frac{\delta\omega}{15} -\frac{\delta\omega^2}{90} - \frac{2\,\delta\omega^3}{45} - \frac{\delta\omega^4}{90} +\frac{\tilde F^2_b}3 -\frac{\omega^2\tilde F_b}3 \right) \nn\\
 &\simeq& \frac{m^2}{2(4\pi)^2} \left(\frac{\delta\omega}{15} - \frac{|\tilde\Phi_b|}3\right)  \,,\\
 \mu \, \frac{\partial }{\partial \mu} \left(\frac{\zeta_{\tilde\ssA\ssR b}}{\tilde g^2}\right) &=& \frac{q}{(4\pi)^2\omega} \Bigg[ \frac{\rho_b\, \cN(1-\omega^4)}{30} - \frac{\cN\tilde\Phi_b}3  \bigg( \tilde F_b^2 + \frac{\tilde\Phi_b^2 \tilde F_b}2 + \frac{\tilde\Phi_b^4}{10} \bigg) - \frac{\omega^2\cN\tilde\Phi_b}6   \, G(|\tilde\Phi_b|)\nn\\
 &&\qquad +\frac{\omega^4\cN\tilde\Phi_b}{30} \Bigg] \simeq -\frac{2\,q}{15(4\pi)^2} \left(\rho_b \,\delta\omega +\tilde\Phi_b\right) \,,\\
 \mu \, \frac{\partial \zeta_{\ssR^2 b}}{\partial\mu} &=& \frac{1}{4(4\pi)^2\omega} \Bigg( \frac{\delta\omega}{42} + \frac{19\,\delta\omega^2}{1260} - \frac{\delta\omega^3}{105} - \frac{23\,\delta\omega^4}{1260} -\frac{\delta\omega^5}{105} - \frac{\delta\omega^6}{630}  \nn\\
 &&\qquad + \frac{\tilde F_b^2}{30} + \frac{\tilde F_b^3}{15} -\frac{\omega^2 \tilde F_b^2}6 - \frac{\omega^4 \tilde F_b}{30} \Bigg) \simeq \frac{1}{4(4\pi)^2}\left(\frac{\delta\omega}{42} - \frac{|\tilde\Phi_b|}{30}\right) \,,\\
  \mu \, \frac{\partial }{\partial \mu} \left(\frac{\zeta_{\ssA b}}{\tilde g^2}\right) &=& \frac{8\, q^2}{(4\pi)^2\omega} \left( -\frac{\delta\omega}{30} + \frac{\delta\omega^2}{180} +\frac{\delta\omega^3}{45}+\frac{\delta\omega^4}{180} -\frac{F^2_b}6 + \frac{\omega^2 \tilde F_b}6 \right)  \nn\\
&\simeq& -\frac{4\, q^2}{3(4\pi)^2}\left( \frac{\delta\omega}{5} - |\tilde\Phi_b| \right)  \,.
\eea
%

\subsection{Gauge fields}

We next state the results for the Casimir coefficient for a gauge field, provided this gauge field is {\em not} the field whose flux stabilizes the background 2D geometry.
We consider in turn the cases where the 6D gauge field is massless or massive (in the 6D sense).

\subsubsection*{Massless 6D gauge fields}

We begin with the massless case. Picking an appropriate gauge (such as light-cone gauge)
allows the 6D gauge field to be decomposed into four components,\footnote{Ghosts also do not contribute to the
one-loop result in this gauge.} each with a spectrum (when evaluated using the north patch of the gauge potential) given by
\be \label{4D-gaugescalar-spectrum}
\lambda_{jn}^{\rm gf\xi} = \left( j + \frac\omega2 \left| n  - \Phi_+ + \frac\xi\omega \right| + \frac\omega2 \left| n - N +\Phi_- - \frac\xi\omega \right| +\frac12\right)^2 - \frac{(1+\cN^2)}4
\ee
where $\xi \in \{0,0,+ 1, -1\}$ for each of the four components. (For more details see Appendix \ref{app:SpecNSum} and \cite{Parameswaran:2007cb}.) From this, we see that two modes have the exact same spectrum as scalars (i.e.~the $\xi=0$ modes), and two modes have almost the same spectrum as scalars (i.e.~the $\xi = \pm 1$ modes). Similar to before, we can identify
\be
 N_{\rm gf\xi} := N  \,,\quad \Phi_b^{\rm gf\xi} := \Phi_b - \xi \Phi_0^{\rm gf} \,, \quad \Phi_0^{\rm gf} = \omega^{-1} = \alpha = 1 - \frac{\delta}{2\pi}
\ee
(and so $\cN_{\rm gf\xi} := \omega(N_{\rm gf\xi} - \Phi_{\rm gf\xi} ) = \cN +2\xi$) to relate these modes to the scalar spectrum:
\be
\lambda_{jn}^{\rm gf\xi}(\omega,N,\Phi_b) + \frac{(1+\cN^2)}4 = \lambda^{\rm s}_{jn} (\omega,N_{\rm gf\xi}, \Phi_b^{\rm gf\xi}) + \frac{(1+\cN_{\rm gf\xi}^2)}4 \,.
\ee
However, since the resulting spectrum for the $\xi= \pm 1$ modes ends up being very similar to that of the scalars, we shall just write
\be \label{eq:gfassplusDelta}
s_i^{\rm gf} = 4 s_i^{\rm s} + \Delta s_i^{\rm gf}
\ee
where $\Delta s_{-1}^{\rm gf} = 0$,
\bea
\Delta s^{\rm gf}_0(\omega,N,\Phi_b) &=& \frac1\omega\left[-2\omega+ \omega^2 \sum_b |\Phi_b| \right] \,, \\
\Delta s^{\rm gf}_1(\omega,N,\Phi_b) &=& \frac1\omega\Bigg[ \cN^2 + \omega \cN \Phi +\frac{\omega^2}3 \sum_b |\Phi_b| -\frac{\omega^3 \cN}2 \sum_b \Phi_b \, |\Phi_b| - \frac{\omega^4}3 \sum_b |\Phi_b|^3 \Bigg] \,,\qquad \\
\Delta s^{\rm gf}_2(\omega,N,\Phi_b) &=& \frac1\omega \Bigg[ -\frac{\omega\cN^2}4 + \omega^2 \bigg( \frac1{15} -\frac{\cN^2}{24}\bigg) \sum_b |\Phi_b|  - \frac{\omega^3\cN}4 \sum_b \Phi_b \, |\Phi_b|  \nn\\
&& \qquad  -\frac{\omega^4 (1-\cN^2)}6 \sum_b |\Phi_b|^3 + \frac{\omega^5\cN}4 \sum_b \Phi_b \, |\Phi_b|^3 +\frac{\omega^6}{10} \sum_b |\Phi_b|^5 \Bigg]  \,,
\eea
so long as $|\Phi_b| \leq \Phi_0^{\rm gf}$. As it turns out, the $|\Phi_b|$--dependent terms seen here in $\Delta s_i^{\rm gf}$ serve to exactly cancel any corresponding terms in $4 s_i^{\rm s}$ in eq.~\pref{eq:gfassplusDelta} that are odd in $|\Phi_b|$ (for the $\xi=\pm1$ modes only).

Even though only $s_2$ contributes to the Casimir energy for massless fields, we nonetheless also follow $s_{-1}$, $s_0$ and $s_1$ since these are useful as intermediate steps when assembling the contributions of a massive vector field.
The corresponding bulk quantities therefore are
\bea
s_{-1}^{\rm sph} = 4 \,,\quad s_0^{\rm sph} &=& -\frac23 \,,\quad s_1^{\rm sph,\,0} = \frac4{15} \,,\quad s_1^{\rm sph,\,2} = \frac{5\,\cN^2}6 \nn\,,\\
 s_2^{\rm sph,\,0} &=& \frac{16}{315} \,,\quad s_2^{\rm sph,\,2} = -\frac{7\,\cN^2}{20}
\eea
and after these are subtracted the brane renormalizations are obtained from
\bea
\delta s_0 &=& \frac1\omega \left( - (\omega-1) +\frac{(\omega^2-1)}3 -2\omega^2 F_b + \omega^2 |\Phi_b| \right) =\frac1\omega \left(-\frac{\delta\omega}3 + \frac{\delta\omega^2}3 -2\omega^2 F_b + \omega^2 |\Phi_b| \right) \nn\\
&\simeq& -\frac{\delta\omega} 3 -|\Phi_b| \,,\\
\delta s_1^0 &=& \frac1\omega \left( \frac{(\omega^2-1)}9 + \frac{\omega^4-1}{90} - \frac{2\,\omega^2F_b}3 +\frac{\omega^2 |\Phi_b|}3 - \frac{\omega^4F_b^2}3 -\frac{\omega^4 |\Phi_b|^3}3  \right)  \nn\\
 &=& \frac1\omega \left(\frac{4\,\delta\omega}{15} + \frac{8\,\delta\omega^2}{45} +\frac{2\,\delta\omega^3}{45} + \frac{\delta\omega^4}{90} - \frac{2\,\omega^2F_b}3 +\frac{\omega^2 |\Phi_b|}3 - \frac{\omega^4F_b^2}3 -\frac{\omega^4 |\Phi_b|^3}3 \right) \nn\\
&\simeq& \frac{4\,\delta\omega}{15} - \frac{|\Phi_b|}3 \,,\\
\delta s_1^{\rm 1} &=& -\frac{\omega^2\cN}3 \Phi_b G(|\Phi_b|) + \cN \Phi_b - \frac{\omega^2\cN}2 \Phi_b |\Phi_b| \simeq \frac{2\,\cN\Phi_b}3 \,,\\
\delta s_2^0 &=& \frac1\omega\bigg( \frac{(\omega^2-1)}{45} + \frac{\omega^4-1}{180} + \frac{\omega^6-1}{630} -\frac{2\, \omega^2 F_b}{15} +\frac{\omega^2 |\Phi_b|}{15} -\frac{\omega^4F_b^2}{6}  - \frac{\omega^4|\Phi_b|^3}6 \nn\\
&&\qquad -\frac{\omega^6 F^2_b}{30} - \frac{\omega^6 F^3_b}{15}   + \frac{\omega^6|\Phi_b|^5}{10} \bigg) \nn\\
 &=& \frac1\omega\bigg(\frac{8\,\delta\omega}{105}+\frac{5\,\delta\omega^2}{63}+\frac{17\,\delta\omega^3}{315} + \frac{37\,\delta\omega^4}{1260}+\frac{\delta\omega^5}{105} + \frac{\delta\omega^6}{630} -\frac{2\, \omega^2 F_b}{15} +\frac{\omega^2 |\Phi_b|}{15} -\frac{\omega^4F_b^2}{6}  \nn\\
 &&\qquad - \frac{\omega^4|\Phi_b|^3}6 -\frac{\omega^6 F^2_b}{30} - \frac{\omega^6 F^3_b}{15}   + \frac{\omega^6|\Phi_b|^5}{10} \bigg)  \simeq \frac{8\,\delta\omega}{105} -\frac{|\Phi_b|}{15} \,,\\
\delta s_2^{\rm 1} &=&  -\frac{\omega^2 \cN}{6}  \Phi_b \, G(|\Phi_b|) -\frac{\omega^4 \cN}{30}  \Phi_b \, G(|\Phi_b|) (1+3F_b) -\frac{\omega^2\cN\Phi_b}4 |\Phi_b| + \frac{\omega^4\cN\Phi_b}4 |\Phi_b|^3 \nn\\
&\simeq& - \frac{\cN\Phi_b}{5} \,,\\
 \delta s_2^{\rm 2} &=& -\frac{\cN^2}\omega \left(\frac{\omega -1}{8} + \frac{\omega^2-1}{72} +
 \frac{\omega^4-1}{180} -\frac{\omega^2 F_b}{12} + \frac{\omega^2|\Phi_b|}{24} -\frac{\omega^4 F_b^2}6 -\frac{\omega^4|\Phi_b|^3}6 \right) \nn\\
 &=& -\frac{\cN^2}\omega\left(  \frac{7\, \delta\omega}{40} +\frac{17\,\delta\omega^2}{360}
 +\frac{\delta\omega^3}{45} +\frac{\delta\omega^4}{180} -\frac{\omega^2 F_b}{12} + \frac{\omega^2|\Phi_b|}{24} -\frac{\omega^4 F_b^2}6 -\frac{\omega^4|\Phi_b|^3}6 \right) \nn\\
 &\simeq& -\cN^2\left(\frac{7\,\delta\omega}{40} -\frac{|\Phi_b|}{24}\right) \,,
\eea
along with $\delta s_{-1} = \delta s_1^{\rm 2} = 0$ (as usual).

Because the renormalizations coming from $s_k$ are proportional to $m^{4-2k}$, where $m$ is the 6D mass, for massless fields we need only follow the contributions of $s_2$, ensuring the only nonzero renormalizations are
\be
 \mu\,\frac{\partial \zeta_{\ssR^3}}{\partial\mu} = -\frac2{315(4\pi)^3} \quad \hbox{and} \quad \mu\,\frac{\partial}{\partial\mu}\left(\frac{\kappa \zeta_{\ssA\ssR}}{\tilde g^2}\right) = \frac{14\,q^2}{5(4\pi)^3} \,,
\ee
in the bulk, and
\bea
 \mu \, \frac{\partial \zeta_{\ssR^2 b}}{\partial\mu} &=& \frac{1}{4(4\pi)^2\omega} \bigg(\frac{8\,\delta\omega}{105}+\frac{5\,\delta\omega^2}{63}+\frac{17\,\delta\omega^3}{315} + \frac{37\,\delta\omega^4}{1260}+\frac{\delta\omega^5}{105} + \frac{\delta\omega^6}{630}    \nn\\
 &&\qquad -\frac{2\, \omega^2 F_b}{15} +\frac{\omega^2 |\Phi_b|}{15} -\frac{\omega^4F_b^2}{6} - \frac{\omega^4|\Phi_b|^3}6 -\frac{\omega^6 F^2_b}{30} - \frac{\omega^6 F^3_b}{15}   + \frac{\omega^6|\Phi_b|^5}{10} \bigg) \nn\\
 &\simeq& \frac1{(4\pi)^2} \left(\frac{2\,\delta\omega}{105} -\frac{|\Phi_b|}{60} \right) \,,\\
 \mu \, \frac{\pd}{\pd\mu}\left(\frac{\kappa \zeta_{\tilde \ssA\ssR b}}{\tilde g^2}\right) &=& -\frac{q}{(4\pi)^2} \left( \frac{\omega^2}{6}  \Phi_b \, G(|\Phi_b|) +\frac{\omega^4}{30}  \Phi_b \, G(|\Phi_b|) (1+3F_b) +\frac{\omega^2\Phi_b}4 |\Phi_b| - \frac{\omega^4\Phi_b}4 |\Phi_b|^3 \right) \nn\\
 &\simeq& -\frac{q \,\Phi_b}{5(4\pi)^2} = -\frac{2\,q^2}{5(4\pi)^3} \cA_b  \,,\\
 \mu \, \frac{\partial }{\partial \mu} \left(\frac{\zeta_{\ssA b}}{\tilde g^2}\right) &=& -
 \frac{8\, q^2}{(4\pi)^2\omega} \bigg(  \frac{7\, \delta\omega}{40} +\frac{17\,\delta\omega^2}{360}
 +\frac{\delta\omega^3}{45} +\frac{\delta\omega^4}{180} -\frac{\omega^2 F_b}{12} + \frac{\omega^2|\Phi_b|}{24}  \nn\\
 &&\qquad -\frac{\omega^4 F_b^2}6 -\frac{\omega^4|\Phi_b|^3}6 \bigg) \simeq -\frac{q^2}{(4\pi)^2} \left(\frac{7\,\delta\omega}{5} -\frac{|\Phi_b|}{3}\right)  \,,
\eea
on the brane.

\subsubsection*{Massive 6D gauge fields}

Let us now turn to massive gauge-field fluctuations, corresponding to those gauge directions that acquire mass because of the nonzero (but constant) values taken by some of the scalar fields. By assumption, these gauge fields are vanishing in the background, both because this would require a more complicated ansatz than assumed here for the rugby-ball backgrounds \cite{RandjbarDaemi:2006gf}, and because it would complicate the diagonalization of the metric and gauge-field fluctuations.

In this case the linearized theory simplifies  \cite{RandjbarDaemi:2002pq} if we choose light-cone gauge, as described in more detail in Appendix \ref{app:SpecNSum}. The result is that a massive gauge field leads
to the 4D spectrum of a massless gauge field, given in eq.~(\ref{4D-gaugescalar-spectrum}), plus that of a scalar provided earlier. It follows that  the $s_i$ coefficient of a massive gauge field are
\bea
 s_i^{\rm mgf} = s_i^{\rm gf} + s_i^{\rm s} = 5 s_i^{\rm s} + \Delta s_i^{\rm gf}
  \, ,
\eea
where the $s_i^s$ are the corresponding quantities for a 6D scalar, those given in eqs.~(\ref{eq:simplescalars1})--(\ref{eq:simplescalars2}), and we use $N \in \{0,\pm 1\}$ to ensure stability (see Appendix \ref{app:SpecNSum}).

Let us now give the contribution of the massive gauge field to the running of the bulk couplings. Since in the sphere limit we have
\bea
 s_{-1}^{\rm sph,\, 0} = 5 \,, \quad
 s_{0}^{\rm sph,\, 0} &=& -\frac13 \,, \quad
 s_{1}^{\rm sph,\,0} = \frac{1}{3} \,, \quad s_{1}^{\rm sph,\,2} = \frac{19\, \cN^2}{24} \,, \nn \\
  s_{2}^{\rm sph,\,0} &=& \frac{4}{63} \,,
 \quad \mbox{and}\quad s_{2}^{\rm sph,\, 2} = - \frac{3\, \cN^2}{8}
\eea
we obtain
\bea
 \mu \, \frac{\partial U}{\partial \mu} = -\frac{5\, m^6}{6 (4 \pi)^3} \,,\quad&&
 \mu \, \frac{\partial}{\partial \mu} \left( \frac{1}{\kappa^2} \right) = \frac{m^4}{6 (4\pi)^3} \,,\nn\\
 \mu\,\frac{\partial }{\partial\mu}\left(\frac{\zeta_{\ssR^2}}\kappa\right) = -\frac{m^2}{12(4\pi)^3} \,,\quad
 && \mu \, \frac{\partial \zeta_{\ssR^3}}{\partial\mu} = -\frac1{126(4\pi)^3}\, , \nn \\
 \mu \, \frac{\partial}{\partial\mu}\left(\frac{1}{\tilde g^2}\right) = -\frac{19\,q^2m^2}{3(4\pi)^3} \,,\quad
 && \mu \, \frac{\partial }{\partial\mu}\left(\frac{\kappa  \zeta_{\ssA \ssR}}{\tilde{g}^2}\right)= \frac{3\, q^2}{(4\pi)^3} \,.\nn
 \eea
The running of the brane couplings is similarly obtained by computing the $\delta s_i$ coefficients:
 \bea
 \delta s_0 &=& \frac{1}{\omega}\left( -\frac{\delta\omega}6 + \frac{5\,\delta\omega^2}{12} - \frac{5\,\omega^2F_b}2 +\omega^2 |\Phi_b|  \right) \simeq -\frac{\delta\omega}6 - \frac{3|\Phi_b|}2\,, \\
 \delta s^0_1 &=& \frac1\omega\left(\frac{\delta\omega}{3} + \frac{2\,\delta\omega^2}{9} + \frac{\delta\omega^3}{18} + \frac{\delta\omega^4}{72} - \frac{5\,\omega^2 F_b}6 +\frac{\omega^2 |\Phi_b|}3 - \frac{5\,\omega^4 F_b^2}{12} -\frac{\omega^4 |\Phi_b|^3}3 \right) \nn\\
 &\simeq& \frac{\delta\omega}{3} -\frac{|\Phi_b|}2 \, , \\
 \delta s^{\rm 1}_1 &=& -\frac{5\,\omega^2\cN}{12} \Phi_b \, G(|\Phi_b|) + \cN \Phi_b - \frac{\omega^2\cN}2 \Phi_b |\Phi_b| \simeq \frac{7\,\cN\Phi_b}{12} \,,\\
 %
 %
 \delta s^0_2 &=& \frac1\omega\bigg( \frac{2\,\delta\omega}{21} + \frac{25\,\delta\omega^2}{252} + \frac{17\,\delta\omega^3}{252}
+ \frac{37\,\delta\omega^4}{1008} + \frac{\delta\omega^5}{84} + \frac{\delta\omega^6}{504} -\frac{\omega^2 F_b}6 +\frac{\omega^2|\Phi_b|}{15} \nn\\
&&\qquad -\frac{5\,\omega^4 F_b^2}{24} - \frac{\omega^4|\Phi_b|^3}6 - \frac{\omega^6 F_b^2}{24} -\frac{\omega^6 F_b^3}{12} +\frac{\omega^6|\Phi_b|^5}{10} \bigg) \simeq
\frac{2\,\delta\omega}{21} - \frac{|\Phi_b|}{10} \, , \\
 \delta s^{\rm 1}_2 &=& -\frac{5\,\omega^2 \cN}{24}  \Phi_b \, G(|\Phi_b|) -\frac{\omega^4 \cN}{24}  \Phi_b \, G(|\Phi_b|) (1+3F_b) -\frac{\omega^2\cN\Phi_b}4 |\Phi_b| + \frac{\omega^4\cN\Phi_b}4 |\Phi_b|^3 \nn\\
&\simeq& - \frac{\cN\Phi_b}{4} \,,\\
\delta s^{\rm 2}_2 &=& -\frac{\cN^2}\omega \left(  \frac{3\,\delta\omega}{16} + \frac{17\,\delta\omega^2}{288} + \frac{\delta\omega^3}{36} + \frac{\delta\omega^4}{144} -\frac{5\,\omega^2 F_b}{48} + \frac{\omega^2|\Phi_b|}{24} -\frac{5\,\omega^4 F_b^2}{24} -\frac{\omega^4|\Phi_b|^3}6 \right) \nn\\
 &\simeq& - \cN^2 \left(\frac{3\,\delta\omega}{16} - \frac{|\Phi_b|}{16} \right)\,.
\eea
These give
\bea
 \mu \, \frac{\partial T_b}{\partial \mu} &=& \frac{m^4}{2 (4\pi)^2 \omega} \left( -\frac{\delta\omega}6 + \frac{5\,\delta\omega^2}{12} - \frac{5\,\omega^2F_b}2 +\omega^2 |\Phi_b|  \right)   \nn\\
 &\simeq& -\frac{m^4}{(4\pi)^2}\left(\frac{\delta\omega}{12} + \frac{3|\Phi_b|}4\right) \,,\\
 \mu\,\frac{\pd}{\pd\mu} \left(\frac{\cA_b}{\tilde g^2}\right) &=& \frac{2\, q m^2}{(4\pi)^2} \left( -\frac{5\,\omega^2}{12} \Phi_b\, G(|\Phi_b|) + \Phi_b - \frac{\omega^2}2 \Phi_b |\Phi_b| \right) \nn\\
 &\simeq& \frac{7\,qm^2 \Phi_b}{6(4\pi)^2} = - \frac{7\,q^2m^2}{3(4\pi)^3} \cA_b \,,\\
  \mu \, \frac{\partial }{\partial \mu} \left(\frac{\zeta_{\ssR b}}{\kappa}\right) &=&
  \frac{m^2}{2(4\pi)^2\omega} \bigg(\frac{\delta\omega}{3} + \frac{2\,\delta\omega^2}{9} + \frac{\delta\omega^3}{18} + \frac{\delta\omega^4}{72} - \frac{5\,\omega^2 F_b}6 +\frac{\omega^2 |\Phi_b|}3 \nn \\
  &&\qquad - \frac{5\,\omega^4 F_b^2}{12} -\frac{\omega^4 |\Phi_b|^3}3 \bigg) \simeq \frac{m^2}{(4\pi)^2} \left( \frac{\delta\omega}6 - \frac{|\Phi_b|}4 \right)  \,,\\
  \mu\,\frac{\pd}{\pd\mu}\left(\frac{\kappa \zeta_{\tilde \ssA\ssR b}}{\tilde g^2}\right) &=& -\frac{q}{(4\pi)^2} \bigg(\frac{5\,\omega^2}{24}  \Phi_b \, G(|\Phi_b|) +\frac{\omega^4}{24}  \Phi_b \, G(|\Phi_b|) (1+3F_b) \nn\\
  &&\qquad +\frac{\omega^2\Phi_b}4 |\Phi_b| - \frac{\omega^4\Phi_b}4 |\Phi_b|^3\bigg) \simeq -\frac{q\,\Phi_b}{4(4\pi)^2} = -\frac{q^2}{2(4\pi)^3} \cA_b \,,\\
 \mu \, \frac{\partial \zeta_{\ssR^2 b}}{\partial\mu} &=&\frac{1}{4(4\pi)^2\omega} \bigg( \frac{2\,\delta\omega}{21} + \frac{25\,\delta\omega^2}{252} + \frac{17\,\delta\omega^3}{252}
+ \frac{37\,\delta\omega^4}{1008} + \frac{\delta\omega^5}{84} + \frac{\delta\omega^6}{504} -\frac{\omega^2 F_b}6 +\frac{\omega^2|\Phi_b|}{15} \nn\\
&&\qquad -\frac{5\,\omega^4 F_b^2}{24} - \frac{\omega^4|\Phi_b|^3}6 - \frac{\omega^6 F_b^2}{24} -\frac{\omega^6 F_b^3}{12} +\frac{\omega^6|\Phi_b|^5}{10} \bigg) \nn\\
 &\simeq& \frac1{(4\pi)^2} \left(\frac{2\,\delta\omega}{84} -\frac{|\Phi_b|}{40} \right) \,,\\
 \mu \, \frac{\partial }{\partial \mu} \left(\frac{\zeta_{\ssA b}}{\tilde g^2}\right) &=&
 -\frac{8\, q^2}{(4\pi)^2\omega} \bigg(  \frac{3\,\delta\omega}{16} + \frac{17\,\delta\omega^2}{288} + \frac{\delta\omega^3}{36} + \frac{\delta\omega^4}{144} -\frac{5\,\omega^2 F_b}{48} + \frac{\omega^2|\Phi_b|}{24} \nn\\
 &&\qquad -\frac{5\,\omega^4 F_b^2}{24} -\frac{\omega^4|\Phi_b|^3}6 \bigg) \simeq -\frac{q^2}{(4\pi)^2} \left(\frac{3\,\delta\omega}{2} -\frac{|\Phi_b|}{2}\right)  \,.
\eea
%


\section{The 4D vacuum energy}
\label{sec:4DVE}

The previous sections show how to compute the 1PI potential, $\Vone = \cV_\infty + \cV_f$, obtained by integrating out low-spin bulk fields, and how these divergences are renormalized into various bulk and brane interactions, $\cV_\ssB$ and $\cV_b$, so that $\cV := \cV_\ssB + \sum_b \cV_b + \cV_f$ is finite. In this section we compute the implication of the above renormalizations for the effective 4D cosmological constant, $\Lambda$, and on-brane curvature as seen by a low-energy 4D observer. In the 4D theory $\Lambda$ would be obtained as the value of the low-energy 4D effective potential, $V$, after minimizing over any light scalar fields in the 4D effective theory.

If no branes had been present, a standard result for the low-energy potential would have been $V = \cV_\ssB + \cV_f$, so it may come as a surprise that once branes are included the potential $V$ is {\em not} simply given by $V := \cV$, suitably renormalized. Instead we must also recompute the classical contribution to the low-energy cosmological constant coming from integrating out KK bulk modes, keeping track of how the bulk back-reacts to the renormalization of the source branes, along the lines of refs.~\cite{localizedflux}. Neglecting this back-reaction would be inconsistent, since for codimension-two systems it is known to be of the same size as the direct effects of the changes to the brane lagrangians themselves \cite{Towards, SLEDrefs, TNCC}. Indeed, it is this back-reaction that allows flat solutions to exist at all at the classical level, despite the large classical positive tensions carried by each brane.

The logic to determining the cosmological constant of the low-energy effective theory is to compute within the 6D theory how perturbations to brane and bulk interactions change the predicted value for the curvature, $R_{\mu\nu}$, along the brane directions, and then to ask what cosmological constant in the effective 4D theory would give this same curvature. This is a special case of a `matching' calculation between the effective theory and its UV completion \cite{GREFTrev, JFDEFT}. It can be carried out fairly explicitly for small changes about a known background solution, as we do below following refs.~\cite{localizedflux}.

An alternative route to the same end is compute the value taken by the loop-corrected (1PI) action, $\Gamma := S + \Sigma$, evaluated at the background configuration that solves the loop-corrected field equations, $\delta \Gamma/\delta \psi = 0$. This lends itself well to the present purposes because the quantity $\Vone$ as computed in earlier sections is precisely such a contribution to the loop-corrected action evaluated at the background solution. For maximally symmetric geometries, the value of $\Lambda$ in the 4D low-energy effective theory can be found by comparing the result for $\Gamma$ obtained using the full 6D theory with the result for $\Gamma$ computed using the effective 4D theory.

\subsection{Classical bulk back-reaction}

It is useful first to review how back-reaction works at the classical level. For the 6D theory we solve the bulk field equations, eqs.~\pref{E:Beom}, using the renormalized bulk couplings. The renormalized brane couplings enter through the near-brane boundary conditions they imply for the bulk fields \cite{uvcaps}. Our interest is in starting with a classical solution for which $R_{\mu\nu} = 0$, and then ask how $R_{\mu\nu}$ changes as the various renormalized couplings change by small amounts. Of particular interest is how $R_{\mu\nu}$ responds to changes of the couplings in the brane action.

Recall that the rugby ball is sourced by identical brane actions, which to first order in a derivative expansion couple to the background fields by $S_b = - \int \exd^4x \sqrt{- \gamma} \; L_b$ with
\be \label{eq:Lbreplay}
 L_b = T_b-\frac{\cA_b}{2 \tilde g^2} \; \epsilon^{mn} F_{mn} + \cdots  \,,
\ee
and where (as before) $\gamma_{\mu\nu} := g_{\ssM\ssN} \partial_\mu x^\ssM \, \partial_\nu x^\ssN$ is the induced metric on the brane, and the ellipses denote terms involving two or more derivatives. As shown in detail in \cite{localizedflux}, the nominally subdominant $\cA_b$-term can play an important role in understanding the low-energy effective 4D curvature in those situations \cite{conical, Towards} where the stabilization of the extra dimensions arises as a competition between brane and bulk flux. In this case the influence of the $\cA$ term gets enhanced by the volume of the extra dimensions through its effect on the flux-quantization condition.

At the classical level, the source branes back-react onto the background solutions in two distinct ways. First, they change the boundary conditions of the bulk fields, schematically relating the near-brane limit, $\lim_{\rho\to0} (\rho  \,\partial\psi/\partial \rho)$, to the (appropriately renormalized) derivative of the brane action, $\delta S_b/\delta \psi$, for any bulk field $\psi$. In the special case that the functions $T_b$ and $\cA_b$ defining the brane action are independent of any bulk scalar fields, then these boundary conditions boil down to the familiar statement that the brane induces\footnote{More generally, if $T_b$ or $\cA_b$ depend on bulk scalars, back-reaction leads to a bulk curvature singularity at the brane positions \cite{uvcaps, localizedflux}.} a conical defect angle at each brane position, of size (see eq.~(\ref{deltab-Lb})): $\delta = \kappa^2  L_b = \kappa^2 \left( T_b - {\cA_b f}/{\tilde g^2} \right)$. The second way back-reaction influences the background is through the flux
quantization condition, eq.~\pref{eq:BLFquantization}, which depends on $T_b$ and $\cA_b$ because: ($i$) the defect angle changes the volume of integration for the flux, and ($ii$) because the $\cA_b$ term directly contributes as flux localized on the branes. Once both effects are included \cite{localizedflux}, the flux-quantization condition generalizes from eq.~\pref{E:fquant} to $2\pi N / q = \left( 4\pi\alpha \, r^2 f + \sum_b \cA_b \right)$, as before.

What must be done is to track how the bulk solutions react to loop-induced changes, $\delta T_b$, $\delta \cA_b$ (and others), to see how the changes in the bulk solutions appear in the low-energy effective theory.

\subsubsection*{A straw man}

Before doing so, it is worth first putting to rest a common misconception that can confuse issues at this point.

Schematically, our goal is to compute the loop-corrected action, $\Gamma = S + \epsilon \Sigma$, evaluated at the loop-corrected background field configuration, $\psi = \psi_0 + \epsilon \delta \psi$, with $\epsilon$ being the small loop-counting parameter. We do so in order to compare the result when performed in 6D and in 4D. Working to order $\epsilon$ gives
\bea
 \Gamma[\psi] &\simeq& S[\psi_0 + \epsilon \delta \psi] + \epsilon \, \Sigma[ \psi_0] + \cO(\epsilon^2) \nn\\
 &\simeq& S[\psi_0] + \epsilon \left( \left. \frac{\delta S}{\delta \psi} \right|_{\psi_0} \delta \psi + \Sigma[\psi_0] \right) + \cO(\epsilon^2) \,,
\eea
and it is tempting to argue that the first of the $\cO(\epsilon)$ terms vanishes because $\psi_0$ satisfies the classical field equations: $(\delta S/\delta \psi)_{\psi_0} = 0$. If so, then $\Gamma[\psi] = S[\psi_0] + \epsilon \, \Sigma[\psi_0] + \cO(\epsilon^2)$.

Although this argument is often true, if it were true in the present instance then there would be no need to know how bulk fields back-react --- {\em i.e.} to compute $\delta \psi$ --- in response to loop corrections to the action --- $\Sigma$ --- since it would suffice to evaluate both $S$ and $\Sigma$ at the uncorrected classical solution, $\psi_0$. If so, then the renormalized potential, $\cV$, computed in previous sections could be directly interpreted as the effective vacuum energy.

To see why this argument fails it is useful to examine a concrete example. To see why $S_{\rm eff}[\psi_0 + \epsilon \delta \psi] \ne S_{\rm eff}[\psi_0]$ at $\cO(\epsilon)$, consider the case
\bea
 S_{\rm eff} &=& - \int \exd^4x \sqrt{-g} \; \left( \Lambda_0 + \frac{R}{2\kappa^2_{4,0}}  \right) \nn\\
 \hbox{and} \quad
 \Gamma &=& - \int \exd^4x \sqrt{-g} \; \left( \Lambda + \frac{R}{2\kappa^2_{4}}  \right) \,,
\eea
where $\Lambda = \Lambda_0 + \delta \Lambda$ and $1/\kappa_4^2 = 1/\kappa^2_{4,0} + \delta(1/\kappa^2_4)$, with both corrections of order $\epsilon$. For simplicity, suppose further that $\Lambda_0 = 0$.

With these choices the unperturbed classical background satisfies $R_0 = 0$ while the background solving the full loop-corrected equations is $R_1 = - 4 \kappa_{4,0}^2 \delta \Lambda$. Consequently the classical action evaluated at the classical background is $S_{\rm eff}[g_0] = 0$, which disagrees with its evaluation at the loop-corrected solution:
\be
 S_{\rm eff}[g_0 + \delta g ] = - \int \exd^4x \sqrt{-g} \;  \frac{R_1}{2\kappa^2_{4,0}}
 = +2 \int \exd^4x \sqrt{-g_0} \; \delta \Lambda \,.
\ee

Formally, the reason $S_{\rm eff}[g_0+\delta g]$ can differ from $S_{\rm eff}[g_0]$ is because $S_{\rm eff}$ is actually proportional to the volume of spacetime for any constant nonzero curvature, and so diverges. Consequently precise statements must be regularized, such as by cutting the geometry off at a large radius. But then the action also contains a boundary, Gibbons-Hawking, term at this radius, and it is this boundary term that need not be stationary when evaluated at a solution to Einstein's equations. A similar phenomenon was noticed for the on-shell gravitational action in another context in ref.~\cite{duff}.

\subsection*{Loop-corrected back-reaction}

We now turn to how the bulk solutions respond to the presence of the loop-generated changes to brane and bulk interactions. Since these changes are small within the semiclassical approximation we can analyze this by following ref.~\cite{localizedflux} and solving the bulk field equations linearized about the classical background, using the brane-bulk boundary conditions to relate integration constants to properties of the brane action.

There are two qualitatively different kinds of loop-generated effects. First, renormalizations of the bulk action captured by $\cV_\ssB$ can change the features of the bulk classical solution, such as (but not restricted to) the generation of new terms in the bulk cosmological constant. Second, the renormalizations of the brane action included in $\cV_b$ change the bulk solutions in the two ways mentioned above: changing the boundary conditions and modifying the flux-quantization condition.

The calculation in \cite{localizedflux} can be used to include the influence of the brane renormalizations, once generalized to the loop corrections to the brane action found earlier in this paper. We now briefly describe the result of such a calculation. We first examine corrections to the lowest-derivative terms in the action, and return to the effects of the higher-derivative terms in the next section.

It suffices to work in a simple case to make the point that back-reaction is required to properly infer the implications of $\Vone$ for the 4D curvature. We therefore assume that all gauge fields that are zero in the classical background remain zero at loop level. The calculation is performed, without loss of generality, using Gaussian normal coordinates in the scalar-field target space centered at the background solution, so that the kinetic terms are canonical at this point. Finally, to avoid the issues of ref.~\cite{selflocalized} we imagine the branes only to couple to bulk scalars with vanishing bulk masses, in order not to have the brane and bulk compete in their implications for $\phi^i$, and thereby preclude the existence of constant background configurations.\footnote{Both examples of \cite{localizedflux} fall into this class}

We then imagine the bulk and brane actions to be chosen to admit a rugby-ball solution, and consider the corrections to this solution due to generic corrections to the coefficients
\ba
 \kappa^2 \to \kappa^2+\delta\kappa^2\,, \qquad U(\phi)\to U(\phi)+\delta U(\phi)\,, \qquad \tilde g^2(\phi)\to\tilde g^2(\phi)+\delta\tilde g^2(\phi)\,,
\ea
in the bulk and
\be
 T_b \to T_b + \delta T_b(\phi) \quad{\rm and}\quad \cA_b \to\cA_b + \delta\cA_b(\phi) \,,
\ee
on the branes. We ignore corrections to the target space metric, $\delta \cG_{ij}(\phi)$, since at linear order these do not contribute to the effective cosmological constant. We discuss higher-derivative corrections to the brane action below, but the bottom line is that they are suppressed by powers of the gravitational scale $\kappa$.

Following the steps laid out in ref.~\cite{localizedflux}, we first compute the on-brane curvature generated by these perturbations, by solving the linearized 6D field equations (assuming the initial rugby ball to be flat in the brane directions). The the effective 4D cosmological constant, $\Lambda$, is read off as that constant that would produce the same curvature using the 4D equations of motion: {\em i.e.} $R_4 = - 4 \kappa_4^2 \Lambda$. This gives\footnote{The first two terms in the last equality are ref.~\cite{localizedflux}'s result, since the corrections to $\delta \kappa$, $\delta U$ and $\delta \tilde g$ were not considered.}
\ba \label{eq:R4pert}
 \Lambda &:=& - \frac{\pi \alpha r^2}{\kappa^2} \, \Bigl( g^{\mu\nu} R_{\mu\nu} \Bigr)  = - \frac{\pi \alpha r^2}{\kappa^2} \left[\frac2{r^2} \(\frac{\delta f}f - \frac{\delta U}U - \frac{2\, \delta \tilde g^2}{\tilde g^2}\) \right]_{\phi_\star} \nn\\
 &=& \delta L_{{\rm tot}} - \frac{f \delta\cA_{\rm tot}}{\tilde g^2} + \frac{\pi \alpha}{\kappa^2} \left( \frac{4\,\delta\kappa^2}{\kappa^2} + \frac{3\delta U}{U} + \frac{\delta\tilde g^2}{\tilde g^2}\right)_{\phi_\star} \,,
\ea
where $f$ is defined (as before) by $F_{mn} = f \, \epsilon_{mn}$, while $\cA_{\rm tot} := \sum_b \cA_b$ and $L_{\rm tot} := \sum_b L_b$ with $L_b$ is as given in \pref{eq:Lbreplay}. Generically the right-hand-side of eq.~\pref{eq:R4pert} is evaluated at the loop-corrected stationary point for the bulk scalar field: $\phi_\star^i = \phi^i_0+\delta \phi^i$. When this is fixed from the classical bulk field equations then it suffices to evaluate eq.~\pref{eq:R4pert} at the classical solution, $\phi^i_0$, since keeping $\delta \phi^i$ would be subdominant in the loop expansion.

A more complicated situation arises if the bulk scalars parameterize flat directions along which there is no scalar potential at all in the absence of the perturbed bulk-brane couplings. This can occur if the brane couplings break an approximate symmetry of the bulk equations of motion, and can be used to generate a Goldberger-Wise \cite{GWise} type stabilization of the bulk geometry \cite{uvcaps}. In this case it is the perturbed brane-bulk couplings that stabilize these flat directions, with the stabilized point, $\phi^i_\star$, satisfying
\be
  \left[\frac{\kappa^2 \delta L_{{\rm tot},j}}{4\pi\alpha} + \frac{U_{,j} \delta f}{U f} + \frac{\delta U_{,j}}{\delta U} - \frac{(\delta\tilde g^2)_{,j}}{\tilde g^2} \right]_{\phi_\star} = 0 \,.
\ee
The first term here corresponds to the case studied in ref.~\cite{localizedflux}. Because we work perturbatively in $\delta \phi^i$, some care is required if $\phi_0^i$ labels a flat direction of the zeroth-order equations. In this case (as is standard for degenerate perturbation theory) $\delta \phi^i$ need only be small if the arbitrary unperturbed point is chosen near the minimum of the loop-corrected potential.

Although reasonably complicated in the general case, it is clear that the above result only agrees with $\V$ as computed earlier in the special case that the first term, $\delta L_{\rm tot}$, dominates eq.~\pref{eq:R4pert}. This is not true, in particular, in the supersymmetric examples that are the subject of a companion paper \cite{Companion} (for which the above formulae simplify considerably).

\subsection{Higher derivative corrections on the brane}

The above discussion is restricted to a tension and a localized flux, but in general loop corrections to the bulk and brane actions also include higher derivative terms, with the dimensions made up by factors of $\kappa$. Given our experience with the flux, which contributes at the same level as the tension despite being down one derivative, we pause to check that the other terms in the derivative expansion on the brane are indeed suppressed with respect to the tension and flux terms. The two ways in which higher derivative terms can contribute is by modifying the brane flux, and by modifying the matching condition that sets the defect angle at the branes.

The matching condition that relates the defect angle to the on-brane Lagrangian is more complicated if higher derivative corrections are taken into account. The precise statement of the matching condition in the general case is \cite{uvcaps}
\be
 \left[e^B\pd_\rho B\right]g^{\mu\nu}=g^{\mu\nu}+\frac{\kappa^2}{\pi\sqrt{-\gamma}} \left( \frac{\delta S}{\delta g_{\mu\nu}} \right) \,,
\ee
where the metric near the brane is assumed to take the form $\exd s^2=\exd\rho^2 + e^{2B(\rho)}\exd\theta^2$. For most fields, maximal symmetry in the brane directions guarantees that the right hand side of this matching condition only gets contributions from the variation of $\sqrt{-\gamma}$ in the action. An important exception to this are curvature terms on the brane, which we now explore.

A brane with the structure
\be
 S_b=-\int \exd^4x\sqrt{-\gamma}\( L_b + \frac{\zeta_{\ssR b}}{\kappa}R \)
\ee
has a metric variation
\be
 \frac1{\sqrt{-\gamma}}\frac{\delta S_b}{\delta g_{\mu\nu}} = \frac{\zeta_{\ssR b}}{\kappa} \left( R^{\mu\nu} - \frac{R}{2} \, g^{\mu\nu} \right) - \frac{L_b}{2} \, g^{\mu\nu}
 = - \left(\frac{\zeta_{\ssR b}R}{4\kappa} +\frac{L_b}{2} \right)g^{\mu\nu}\,,
\ee
where the last equality uses the maximal symmetry of the on-brane directions. Effectively this means that in eq.~\pref{eq:R4pert} we should replace $\delta L_{\rm tot} \to \delta L_{\rm{tot}} + (\zeta_{\ssR b} R_4)/(2\kappa)$, and re-solve for $R_4$. The result is
\ba
  \(1+\frac{\kappa \zeta_{\ssR b}}{2\pi\alpha r^2}\)R_4 &=& \left[ \frac{\kappa^2}{\pi\alpha \, r^2}\(\frac{f \delta\cA_{\rm tot}}{\tilde g^2} - \delta L_{{\rm tot}}\) - \frac{4\,\delta\kappa^2}{r^2\kappa^2} - \frac{3\delta U}{2U} -  \frac{\delta\tilde g^2}{2\tilde g^2}\right]_{\phi_\star} \,.
\ea
{}From this we see that the correction that is associated with a brane curvature term is of higher order in the expansion in $\kappa/r^2$.

The other contribution higher derivative terms can have is to modify the brane localized flux. In particular, the brane localized Maxwell term has this effect. Following the regularization of the brane flux in \cite{localizedflux} we find a divergent contribution to the gauge potential at a small distance, $\delta$, from the brane. Such divergences are normal for codimension-two matching \cite{uvcaps, GWise}, and the limit $\delta \to 0$ is to be taken after renormalization of the brane-bulk couplings. The divergent near-brane form for the bulk gauge field is in this case
\ba
 A_\theta(\delta) &=& \frac{\delta^2}{2} \left( \frac{f\tilde g^2 + \cA_b/\pi\delta^2}{1+\kappa\zeta_{\ssA b}/\pi\delta^2} \right) \nn\\
 &\approx& \frac{\delta^2 f \tilde g^2}{2} + \left( \frac{\cA_b-f\tilde g^2\kappa\zeta_{\ssA b}}{2\pi} \right) -\frac{\kappa}{\delta^2} \left( \frac{\cA_b\zeta_{\ssA b}}{2\pi^2} \right) \,.
\ea
The first term vanishes as $\delta \to 0$ and so can be ignored. The other correction terms are also small because consistency of the semiclassical approximation requires both $\kappa \tilde g^2 \ll 1$ and $\kappa\delta^{-2} \ll 1$. Again, we find corrections to the cosmological constant due to the higher derivative corrections to the brane are subleading in $\kappa$.

\section{Conclusions}
\label{sec:concl}

To repeat the summary of the introduction, in this paper we present an explicit calculation of the divergent part of the Casimir energy obtained from loops of low-spin bulk fields in a 6D geometry compactified on a flux-stabilized 2D rugby ball. We explicitly show how the UV divergences can be disentangled in order to identify separately how both bulk and brane interactions renormalize, and we compute the renormalization-group beta functions for each of the corresponding couplings.

Although our results are general, the first applications we envisage are to the cosmological constant problem within the framework of supersymmetric large extra dimensions \cite{Companion}. There are several features about this paper's results that are noteworthy for such applications.

\begin{itemize}
\item The first observation is that extra dimensions are not a free lunch in themselves. For large $m$, in general all positive powers of $m$ up to $m^6$ appear in the bulk loops, $\Vone \propto m^6 r^2$. This is as expected on dimensional grounds for a contribution from a 6D cosmological constant. Although the $m^6$ term cancels whenever there are equal numbers of bosons and fermions, in general the subdominant $\Vone \propto m^4$ term is still present. In the absence of something special (like unbroken supersymmetry \cite{UVsensitivity}), the UV sensitivity we find is precisely what would have been expected generically on dimensional grounds.
\item Second, although loops of heavy particles are dangerous in the bulk, in a world where all Standard Model particles live on a brane nothing really requires there to be any heavy bulk particles. If all bulk masses were generically of order $1/r$ then a single bulk loop need not be so dangerous, since $V$ is of generic order $V \sim C/(4 \pi r^2)^2$. In such a case it would be two-loop and higher contributions that would instead be worrisome, since these could introduce the larger brane-localized masses of the Standard Model back into the result. What is required is a mechanism that suppresses higher loops, and here again supersymmetry is likely to be relevant, as described in Refs. \cite{Towards,TNCC}.
\item Should a mechanism ensure that the 4D vacuum energy is indeed of order $1/(4 \pi r^2)^2$ (as we believe to be the case for a bulk supergravity\footnote{Although it is tempting to ask what the 4D picture is for this mechanism, it is not clear that such a picture must exist given the central role played by back reaction (whose dynamics occurs above the KK scale \cite{laterconical}, where the low energy 4D effective theory breaks down). What the 4D effective theory {\em should} be able to do, however, is to explain why the 4D curvature remains precisely zero when working purely at the classical level for the bulk, regardless of the energy density on the branes. A proper description of how this happens is under development, but goes well beyond the scope of the present article.} \cite{Companion}), then it is the Kaluza-Klein scale, $1/r$, that would set the size of the cosmological constant. This could be acceptably small if $r$ is chosen not to be far below its current upper limit, $r \lsim 45$ microns, arising from tests of Newton's laws at short distances \cite{InvSqTests}. (Modifications of Newton's law from supersymmetric large extra dimensions have been studied in \cite{SLEDInvSqLaw}.) Notice that the numerical factor of $16 \pi^2$ in the vacuum energy is an important part of why these scales can be compatible.
\item Finally, even should the numerical size of the Casimir energy be right, its sign is also important. It is in this context that the discussion of \S\ref{sec:4DVE} is most important, since it shows that the sign of the low-energy cosmological constant need not simply be the sign of $\V$ as computed in \S\ref{sec:genloops}.
\end{itemize}

In any event, there is no substitute for a real calculation for which we regard the results of this paper as a crucial first step.

\newpage
\section*{Acknowledgements}

We thank Riccardo Barbieri, Gregory Gabadadze, Hyun-Min Lee, Susha Parameswaran, Oriol Pujol\`as, Fernando Quevedo, Seifallah Randjbar-Daemi, George Thompson and Itay Yavin for useful discussions. Various combinations of us are grateful for the support of, and the pleasant environs provided by, the Abdus Salam International Center for Theoretical Physics, and AS thanks the Perimeter Institute and McMaster University for its hospitality, while thinking about these problems. CB's research was supported in part by funds from the Natural Sciences and Engineering Research Council (NSERC) of Canada. Research at the Perimeter Institute is supported in part by the Government of Canada through Industry Canada, and by the Province of Ontario through the Ministry of Research and Information (MRI). The work of AS was supported by the EU ITN ``Unification in the LHC Era", contract PITN-GA-2009-237920 (UNILHC) and by MIUR under contract 2006022501.

\appendix

\section{Heat kernels and bulk renormalization}
\label{app:gilkeydewitt}

In this appendix we collect for convenience the explicit
expressions for the heat-kernel coefficients for manifolds without singularities and boundaries, and specialize the results to the case where the two extra dimensions are a 2-sphere.

The small-$t$ expansion for the heat-kernel representation of the one-loop vacuum energy described in the main text can be evaluated for a broad class of geometries in great generality \cite{GdWrev}. Such generality is possible because the small-$t$ limit physically corresponds to the coincidence limit of the corresponding propagator, and this does not `know' about the boundary conditions and topology of the space if the coincident points are taken far from any boundaries or branes.

\subsection*{Gilkey-de Witt coefficients}

This section collects the results for the ultraviolet-divergent parts of the one-loop action obtained by integrating out various kinds of particles in 6 dimensions. To this end, consider a collection of $N$ fields, assembled into a column vector, $\Psi$, and coupled to a background spacetime metric, $g_{\ssM\ssN}$, scalars, $\varphi^i$, and gauge fields, $A^a_\ssM$. We suppress the gauge and/or Lorentz indices to which these fields couple, leading to a background-covariant derivative, $D_\ssM$, of the form
\be
    D_\ssM \Psi = \partial_\ssM \Psi + \omega_\ssM \, \Psi
    - i A^a_\ssM \, t_a \Psi \,,
\ee
where $\omega_\ssM$ is the matrix-valued spin connection, and the gauge group is represented by the hermitian matrices $t_a$. For real fields the $t_a$ are imaginary antisymmetric matrices, which (for canonically-normalized gauge bosons) include a factor of the corresponding gauge coupling, $g_a$. The commutator of two such derivatives defines the matrix-valued curvature, $Y_{\ssM\ssN} \Psi = [D_\ssM, D_\ssN] \Psi$, which has the following form:
\be
    Y_{\ssM\ssN} = \cR_{\ssM\ssN} -i F^a_{\ssM\ssN} \, t_a \,.
\ee
Here $\cR_{\ssM\ssN}$ is the curvature built from the spin connection $\omega_\ssM$, which is also related to the Riemann curvature of the background spacetime in a way which is made explicit in what follows.

Integrating out the fields $\Psi$ at one loop often leads to the following contribution to the quantum action
\beq     
    i\Sigma = - (-)^F \, \frac{1}{2} \, \Tr \log
    \Bigl( - \Box + X + m^2 \Bigr) \,,
\eeq
where $(-)^F = +$ for bosons and $-$ for fermions. As before, $\Box = g^{\ssM\ssN} D_\ssM D_\ssN$ and the quantity $X$ is some local quantity built from the background fields. The mass matrix, $m^2$, are constants.

Our interest is in that part of $\Sigma$ for which the functional trace is ultraviolet divergent. To identify the divergent part we work within dimensional regularization and so continue the spacetime dimension to complex values, $n$, which are slightly displaced from the actual integer spacetime dimension, $6$, which is of interest: $n = 6 - 2\epsilon$. We then follow the poles in $\Sigma$ as $\epsilon \to 0$, in the usual fashion. Notice that this continuation to $d = 6$ differs from the regularization used for the Casimir energy calculation, for which the limit $d \to 4$ was taken.

For 6D spaces without boundaries and singularities the resulting divergent terms are simply characterized. They can be written as \cite{GdWrev}
\beq \label{sigmainfty}
    \Sigma_\infty =
    \frac{1}{2(4 \pi)^3} \, (-)^F
    \sum_{k=0}^{3} \Gamma ( k - 3 + \epsilon)
    \int d^6 x \sqrt{-g} \; \tr[ m^{6-2k}\, a_k ]
\eeq
where $\Gamma(z)$ denotes Euler's gamma function. The divergence as $\epsilon \to 0$ is contained within the gamma function, which has poles at non-positive integers of the form $\Gamma(-r +\epsilon) = (-)^r/(r!\epsilon)+\cdots$, for $\epsilon$ an infinitesimal and $r$ a non-negative integer. The coefficients, $a_k$, are known matrix-valued local quantities constructed from the background fields, to which we return below. The trace is over the matrix indices of the $a_{k}$, of which there are $N' = N \, d$ with $N$ counting the number of fields and $d$ being the dimension of the relevant Lorentz representation.

The above expression shows that for massless fields ($m = 0$) in compact spaces without boundaries and singularities in 6
dimensions the divergent contribution is proportional to
$\tr[a_{3}]$, so the problem reduces to the construction of this coefficient. By contrast, for massive fields there are also divergences proportional to $\tr [ m^6 a_0]$, $\tr[ m^4 a_1]$, $\tr[ m^2 a_2]$ and $\tr[a_3]$. (Notice that the freedom to keep $m^2$ within or separate from $X$ implies that the divergence obtained from computing just $a_3$ using $X_m = X + m^2$ gives the same result as computing $a_0$ through $a_3$ using only $X$.)

An algorithm for constructing the coefficients $a_k$ is known for general $X$ and $D_\ssM$, involving and the result for the first few has been computed explicitly \cite{GdWrev} and can be given as a closed form in terms of $X$, background curvatures and the generalized curvature $Y_{\ssM\ssN}$. The first few coefficients are given explicitly by \cite{GdWrev}:\footnote{In comparing with this
reference recall that our metric is `mostly plus' and we adopt Weinberg's curvature conventions \cite{GandC}, which for the Riemann tensor agree with those of ref.~\cite{GdWrev}, but disagree with this reference by a sign for the Ricci tensor and scalar.}
\begin{eqnarray}
  \label{eqn: gilkey}
  a_0 &=& I \\
  a_1 &=& -\frac{1}{6}(R+6X)  \\
  a_2 &=& \frac{1}{360} \left( 2 \Riem2 - 2 \Ricci2 + 5 R^2 -12\, \Box R \right)
   \nonumber \\  &&+ \frac{1}{6} R X + \frac{1}{2} X^2 - \frac{1}{6} \Box X +
  \frac{1}{12} \y2 \\
  a_3 &=& \frac{1}{7!} \left( - 18 \, \Box^2 R + 17 D_\ssM R D^\ssM R
  -2 D_\ssL R_{\ssM\ssN} D^\ssL R^{\ssM\ssN}
  -4 D_\ssL R_{\ssM\ssN} D^\ssN R^{\ssM\ssL} \phantom{\frac12} \right. \nonumber\\
  && + 9 D_\ssK R_{\ssM\ssN\ssL\ssP} D^\ssK
    R^{\ssM\ssN\ssL\ssP} +28 R \Box R - 8 R_{\ssM\ssN} \Box R^{\ssM\ssN}
    +24 {R^\ssM}_{\ssN} D^\ssL D^\ssN R_{\ssM\ssL} \nonumber\\
    &&+ 12 R_{\ssM\ssN\ssL\ssP} \Box R^{\ssM\ssN\ssL\ssP}
    - \frac{35}{9} \, R^3 + \frac{14}{3} \, R \,\Ricci2
    - \frac{14}{3} \, R \, \Riem2 \nonumber \\
    && + \frac{208}{9} \, {R^\ssM}_\ssN \, R_{\ssM\ssL} \, R^{\ssN\ssL}
    - \frac{64}{3} \, R^{\ssM\ssN} \, R^{\ssK\ssL} \, R_{\ssM\ssK\ssN\ssL}
    + \frac{16}{3} \, {R^\ssM}_\ssN \, R_{\ssM\ssK\ssL\ssP} \, R^{\ssN\ssK\ssL\ssP} \nonumber\\
    && \left. - \frac{44}{9} \, {R^{\ssA\ssB}}_{\ssM\ssN} \, R_{\ssA\ssB\ssK\ssL}
    \, R^{\ssM\ssN\ssK\ssL} - \frac{80}{9} \, {{{R^\ssA}_\ssB}^\ssM}_\ssN \,
    R_{\ssA\ssK\ssM\ssP} \, R^{\ssB\ssK\ssN\ssP} \right)  \nonumber \\
    &&+ \frac{1}{360} \left( 8 D_\ssM Y_{\ssN\ssK} \, D^\ssM Y^{\ssN\ssK}
    +2 D^\ssM Y_{\ssN\ssM} \, D_\ssK Y^{\ssN\ssK} + 12 Y^{\ssM\ssN} \Box Y_{\ssM\ssN}
    \phantom{\frac12} \right.  \\
    && - 12 {Y^\ssM}_\ssN \, {Y^\ssN}_\ssK \, {Y^\ssK}_\ssM - 6 R^{\ssM\ssN\ssK\ssL} \, Y_{\ssM\ssN}
    \, Y_{\ssK\ssL} +4 {R^\ssM}_\ssN \, Y_{\ssM\ssK} \, Y^{\ssN\ssK} \nonumber \\
    && - 5 R \, Y^{\ssM\ssN} \, Y_{\ssM\ssN} - 6 \Box^2 X + 60 X \Box X
    +30 D_\ssM X \, D^\ssM X - 60 X^3
    \nonumber \\
    && - 30 X \, Y^{\ssM\ssN} \, Y_{\ssM\ssN} + 10 R \, \Box X + 4 R^{\ssM\ssN}
    \, D_\ssM D_\ssN X +12 D^\ssM R \, D_\ssM X
    -30 X^2 \, R \nonumber \\
    && \left. \phantom{\frac12}
    + 12 X \, \Box R - 5 X \, R^2 + 2 X \, \Ricci2
    -2 X \, \Riem2 \right) \,. \nonumber
\end{eqnarray}
Here $I$ is the $N \times N$ identity matrix and $\Y$ is the matrix-valued quantity defined above in terms of the commutator to two covariant derivatives.

\subsubsection*{Bulk Counterterms for Spheres}

Since our applications are to compactifications on spaces which are spheres, it suffices to specialize the general results of the appendix to these simpler background field configurations.

Consider therefore 6D spacetime geometries which are the product of 4D Minkowski space with a maximally-symmetric 2D manifold:
\be
    \exd s^2 = \eta_{\mu\nu} \, \exd x^\mu \exd x^\nu + g_{mn} \,
    \exd y^m \exd y^n \,,
\ee
where maximal symmetry for the 2D metric implies $R_{mnpq} =
\frac12 \, R \, (g_{mp} g_{nq} - g_{mq} g_{np} )$, $R_{mn} =
\frac12 \, R \, g_{mn}$ and $D_m R = 0$, and so $R_{mnpq} R^{mnpq} = 2 R_{mn} R^{mn} = R^2$. We also take any background scalars to be constants, and allow only a single background gauge field to be nonzero, and take it to be proportional to the 2D volume form: $F_{mn} = f \, \epsilon_{mn}$, for some scalar $f$.

With these choices the only nonzero components of $Y_{\ssM \ssN}$ lie in the 2 dimensions, $Y_{mn}$, and all of the curvatures are covariantly constant. The coefficients $a_1$ through $a_3$ simplify considerably, reducing to \cite{GdWrev}:
\bea \label{gilkeysc}
   a_0 &=& I  \\
   a_1 &=& -\frac{1}{6} \, R - X \\
   a_2 &=& \frac{1}{60} \, R^2 + \frac{1}{6} \, R X +
   \frac{1}{2} \, X^2  +
   \frac{1}{12} \, Y_{mn} Y^{mn}  \\
   a_3 &=& - \, \frac{1}{630} \, R^3  - \frac{1}{30} \,
   {Y^m}_n \, {Y^n}_l \, {Y^l}_m - \frac{1}{40}
      R \, Y_{mn} Y^{mn}  - \frac{1}{12} \, X \,
      Y_{mn} Y^{mn} \nonumber \\
   && \qquad - \frac16 \, X^3 -\frac{1}{12} \, X^2 \, R
      - \frac{1}{60} \, X \, R^2  \,.
\eea
Here $I$ is the $N' \times N'$ identity matrix, and $X$ and
$Y_{mn}$ are the $N' \times N'$ matrix-valued quantities defined above. These expressions may be used to compare with the bulk part of the ultraviolet divergences encountered in the explicit calculations of the main text.

\subsection*{Scalars, spinors and gauge fields}

This section collects the results for $X$ and $Y_{\ssM \ssN}$, and so also for the ultraviolet-divergent parts of the one-loop action, for several kinds of particles in 6 dimensions. Attention is restricted to those fields that do not mix appreciably with the gravity sector, and this means in particular that any gauge fields considered cannot be those whose background flux stabilizes the extra dimensions.

\subsubsection*{Scalars}

For spinless fields we begin with the following general
scalar-field action involving two derivatives or less,
\be \label{scalaraction}
     S = - \int \exd^6 x \sqrt{-g} \; \left[ \frac12 \, g^{\ssM \ssN} \,
     G_{ij} \, D_\ssM \Phi^i D_\ssN \Phi^j
     + V  + \frac12 \, U \, R +
     \frac14 \, \cH \, F^a_{\ssM \ssN} F_a^{\ssM \ssN}
     \right] \,,
\ee
for a collection of $N$ real scalar fields, $\Phi^i$, coupled to a background metric, $g_{\ssM \ssN}$, and gauge fields, $A^a_\ssM$. The functions $U$, $V$, $W$ and the target-space metric, $G_{ij}$, are imagined to be known functions of the $\Phi^i$. The background-covariant derivative appropriate to this case is:
\be
    D_\ssM \Phi^i = \partial_\ssM \Phi^i - i A^a_\ssM \, {(t_a)^i}_j \Phi^j
     \,,
\ee
where the matrices ${(t_a)^i}_j$ represent the gauge group on the scalars.

To compute the one-loop quantum effects of scalar fluctuations we linearize this action about a particular background configuration, $\varphi^i$, according to: $\Phi^i = \varphi^i + \phi^i$, where $\partial_\ssM \varphi = 0$. Expanding the classical action to quadratic order in $\phi^i$ reveals the kinetic operator ${\Delta^i}_{j}$, which is given by
\be
    {\Delta^i}_{j} = - {\delta^i}_{j} \, \Box  + {X^i}_{j}
     \,,
\ee
with ${X^i}_{j}$ given by
\be
    {X^i}_{j} = G^{ik} \Bigl[ V_{kj}(\varphi) + \frac12 \, R \, U_{kj}(\varphi) + \frac14 \, F^a_{\ssM\ssN} F_a^{\ssM\ssN} \, \cH_{kj}(\varphi) \Bigr] \,.
\ee
In this last expression the subscripts on $U$, $V$ and $\cH$ denote differentiation with respect to the background field $\varphi^i$. Specializing to the simple 2D geometries and Maxwell fields discussed earlier, these simplify to ${X^i}_{j} = G^{ik}[ V_{kj} + \frac12 \, R \,U_{kj} + \frac{1}{2} \, f^2 \, \cH_{kj}]$ and $Y_{mn} = -i\tilde{g} f \, Q \, \epsilon_{mn}$, where $\tilde{g}$ is the gauge coupling and $\tilde{g}Q = t_a$ is the hermitian, antisymmetric charge matrix for the background gauge field. Notice that these imply $Y_{mn} Y^{mn} = -2 \tilde{g}^2 f^2 \, Q^2$ and ${Y^m}_n \, {Y^n}_l \, {Y^l}_m = 0$.

Since scalars transform trivially under Lorentz transformations, $d = 0$ and so $N' = N$. The $a_k$ are then $N \times N$ matrices, with the trace of $a_0$ through $a_3$ given by
\be \label{gilkeyscalar0}
   \tr a_0 = N \,, \qquad
   \tr a_1 = -\frac{N}{6} \, R  - \tr X \,,
\ee
and
\ba \label{gilkeyscalar}
   \tr a_2 &=& \frac{N}{60} \, R^2  + \frac{1}{6} \, R \, \tr X + \frac{1}{2} \, \tr X^2  -
   \frac{1}{6} \, \tilde{g}^2 f^2 \, \tr Q^2  \\
   \tr a_3 &=& - \, \frac{N}{630} \, R^3  + \frac{1}{20}
      R \, \tilde{g}^2 f^2 \, \tr Q^2  + \frac{1}{6} \, \tilde{g}^2 f^2 \, \tr (X Q^2) \nn\\
      && \qquad\qquad\qquad\qquad
      - \frac16 \,\tr X^3 -\frac{1}{12} \, R \, \tr X^2
      - \frac{1}{60} \, R^2 \, \tr X  \,.
\ea
These give explicit functions of $\varphi$ once the above expression for $X$ is used.

\subsubsection*{Fermions}

For $N$ 6D massless spinors, $\psi^a$ with $a = 1,...,N$, we take the following action
\be
     S = - \int \exd^6x \sqrt{-g} \;  \frac12 \,
     G_{ab}(\varphi) \, \overline\psi^{\;a} \nott{D} \psi^b \,,
\ee
where $\nott{D} = {e_\ssA}^{\ssM} \, \gamma^\ssA D_\ssM$ with
\be
 D_\ssM \psi^a = \partial_\ssM \psi^a - \frac14 \, \omega_\ssM^{\ssA\ssB} \gamma_{\ssA\ssB} \psi - i A_\ssM^a t_a \psi \,,
\ee
with $\gamma^\ssA$ being the 6D Dirac matrices and ${e_\ssA}^\ssM$ the inverse sechsbein, $\gamma_{\ssA\ssB} = \frac12 [ \gamma_\ssA, \gamma_\ssB]$, and $t_a$
denotes the gauge-group generator acting on the spinor fields. Since 6D Weyl spinors have 4 complex components their representation of the 6D Lorentz group has $d = 8$ real dimensions.

The differential operator which governs the one-loop contributions is in this case $\nott{D} = {e_\ssA}^\ssM \gamma^\ssA D_\ssM$ and so in order to use the general results of the previous section we write (assuming there are no gauge or Lorentz anomalies) $\log \det \nott{D} = \frac{1}{2} \log \det(-\nott{D}^2)$, which implies
\begin{eqnarray}
\label{eqn: sigmaspin1/2}
    i \Sigma_{1/2} &=& \frac12 \, \Tr \log \nott{D}  =
    \frac{1}{4} \Tr \log \left( -
    {\nott{D}}^2 \right) \nonumber \\
        &=& \frac{1}{4} \Tr \log \left(-\Box - \frac{1}{4} R + \frac{1}{4} \gamma^{\ssA \ssB} F^a_{\ssA \ssB} t_a \right) \,.
\end{eqnarray}
This allows us to adopt the previous results for the ultraviolet divergences, provided we divide the result by an overall factor of 2 (and so effectively $d=4$ instead of 8), and use
\be
    X =  -\frac{1}{4} \, R  + \frac{1}{4} \,
    \gamma^{\ssA \ssB} \, F^a_{\ssA \ssB} \, t_a  \,.
\ee
Similarly, we find
\be
    Y_{\ssM\ssN} = -\, \frac{i}{2} \, R_{\ssM\ssN\ssA\ssB} \gamma^{\ssA\ssB} -i {F^a}_{\ssM\ssN} t_a
    \,,
\ee
and so\footnote{We adopt the convention of using $\Tr[...]$ to denote a trace which includes the Lorentz and/or spacetime indices, while reserving $\tr[...]$ for those which run only over the `flavor' indices which count the fields of a given spin.}
\ba
    \Tr_{1/2}[\y2] &=& - 4 \, \tr_{1/2}(t_a t_b) \, F^a_{\ssM\ssN} F^{b\ssM\ssN}
    -\frac{N}{2} \, \Riem2 \nonumber \\
    &=& - 8\, \tilde{g}^2 f^2 \, \tr_{1/2}(Q^2) - \frac{N}{2} \, R^2  \,,
\ea
where the second line specializes to the simple spherical 2D
geometry and background gauge fields discussed earlier.

Keeping explicit the sign due to statistics, and dropping terms
which vanish when traced, this leads to the following expressions
for the divergent contributions of $N$ 6D Weyl fermions:
\begin{eqnarray}
    \label{gilkeyspinor}
    &&(-)^F \, \Tr_{1/2}[a_0] = - 4N \,, \qquad
    (-)^F \, \Tr_{1/2}[a_1] = -\frac{N}{3} \, R   \\
    &&(-)^F \, \Tr_{1/2}[a_2] =  \frac{N}{60} \, R^2 - \frac{4}{3} \,
    \tilde{g}^2 f^2 \, \tr_{1/2}(Q^2) \\
    &&(-)^F \, \Tr_{1/2}[a_3] = -\, \frac{N}{504} \, R^3 + \frac{2}{15}
    \, \tilde{g}^2 f^2 \, R\, \tr_{1/2}(Q^2)  \,.
\end{eqnarray}

\subsubsection*{Gauge bosons}

For $N$ gauge bosons, ${\cA}^a_\ssM$, with field strength
$\cF^a_{\ssM\ssN}$ and $a = 1,...,N$, we use the usual Yang-Mills action
\be
     S = - \int \exd^6x \sqrt{-g} \;  \frac14 \, \cH(\varphi)
     \, \cF^a_{\ssM\ssN} \cF_a^{\ssM\ssN}  \,,
\ee
expanded to quadratic order about the background fields: $\cA^a_\ssM = A_\ssM^a + \delta A^a_\ssM$. For an appropriate choice of gauge the differential operator which governs the loop contributions becomes
\be
    {\Delta^{a\ssM}}_{b\ssN} = - {\delta^a}_b \, {\delta^\ssM}_\ssN \Box
    + {X^{a\ssM}}_{b\ssN} \,,
\ee
with
\be
    {X^{a\ssM}}_{b \ssN} = -
    {R^\ssM}_{ \ssN} {\delta^a}_{ b} + 2i
    {({t}_c)^a}_{b} {F^{c \ssM}}_{\ssN}  \,,
\ee
where $t_c$ here denotes a gauge generator in the adjoint
representation.

The dimension of the 6-vector representation of the Lorentz group is in this case $d = 6$. We therefore find $\Tr_\ssV I= 6 N$, $\Tr_\ssV(X) = - N \, R$, and
\ba
    \Tr_\ssV(X^2) &=& N\, \Ricci2 + 4 \, C(A) \, \f2 \\
    \Tr_\ssV(\y2) &=& - N \, \Riem2 - 6 \, C(A) \, \f2 \,,
\ea
where $C(A)$ is the Dynkin index for $N$ fields in the adjoint representation, defined by $\tr(t_a t_b) = C(A) \, \delta_{ab}$. The subscript `$V$' in these expressions is meant to emphasize that the trace has been taken over a vector field (as opposed to the physical spin-1 field, including ghosts).

These expressions suffice to compute $\Tr_\ssV[a_k]$, for the vector field. Once specialized to the spherical geometries and background gauge field of interest we find
\begin{eqnarray} \label{gilkeyvector}
 &&(-)^F \, \Tr_\ssV[a_0] = 6 N \,, \qquad
 (-)^F \, \Tr_\ssV[a_1] = 0  \\
 &&(-)^F \, \Tr_\ssV[a_2] = \frac{N}{10} \, R^2
 + 3 \tilde{g}^2 f^2 \, \tr(Q^2)  \\
 && (-)^F \, \Tr_\ssV[a_3] = - \frac{N}{105} \, R^3 + \frac{4}{5}
 \, \tilde{g}^2 f^2 \, R \, \tr(Q^2) \,.
\end{eqnarray}

To this we must add the ghost contribution, which consists of $N$
complex scalar fields having fermionic statistics and transforming
in the adjoint representation of the gauge group. The
contributions to the $a_k$ may be read off from our
previously-quoted expressions for scalar fields in the special
case $X = 0$. For such fields we have
\ba \label{gilkeyghost}
   &&(-)^F \, \Tr_{gh}[a_0] = -2N \,, \qquad
   (-)^F \, \Tr_{gh}[a_1] = \frac{N}{3} \, R    \\
   &&(-)^F \, \Tr_{gh}[a_2] = -\, \frac{N}{30} \, R^2
      + \frac{1}{3} \, \tilde{g}^2 f^2 \, \tr(Q^2)  \\
   &&(-)^F \, \Tr_{gh}[a_3] =  \frac{N}{315} \, R^3 - \frac{1}{10}
       \, \tilde{g}^2 f^2 \, R \, \tr(Q^2)  \,.
\ea

Summing the contributions of eqs.~(\ref{gilkeyvector}) and
(\ref{gilkeyghost}) gives the contribution of $N$ physical 6D
massless gauge bosons:
\begin{eqnarray} \label{gilkeyspin1}
 &&(-)^F \, \Tr_1[a_0] = 4 N \,, \qquad
 (-)^F \, \Tr_1[a_1] =  \frac{N}{3} \, R \\
 &&(-)^F \, \Tr_1[a_2] = \frac{N}{15} \, R^2
 + \frac{10}{3} \, \tilde{g}^2 f^2 \, \tr_1(Q^2)  \\
 && (-)^F \, \Tr_1[a_3] = - \frac{2N}{315} \, R^3 + \frac{7}{10}
 \, \tilde{g}^2 f^2 \, R \, \tr_1(Q^2) \,.
\end{eqnarray}

\section{Sums and zeta functions}

This appendix has two purposes. The first section provides a more secure theoretical foundation for many of the zeta-function calculations of the main text; while the second section provides formulae for several useful sums encountered in the calculations.

\subsection*{Justifying zeta function regularization}
\label{app:honestcalc}

The purpose of this section is to compute the sum
\be
d^{\mathrm{reg}}(\omega,t,a_0) := \sum_{j=0}^\infty \sum_{k\neq 0} \cG(2\pi k) \,,
\ee
where
\be
\cG(q) := \int_{-\infty}^\infty \! \exd x \, \exp[-t(j + \omega |x| + a_0)^2]\, e^{-iqx} \,,
\ee
\emph{without} resorting to the use of zeta function regularization. Pursuing this exercise will show that the zeta function approach used later is valid.

Computing this Fourier transform gives
\be \label{cG}
\cG(q) = \frac{1}{2\omega} \sqrt{\frac{\pi}{t}}\, e^{-\bar q^2} \left[ e^{2i\bar j \bar q} \bigg( 1 - \erf(\bar j +i \bar q) \bigg) + c.c. \right]
\ee
where
\be
\bar j:= \sqrt{t} (j+a_0) \,,\quad \bar q:= \frac{q}{2\omega\sqrt{t}} \,.
\ee
Since $\cG(-q) = \cG(q)$, we can write
\be \label{sumcGfromk1}
d^\mathrm{reg}(\omega,t,a_0) = 2 \sum_{j=0}^\infty \sum_{k=1}^\infty \cG(2\pi k) \,.
\ee

When evaluating $\cG(q)$ for non-zero $q$, it is important to note that the overall factor of $\exp(-q^2/(4\omega^2 t))$ in eq.~\pref{cG} will exponentially suppress any quantity multiplying it that does not diverge in the $t\rightarrow 0$ limit. Since we are only interested in the $t\rightarrow 0$ limit on this sum, we can therefore approximate $\cG(q)$ as follows:
\bea
\cG(q) &\simeq& -\frac{1}{2\omega} \sqrt{\frac{\pi}{t}}\, e^{-\bar q^2} \left[ e^{2i\bar j \bar q} \, \erf(\bar j +i \bar q) + c.c. \right] \nn\\
&=& -\frac{1}{\omega\sqrt{t}} \, e^{-\bar q^2} \left[ e^{2i\bar j \bar q} \left( \int_0^{\bar j}\! dx\, e^{-x^2} + \int_{\bar j}^{\bar j+i\bar q}\! dx\, e^{-x^2} \right) + c.c. \right] \nn\\
&\simeq& -\frac{1}{\omega\sqrt{t}} \, e^{-\bar q^2} \left[ i \, e^{2i\bar j \bar q}  \int_{0}^{\bar q}\! du\, e^{-(iu+\bar j)^2}  + c.c. \right] \,.
\eea
(Here, we have taken $u = -i(x-\bar j)$.) It is also helpful to make the following substitutions:
\bea
\cG(q)&=& \frac{2}{\omega\sqrt{t}} \, e^{ - \bar q^2} \int_{0}^{\bar q}\! du\, e^{u^2-\bar j^2} \sin [2\bar j (\bar q-u)] \nn\\
&=& \frac{2\bar q}{\omega\sqrt{t}} \, e^{ - \bar q^2} \int_{0}^{1}\! dy\, e^{\bar q^2(1-y)^2} \left( e^{-\bar j^2} \sin (2\bar j \bar q \, y) \right)
\eea
where $y := (\bar q-u)/\bar q$. Since all of the $j$-dependence is in the bracketed term within the above integral, we can consider first carrying out the $j$-sum and then performing the integral over $y$:
\be \label{Gjsum}
\sum_{j=0}^\infty \cG(2\pi k) = \frac{2\bar k}{\omega t} \, e^{ - \bar k^2/t} \int_{0}^{1}\! dy\, e^{\bar k^2(1-y)^2/t} \underbrace{\left( \sum_{j=0}^\infty e^{-\bar j^2} \sin (2\bar j \bar k \, y/\sqrt{t}) \right) }_{:=R(\bar k y,t)}
\ee
where we have introduced the notation $\bar k = \pi k /\omega$. Calculating $R(\bar k y,t)$ can be done as before, using the Euler-Maclaurin formula in eq.~\pref{EulerMac}. Defining
\be
g(x+a_0;\bar k y,t) := e^{-t (x+a_0)^2} \sin \big(2 (x+a_0) \bar k y \big) \,,
\ee
this gives
\be
R(\bar k y,t) = \int_{0}^\infty \! dx \, g(x+a_0; \bar k y,t) - \sum_{i=1}^\infty \frac{B_i}{i!} g^{(i-1)}(a_0;\bar k y,t)
\ee
where $g^{(i-1)}(a_0;\bar k y,t)$ denotes the $(i-1)$-th derivative of $g(x;\bar k y,t)$ with respect to $x$ evaluated at $x=a_0$. The integral can be simplified as follows:
\bea
\int_{0}^\infty \! dx \, g(x+a_0;\bar k y,t) &=& \int_{0}^{-a_0} \! dx  \, g(x+a_0;\bar k y,t) + \int_{-a_0}^\infty \! dx  \, g(x+a_0;\bar k y,t) \nn\\
&=& \int_0^\infty \! dx \, g(x;\bar k y,t) -\int_0^{a_0} \! dx  \, g(x;\bar k y,t) \nn\\
&=&\frac{1}{\sqrt{t}} \left(D\left(\frac{\bar k y}{\sqrt{t}}\right) - \int_0^{\sqrt{t} a_0} \! dx  \, g(x;\bar k y/\sqrt{t},1)\right)
\eea
%
%
where
\be \label{DawsonInt}
D(w) := \int_0^\infty \! dx \, e^{-x^2} \sin (2 x w) 
\ee
is known as Dawson's integral -- a real, analytic function with known series expansions about $w=0$ and $w\rightarrow \infty$. The second integral can be reliably Taylor expanded in its upper limit, since the $t$-dependence of the integrand is contained within the sine function (which is guaranteed not to diverge). Furthermore, since the integrand is an odd function of $x$, only even powers of $\sqrt{t} a_0$ will appear. The coefficients in such an expansion will be related to derivatives of $g(x;\bar k y,t)$ (with respect to $x$):
\be
\int_0^{\sqrt{t} \, a_0} \! dx \, g(x;\bar k y/\sqrt{t},1) = \sum_{i=1}^\infty \frac{g^{(2i-1)}(0;\bar k y/\sqrt{t},1)}{(2i)!} (t \, a_0^2)^i \,.
\ee
%

Given all of this, we see that the sum over $\cG(2\pi k)$ can be divided into three contributions:
\be
\sum_{j=0}^\infty \cG(2\pi k) = \frac{2}{\omega t} \bigg[ G_1(\bar k,t) + G_2(\bar k,t) + G_3(\bar k,t) \bigg]
\ee
where
\bea
G_1(\bar k,t) &=& \frac{\bar k}{\sqrt{t}} e^{-\bar k^2/t} \int_0^1 \! dy \, e^{\bar k^2 (1-y)^2/t} D\left(\frac{\bar k y}{\sqrt{t}}\right) \\
G_2(\bar k,t) &=& - \frac{\bar k}{\sqrt{t}}  \sum_{i=1}^\infty \frac{(t \, a_0^2)^i}{(2i)!} \,e^{-\bar k^2/t} \!\int_0^1 \! dy \, e^{\bar k^2 (1-y)^2/t} g^{(2i-1)}(0;\bar k y/\sqrt{t},1) \\
G_3(\bar k,t) &=& - \bar k \sum_{i=1}^\infty \frac{B_i}{i!} \,e^{-\bar k^2/t} \!\int_0^1 \! dy \, e^{\bar k^2 (1-y)^2/t} g^{(i-1)}(a_0;\bar k y,t) \,.
\eea
The small-$t$ limit of these expressions can be obtained by making extensive use of the definition of the Dawson integral in eq.~\pref{DawsonInt}, as well as its asymptotic expansion
\be
D(w) \simeq \frac{1}{2w} + \frac1{4w^3} + \frac3{8w^5} + \frac{15}{16w^7}+\cO\left(\frac1{w^9} \right) \,.
\ee
(This is most easily done using computing software.) And, of course, any term in the final result that is suppressed by a factor of $\exp(-\bar k^2/t)$ can be safely dropped in this limit. Performing this expansion yields
\bea
G_1(\bar k,t) &\simeq& \frac{t}{4 \bar k^2} + \frac{t^2}{8 \bar k^4} + \frac{3 t^3}{16 \bar k^6} + \ldots \\
G_2(\bar k,t) &\simeq& - \frac{t a_0^2}{2!} \times \left( \frac{t}{2\bar k^2} +\frac{3t^2}{4\bar k^4} +\ldots \right) - \frac{t^2 a_0^4}{4!} \times \left( -\frac{3t}{\bar k^2} + \ldots \right) - \ldots \\
G_3(\bar k,t) &\simeq& - \frac{B_1}{1!} \left(\frac{a_0 t^2}{2\bar k^2}+\left(-\frac{a_0^3}{2 \bar k^2}+\frac{3 a_0}{4 \bar k^4}\right)t^3 +\ldots \right) \nn\\
&&\quad- \frac{B_2}{2!} \left( \frac{t^2}{2 \bar k^2}+\left(-\frac{3a_0^2}{2 \bar k^2}+\frac{3}{4 \bar k^4} \right) t^3\ldots\right) \nn\\
&&\quad - \frac{B_3}{3!} \left( -\frac{3 a_0 t^3}{\bar k^2}   \ldots\right) - \frac{B_4}{4!} \left( -\frac{3 t^3}{\bar k^2}  \ldots\right) -\ldots
\eea
In the above, the $\ldots$ indicate terms that will contribute to $S_0(\omega,t)$ at $\cO(t^3)$ or higher, and so will give no contribution to the Casimir energy.

Finally, we can substitute this result into eq.~\pref{sumcGfromk1} and, with the use of
\bea
&&\sum_{k=1}^\infty \frac{1}{\bar k^{n}} = \left(\frac{\omega}{\pi}\right)^n \zeta(n) \quad\mathrm{with}\quad \zeta(2) = \frac{\pi^2}{6} \,,\quad \zeta(4) = \frac{\pi^4}{90} \,,\quad \zeta(6) = \frac{\pi^6}{945} \nn\\
&&\rightarrow\quad \sum_{k=1}^\infty \frac{1}{\bar k^{2}} = \frac16 \omega^2 \,,\quad \sum_{k=1}^\infty \frac{1}{\bar k^{4}} = \frac1{90} \omega^4 \,,\quad \sum_{k=1}^\infty \frac{1}{\bar k^{6}} = \frac1{945} \omega^6 \,,
\eea
we find that
\bea
4 \sum_{k=1}^\infty G_1(\bar k,t) &\simeq& \frac{\omega^2 t}{6} + \frac{\omega^4 t^2}{180} + \frac{\omega^6 t^3}{1260} + \ldots \\
4 \sum_{k=1}^\infty G_2(\bar k,t) &\simeq& - \frac{a_0^2 \omega^2 t^2}{6} - \frac{a_0^2 \omega^4 t^3}{60} +\frac{a_0^4 \omega^2 t^3}{12} + \ldots \\
4 \sum_{k=1}^\infty G_3(\bar k,t) &\simeq& - \left(-\frac12\right)\left(\frac{a_0 \omega^2 t^2}{3} - \frac{a_0^3 \omega^2 t^3}{3} + \frac{a_0 \omega^4 t^3}{30}\right) \nn\\
&&\quad- \left(\frac16\right) \left( \frac{\omega^2 t^2}{6} - \frac{a_0^2\omega^2 t^3}{2} + \frac{\omega^4 t^3}{60}\right) \nn\\
&&\quad - (0) - \left(-\frac1{30}\right)\left( -\frac{\omega^2 t^3}{12}\right) -\ldots
\eea
and so
\bea
d^{\mathrm{reg}}(\omega,t,a_0) &=& 2 \times \sum_{k=1}^\infty \left( \sum_{j=0}^\infty \cG(2\pi k) \right) \nn \\
&=& \frac{4}{\omega t} \sum_{k=1}^\infty \Big[ G_1(\bar k,t) + G_2(\bar k,t) + G_3(\bar k,t) \Big]\nn\\
&=&\frac{1}{\omega}  \Bigg\{ \frac16 \omega^2 + \left[\left( -\frac{1}{36}+ \frac{1}{6}a_0 -\frac{1}{6} a_0^2\right) \omega^2 + \frac{1}{180}\omega^4 \right] t \nn\\
&&\quad\qquad\qquad+ \left[\left( -\frac{1}{360} + \frac{1}{12} a_0^2 -\frac{1}{6} a_0^3 + \frac{1}{12} a_0^4 \right) \omega^2 \right. \nn\\
&&\quad\qquad\qquad+\left.\left(-\frac{1}{360} +\frac{1}{60}a_0 -\frac{1}{60}a_0^2\right)\omega^4 + \frac{1}{1260} \omega^6 \right] t^2 \Bigg\}
\eea
as found later, using the more efficient zeta--function regularization method.

\subsection*{Some useful sums}
\label{app:usefulsums}

Next, we will explicitly evaluate the small-$t$ limit of two often-encountered sums.

\subsubsection*{The sum $c(t,a_0)$} \label{csum}

This sum is defined as follows:
\bea
c(t,a_0) &:=& \sum_{j=0}^\infty \exp [-t (j+ a_0)^2] \,.
\eea
%
The Euler-Maclaurin formula states that, for any analytic function $f(x)$, we can write
\be
\sum_{j=0}^\infty f(j) = \int_0^\infty \! dx \, f(x) - \sum_{i=1}^\infty \frac{B_i}{i!} f^{(i-1)}(0)
\ee
where $f^{(i-1)}(x)$ denotes the $(i-1)$-th derivative of $f(x)$ and where the $B_i$ are the Bernoulli numbers. (The first few are $B_1 = -1/2$, $B_2 = 1/6$, $B_4 = -1/30$, and $B_{2i+1}=0$ for $i\geq 1$.) Identifying $f(x) = \exp[-t(x+a_0)^2]$, the first few derivative terms are
\bea
f^{(0)}(0) &=& e^{-t a_0^2} \,,\quad f^{(1)}(0) = -2ta_0 \, e^{-t a_0^2} \,,\quad f^{(2)}(0) = (-2t + 4 t^2 a_0^2) \, e^{-t a_0^2} \nn\\
f^{(3)}(0) &=& (12 t^2a_0 -8t^3a_0^3 ) \, e^{-t a_0^2} \,,\quad f^{(4)}(0) = (12 t^2-48 t^3 a_0^2+16 t^4a_0^4) \, e^{-t a_0^2} \,,
\eea
and $f^{(5)}(0) \sim \cO(t^3)$. (We only need powers of $t$ smaller than 3, for reasons discussed in the text.) The integral can be written in terms of the error function, and has the following series expansion:
\bea \label{intexp}
\int_0^\infty \! dx \, e^{-t(x+a_0)^2} &=& \frac{1}{2}\sqrt{\frac{\pi}{t}}  \left( 1- \erf(a_0\sqrt{t})\right) \nn\\
&=& \frac12 \sqrt{\frac{\pi}{t}}-a_0+\frac13 a_0^3 \, t-\frac{1}{10} a_0^5 \, t^2 + \cO(t^3) \,.
\eea
Upon Taylor expanding the derivative terms, we find that
\be \label{csumresult}
c(t,a_0) = \frac{c_{-1/2}}{\sqrt{t}}+c_{0}+c_{1} \, t + c_{2} \, t^2 + \cO(t^3)
\ee
where
\bea
c_{-1/2} = \frac{\sqrt{\pi}}2\,, &&\quad c_{0} := \frac12 - a_0 \,,\qquad c_{1} := \frac16 a_0 -\frac12 a_0^2 + \frac13 a_0^3 \,,\nn\\
&&c_{2} := \frac{1}{60} a_0 - \frac16 a_0^3 + \frac14 a_0^4 -\frac{1}{10} a_0^5 \,. \label{csumcoeffs}
\eea

We will usually need a ``tilded'' version of the above sum. This is defined in the following way:
\be
\tilde c(t,a_0,\tau) := e^{\tau t}\, c(t,a_0) = \frac{\tilde c_{-1/2}}{\sqrt{t}} + \tilde c_0 + \tilde c_{1/2} \, \sqrt{t} + \tilde c_1 \, t + \tilde c_{3/2} \, t^{3/2} + \tilde c_2 \, t^2 + \cO(t^{5/2}) \,.
\ee
In terms of components,
\bea
\tilde c_{-1/2} = c_{-1/2} \,,&& \tilde c_0 = c_0 \,,\quad \tilde c_{1/2} = c_{-1/2} \, \tau \,,\quad \tilde c_1 = c_1 + c_0 \, \tau \,, \nn\\
 \tilde c_{3/2} &=& \frac{c_{-1/2}}2 \, \tau^2 \,,\quad \tilde c_2 = c_2 + c_1 \, \tau + \frac{c_0}2 \, \tau^2 \,.
\eea

%

\subsubsection*{The double sum $d(\omega,t,a_0)$}
\label{app:usefuldsum}

This sum is defined as follows:
\bea \label{dsum}
d(\omega,t,a_0) &:=& \sum_{j=0}^\infty \sum_{n=-\infty}^\infty \exp [-t (j+\omega|n| + a_0)^2] \,.
\eea
%
The $j$-sum can be performed as before, but the $n$-sum will pose a difficulty in the $\omega,t \rightarrow 0$ limit, where it converges very slowly. This is because the sum in fact diverges in this limit. However, by using Poisson resummation, we can compute these divergent contributions efficiently.

With this in mind, let's write
\be
d(\omega,t,a_0) = d^{\mathrm{sing}} (\omega,t,a_0) + d^{\mathrm{reg}} (\omega,t,a_0)
\ee
where we require that $d^{\mathrm{reg}}(\omega,t,a_0)$ contain all terms that vanish as $\omega,t \rightarrow 0$. Since the calculation of $d^{\mathrm{sing}}(\omega,t,a_0)$ is more subtle (for the above-mentioned reasons), we will compute it first and then apply a more cavalier approach to $d^{\mathrm{reg}}(\omega,t,a_0)$.

\subsubsection*{Poisson resummation and $d^{\mathrm{sing}}(\omega,t,a_0)$}

A method for finding the behaviour of $d^{\mathrm{sing}}(\omega,t,a_0)$ as $\omega,t \rightarrow 0$ is to use Poisson resummation. Defining $g(n):=\exp[-\omega^2 t( |n| + b_j)^2]$ where $b_j := (j+a_0)/\omega$, we can exchange
\be \label{Poissonsum}
\sum_{n=-\infty}^\infty g(n) = \sum_{k=-\infty}^\infty \cG(2\pi k)
\ee
where
\be
\cG(q) := \int_{-\infty}^\infty \! \exd x \, g(x) e^{-iqx}
\ee
is the Fourier transform of $g(x)$. For example, in the special case where $b_j=0$ (i.e. $j=a_0=0$), we find that $g(n) = \exp(-\omega^2 t \,n^2)$ and so
\be
\sum_{n=-\infty}^\infty e^{-\omega^2 t\, n^2} =\sqrt{\frac{\pi}{\omega^2 t}} \sum_{k=-\infty}^\infty e^{-\pi^2 k^2/(\omega^2 t)}
\ee
since
\be
\cG(q) = \int_{-\infty}^\infty \! \exd x \, e^{-\omega^2 t \,x^2-iqx} = \sqrt{\frac{\pi}{\omega^2 t}} \, e^{-q^2/(4\omega^2 t)} \,.
\ee
This special case is sufficient to illustrate the value of Poisson resummation: a sum that converges slowly in the $\omega,t \rightarrow 0$ limit is transformed into one that converges quickly in this same limit. In fact, the part of the $n$--sum that diverges in this limit gets mapped entirely onto the $k=0$ term in the $k$--sum. Therefore, we can write
\be
d^{\mathrm{sing}}(\omega,t,a_0) = \sum_{j=0}^\infty \,\cG(0)
\ee
where now, for $b_j \neq 0$,
\be
\cG(0) = \int_{-\infty}^\infty \! \exd x \, e^{-\omega^2 t (|x|+b_j)^2} = \sqrt{\frac{\pi}{\omega^2 t}} [1-\erf(\omega \sqrt{t} b_j)] \,.
\ee
This sum can easily be evaluated in the same way as in section \ref{csum} using the Euler-Maclaurin formula. Integrating and differentiating $\cG(0)$ as prescribed gives
\bea
d^{\mathrm{sing}}_0(\omega,t,a_0) &=& \frac1{\omega} \sqrt{\frac{\pi}{t}} \sum_{j=0}^\infty \left[ 1- \erf\left(\sqrt{t}(j+a_0)\right)\right] \\
&=& \frac{d^{\mathrm{sing}}_{-1}}t + \frac{d^{\mathrm{sing}}_{-1/2}}{\sqrt{t}} + d^{\mathrm{sing}}_0 + d^{\mathrm{sing}}_1 \, t + d^{\mathrm{sing}}_2 \, t^2 + \cO(t^3) \nn
\eea
in the small--$t$ limit, where
\bea
d^{\mathrm{sing}}_{-1} := \frac1\omega \,,\quad d^{\mathrm{sing}}_{-1/2} := \frac{\sqrt{\pi}}\omega \left(\frac12-a_0\right) &\,,&\quad d^{\mathrm{sing}}_0 := \frac1\omega \left(\frac16 -a_0 + a_0^2 \right) \,, \\
d^{\mathrm{sing}}_1 := \frac1\omega \left(\frac{1}{180} - \frac{a_0^2}{6}  +\frac{a_0^3}3 -\frac{a_0^4}6 \right) \,,\quad d^{\mathrm{sing}}_2 &:=& \frac1\omega \left(\frac{1}{1260} - \frac{a_0^2}{60} + \frac{a_0^4}{12} - \frac{a_0^5}{10} + \frac{a_0^6}{30} \right) \nn\,.
\eea
As promised, each of these contributions diverges in the limit where $\omega,t \rightarrow 0$.

\subsubsection*{Calculating $d^{\mathrm{reg}}(\omega,t,a_0)$ using a zeta function approach}

First, we should mention that the use of zeta function method here is not necessary. In fact, we could continue as before, and compute $d^{\mathrm{reg}}(\omega,t,a_0)$ using Poisson resummation:
\be
d^{\mathrm{reg}}(\omega,t,a_0) = 2 \sum_{j=0}^\infty \sum_{k=1}^\infty \, \cG(2\pi k) \,.
\ee
Here, we have identified $d^{\mathrm{reg}}(\omega,t,a_0)$ as the left--over $k\neq 0$ terms in the Poisson sum, eq.~\pref{Poissonsum}. (The factor of two arises because $\cG(q)$ is even: $g(-x)=g(x) \Rightarrow \cG(-q) = \cG(q)$.) This more ``honest'' approach is taken in Appendix \ref{app:honestcalc}. However, since this calculation is significantly less tiresome if we take advantage of zeta function regularization, we will present this version of the derivation herein. (The two calculations give the same result, as can be seen by comparison with the result in Appendix \ref{app:honestcalc}.) Much of the simplification arises because we can take advantage of the previous result for $c(t,a_0)$.

Since we are guaranteed to have captured all of the divergent terms (as $\omega,t \rightarrow 0$) in $d^{\mathrm{sing}}(\omega,t,a_0)$, computing the full sum in a na\"ive way --- one that does not capture these divergent terms --- will give us $d^{\mathrm{reg}}(\omega,t,a_0)$. The zeta function method is just such a na\"ive approach! Let's start by interchanging the $j$-- and $n$--sums in eq.~\pref{dsum} and write
\be
d^{\mathrm{reg}}(\omega,t,a_0) = \sum_{n=-\infty}^\infty c(t, \omega |n| + a_0) = c(t,a_0) + 2 \sum_{n=1}^\infty c(t, \omega n + a_0) \,.
\ee
Since, the zeta function is defined as
\be
\zeta (s) = \sum_{n=1}^\infty \frac1{n^s} \,,
\ee
we have that $\sum_{n=1}^\infty 1 = \sum_{n=1}^\infty 1/(n^0) = \zeta(0) = -1/2$ and so we can write
\be
d^{\mathrm{reg}}(\omega,t,a_0) = 2 \sum_{n=1}^\infty \Big( c(t,\omega n +a_0) - c(t, a_0) \Big) \,.
\ee
Given the result in eqs.~\pref{csumresult} and \pref{csumcoeffs}, we find that the summand --- expanded out in powers of $t$ --- is given by
\be
\Delta c := c(t,\omega n + a_0) - c(t,a_0) = \sum_r \, t^r \,\Delta c_r
\ee
where $r \in \{-1/2,0,1,2\}$ and where
\bea
\Delta c_{-1/2} = 0 \,,&& \Delta c_0 = -\omega n \,,\quad \Delta c_1 = \left(\frac16 -a_0 +a_0^2\right) \omega n + \left(-\frac12 + a_0\right) \omega^2 n^2 + \frac13 \omega^3 n^3 \nn\\
\Delta c_2 &=& \left(\frac1{60} -\frac{a_0^2}2 +a_0^3 - \frac{a_0^4}2 \right) \omega n + \left( -\frac{a_0}2 + \frac{3a_0^2}2 - a_0^3 \right) \omega^2 n^2 \\
&&\quad+ \left( -\frac16 + a_0 - a_0^2 \right) \omega^3 n^3 + \left( \frac14 - \frac{a_0}2  \right) \omega^4 n^4 - \frac1{10} \omega^5 n^5 \,.\nn
\eea
Each term in $\Delta c$ proportional to $n^s$ will get a corresponding factor of $\zeta(-s)$ once the sum over $n$ is performed. (Here, $\zeta(-1) = -1/12$, $\zeta(-3)=1/120$, $\zeta(-5)=-1/252$ and $\zeta(-2s) = 0$ for $s\geq 1$.) The end result is
\be
d^{\mathrm{reg}}(\omega,t,a_0) = d^{\mathrm{reg}}_0 + d^{\mathrm{reg}}_1 \, t + d^{\mathrm{reg}}_2 \, t^2 + \cO(t^3)
\ee
where
\bea
&&d^{\mathrm{reg}}_0 = \frac{\omega}6 \,,\quad d^{\mathrm{reg}}_1 = \left(-\frac{1}{36}+ \frac{a_0}{6} -\frac{a_0^2}{6}\right) \omega + \frac{\omega^3}{180} \,, \\
d^{\mathrm{reg}}_2 &=& \left( -\frac{1}{360} + \frac{a_0^2}{12} -\frac{a_0^3}{6} + \frac{a_0^4}{12} \right) \omega  +\left(-\frac{1}{360} +\frac{a_0}{60} -\frac{a_0^2}{60} \right)\omega^3 + \frac{\omega^5}{1260}  \nn\,.
\eea
As promised, the above result vanishes in the $\omega,t \rightarrow 0$ limit.

Combining the results for $d^{\mathrm{sing}}(\omega,t,a_0)$ and $d^{\mathrm{reg}}(\omega,t,a_0)$, we find that
\be
d(\omega,t,a_0) = \frac{d_{-1}}t + \frac{d_{-1/2}}{\sqrt{t}} + d_0 + d_1 \, t + d_2 \, t^2 + \cO(t^3)
\ee
where
\bea
d_{-1} &=& \frac1{\omega} \,,\quad d_{-1/2} = \frac{\sqrt{\pi}}\omega \left(\frac12-a_0\right) \,,\quad d_0 = \frac1\omega \left( \frac16 -a_0 + a_0^2 + \frac{\omega^2}6 \right) \,, \nn\\
d_{1} &=& \frac1\omega \left[ \frac{1}{180} - \frac{a_0^2}{6}  +\frac{a_0^3}3 -\frac{a_0^4}6 + \left(-\frac{1}{36}+ \frac{a_0}{6} -\frac{a_0^2}{6}\right) \omega^2 + \frac{\omega^4}{180} \right] \,, \\
d_2 &=& \frac1\omega \left[ \frac{1}{1260} - \frac{a_0^2}{60} + \frac{a_0^4}{12} - \frac{a_0^5}{10} + \frac{a_0^6}{30} + \left(-\frac{1}{360} + \frac{a_0^2}{12} -\frac{a_0^3}{6} + \frac{a_0^4}{12} \right) \omega^2 \right. \nn\\
&&\qquad\left. +\left(-\frac{1}{360} +\frac{a_0}{60} -\frac{a_0^2}{60} \right) \omega^4 + \frac{\omega^6}{1260} \right] \,.\nn
\eea

As with $c(t,a_0)$, we will often need a ``tilded'' version of the above sum. This is defined in a similar way as before:
\bea
\tilde d(\omega,t, a_0,\tau) &:=& e^{\tau t} \, d(\omega,t,a_0) \\
&=& \frac{\tilde d_{-1}}t + \frac{\tilde d_{-1/2}}{\sqrt{t}} + \tilde d_0 + \tilde d_{1/2} \, \sqrt{t} + \tilde d_1 \, t + \tilde d_{3/2} \, t^{3/2} + \tilde d_2 \, t^2 + \cO(t^{5/2}) \,.\nn
\eea
In terms of components,
\bea
\tilde d_{-1} = d_{-1} \,,\quad \tilde d_{-1/2} = d_{-1/2} \,,\quad \tilde d_0 &=& d_0 + d_{-1} \, \tau \,,\quad \tilde d_{1/2} = d_{-1/2} \,\tau \,, \\
 \tilde d_1 = d_1 + d_0 \, \tau + \frac{d_{-1}}2 \, \tau^2 \,,\quad \tilde d_{3/2} = \frac{d_{-1/2}}2 \, \tau^2 \,,&& \tilde d_2 = d_2 + d_1 \, \tau + \frac{d_0}2 \,\tau^2 + \frac{d_{-1}}6 \, \tau^3 \,.\nn
\eea

\section{Spectra and mode sums}

This appendix computes the KK spectra and mode sums that are quoted in the main text, for spins zero, half and one.

\section*{Spectrum and Mode Sum for the Scalar Field}
\label{app:SpecNSum}

This appendix computes the spectrum $\lambda^\mathrm{s}_{jn}$ and the corresponding small-$t$ limit of the mode sum
\be
S_\mathrm{s}(t):= \sum_{j,n} e^{-t\lambda^\mathrm{s}_{jn}} = \sum_{i=-1}^2 s^\mathrm{s}_i \, t^i + \cdots
\ee
for a 6D charged scalar $\phi$, with mass $m$ and charge $q \tilde g$, on the rugby ball.

\subsection*{Eigenvalues}

The scalar equation of motion is
\be
\Big( g^{\ssM\ssN} D_\ssM D_\ssN - m^2 \Big) \phi =0
\ee
where $D_\ssM$ is the covariant derivative, and where
\bea
ds^2 &=& \eta_{\mu\nu} \exd x^\mu \, \exd x^\nu + r^2(\exd \theta^2 + \alpha^2\sin^2\theta \, \exd \varphi^2) \\
q A_m \, \exd x^m &=& q A_\varphi \, \exd \varphi = \left[ -\frac{(N-\Phi)}2 (\cos\theta -b) + b\, \Phi_b \right] \exd\varphi \,.
\eea
In the above expression for $A_\varphi$, the variable $b$ is used to distinguish two patches of the gauge potential: $b=+1$ ($-1$) corresponds to the patch encompassing $\cos\theta=+1$ ($-1$). To verify that this gauge potential is a correct one, it is sufficient to check that:
\bea
\frac{\exd A_\varphi}{\exd \theta} = F_{\theta\phi} = f \epsilon_{\theta\phi} &=& \left(\frac{\cN}{2qr^2}\right) (\alpha r^2 \sin\theta) \,, \\
A_\varphi (\cos\theta=b) &=& b \frac{A_b}{2\pi} \,.
\eea
(Recall the definitions $\cN = \omega(N-\Phi)$, $\Phi_b = q A_b /(2\pi)$ and $\Phi = \sum_b \Phi_b$ from the text.)

Expanding out this equation of motion using the ansatz $\phi(x) = e^{ik_4\cdot x}\,\Theta(\theta) e^{in\varphi}$ gives
\bea
\frac{\exd^2\Theta}{\exd\theta^2} + \cot \theta \,\frac{\exd\Theta}{\exd\theta} + \left[ \lambda_{jn} - \frac{(n-q \, A_\phi)^2}{\alpha^2 \sin^2\theta}\right] \Theta =0
\eea
where $\lambda_{jn} = -r^2 (k_4^2+m^2)$ are the required eigenvalues (in units of the KK scale $1/r$). This is just the result for the case without a background field, but with $n \rightarrow n-q \, A_\varphi$. This difference is expected given that the wave equation depends on the azimuthal covariant derivative $D_\varphi = \partial_\varphi - i q \, A_\varphi = i (n -q \, A_\varphi)$ (since our ansatz is that  $\phi(x) \propto e^{in\varphi}$).

The standard approach from here is to change variables to $x:=\cos\theta$, giving
\be \label{ThetaEq}
(1-x^2) \frac{\exd^2\Theta}{\exd x^2} -2 x \frac{\exd\Theta}{\exd x} + \left[ \lambda_{jn} - \frac{(n-q\, A_\phi)^2}{\alpha^2 (1-x^2)}\right] \Theta =0 \,,
\ee
and to let $\Theta(x) = (1-x^2)^y f(x)$ for some convenient power $y$ that removes the singular behaviour of \pref{ThetaEq} near $x= \pm 1$ ($\leftrightarrow \theta = 0,\pi$). However, since the gauge field is defined differently at each pole, it will help to treat the divergences at each pole separately. Therefore, we will instead take
\[
\Theta(x) = (1-x)^y (1+x)^z f(x) \,,
\]
which yields
\be
(1-x^2) \frac{\exd^2 f}{\exd x^2} -2 [(1+y+z)x+y-z] \frac{\exd f}{\exd x} + \left[ \lambda_{jn} - \frac{\omega^2(n-q\, A_\phi)^2+k(x,y,z)}{(1-x^2)}\right] f(x) =0
\ee
where $\omega=\alpha^{-1}$ and
\be \label{kxyzdef}
k(x,y,z) := y+z - (y-z)^2 -2(y^2-z^2)x -(y+z)(y+z+1) x^2 \,.
\ee
Requiring that the numerator $[\omega^2(n-qA_\varphi)^2+k]$ be proportional to $(1-x^2)$ gives the conditions
\be
y=\frac\omega2 \left| n + (N-\Phi)\epsilon_{-b} -b \,\Phi_b \right| \,, \quad z=\frac\omega2 \left| n - (N-\Phi)\epsilon_b - b \, \Phi_b\right|
\ee
where $\epsilon_+ = 1$ and $\epsilon_- = 0$, and the resulting regular ODE is
\be \label{fODE}
(1-x^2) \frac{d^2f}{dx^2} - 2 [(1+y+z)x+y-z] \frac{df}{dx} + \left[ \lambda_{jn} + \frac{\cN^2}{4} - (y+z)(y+z+1) \right] f(x) = 0 \,.
\ee
In passing, it should be mentioned that it is equally valid to take $y$ and/or $z$ to be definitely negative, rather than positive, in eliminating the singularities from eq.~\pref{ThetaEq}. However, in the absence of couplings of the scalar to the brane, we should discard these solutions since they cause $\phi(x)$ to diverge at one or both of the branes (in conflict with the boundary conditions).

It is instructive to consider the special case in which $q=0$ and $\omega=1$. In this case, we find that $y=z=|n|/2$ (thereby justifying the usual choice of $\Theta(x) = (1-x^2)^{|n|/2} f(x)$) and
\[
(1-x^2) \frac{d^2f}{dx^2} - 2 (1+|n|)x \frac{df}{dx} + \left[ \lambda_{jn}  - |n|(|n|+1) \right] f(x) = 0 \,.
\]
This is the standard differential equation that is satisfied by the $|n|$th derivative of a Legendre polynomial (so long as $\lambda_{jn} = (j+|n|)(j+|n|+1)$ for some $j\geq 0$, allowing the corresponding power series solution to terminate). Therefore, $\Theta(x)$ can be any (linear combination) of the associated Legendre polynomials and $\phi(x)$ are (linear combinations of) spherical harmonics. It is conventional to define $\ell := j+ |n|$ so that $\lambda = \ell(\ell+1)$ with degeneracy $(2\ell+1)$, representing the number of different ways of choosing combinations $(j,n)$ that give the same $\ell$.

In the general case, the condition for termination of the power series solution to eq.~\pref{fODE} is
\be
\lambda_{jn}  = (j+y+z)(j+y+z+1) -\frac{\cN^2}{4} \,.
\ee

\subsection*{Mode Sum}

From here, we wish to compute the small--$t$ coefficients of $S_{\rm s}(\omega,N,\Phi_b,t)$, where
\bea
S_{\rm s}(\omega,N,\Phi_b,t) &=& e^{t\cN^2/4} \sum_{j=0}^\infty \sum_{n=-\infty}^\infty \exp \big[ -t(j+y+z)(j+y+z+1)\big] \\
&=& e^{\tau t} \sum_{j=0}^\infty \sum_{n=-\infty}^\infty \exp \big[ -t(j+y+z+1/2)^2\big] \,. \label{bigsum}
\eea
In the last line, we have introduced $\tau := (1+\cN^2)/4$.

In order to use our previously-derived results for the sums
\bea
\tilde c(t,a_0,\tau) &:=& e^{\tau t} \sum_{j=0}^\infty \exp [-t(j+a_0)^2] \\
\tilde d(\omega,t,a_0,\tau) &:=& e^{\tau t} \sum_{j=0}^\infty \sum_{n=-\infty}^\infty \exp [-t(j+\omega |n| + a_0)^2] \,,
\eea
we will first have to do some massaging. We will assume throughout that $|\Phi_b|<1$, since any brane--localized flux larger than this can be absorbed (one integer at a time) into the bulk flux, $N$. Also, we will choose to work with the north ($\cos\theta=+1$) patch of the potential, for which
\be
y+z = \frac{\omega}2 |n-\Phi_+| + \frac{\omega}2 |n-N + \Phi_-|  \,.
\ee
To see the result for the south ($\cos\theta=-1$) patch, where
\be
y+z = \frac{\omega}2 |n+N-\Phi_+| + \frac{\omega}2 |n +\Phi_-| \,,
\ee
we can simply make the replacements $N\rightarrow -N$ and $\Phi_b \to -\Phi_{-b}$ in the final result (since the spectrum only depends on the combined quantity $y+z$). Of course, such a transformation should be a symmetry of the result, since it is just a gauge transformation.

It is helpful to first consider the case when $N>0$ and break up the $n$--sum in eq.~\pref{bigsum} into three parts:
\be
S_{\rm s}(\omega,N,\Phi_b,t) = S^{(+)}_{\rm s}(\omega,N,\Phi_b,t) + S^{(0)}_{\rm s}(\omega,N,\Phi_b,t) + S^{(-)}_{\rm s}(\omega,N,\Phi_b,t)
\ee
where:
\begin{itemize}
\item for $\Phi_->0$, $\Phi_+>0$,
\bea
S^{(+)}_{\rm s} &=& e^{\tau t} \sum_{j=0}^\infty \sum_{n=N}^\infty \exp \left[-t\left(j+ \frac{\omega}2 (n-\Phi_+) + \frac{\omega}2 (n-N+\Phi_-) + \frac12\right)^2 \right] \,,\nn\\
S^{(0)}_{\rm s} &=& e^{\tau t} \sum_{j=0}^\infty \sum_{n=1}^{N-1} \exp \left[ -t\left(j+ \frac{\omega}2 (n-\Phi_+) + \frac{\omega}2 (N-\Phi_- -n) + \frac12\right)^2 \right] \nn \,,\\
S^{(-)}_{\rm s} &=& e^{\tau t} \sum_{j=0}^\infty \sum_{n=0}^{-\infty} \exp \left[-t\left(j + \frac{\omega}2 (\Phi_+ -n) + \frac{\omega}2 (N-\Phi_--n) + \frac12\right)^2  \right] \,;\nn
\eea

\item for $\Phi_->0$, $\Phi_+<0$,
\bea
S^{(+)}_{\rm s} &=& e^{\tau t} \sum_{j=0}^\infty \sum_{n=N}^\infty \exp \left[-t\left(j+\frac{\omega}2 (n-\Phi_+) + \frac{\omega}2 (n-N+\Phi_-) + \frac12\right)^2 \right] \,,\nn\\
S^{(0)}_{\rm s} &=& e^{\tau t} \sum_{j=0}^\infty \sum_{n=0}^{N-1} \exp \left[ -t\left(j+ \frac{\omega}2 (n-\Phi_+) + \frac{\omega}2 (N-\Phi_- -n) + \frac12\right)^2 \right] \nn \,,\\
S^{(-)}_{\rm s} &=& e^{\tau t} \sum_{j=0}^\infty \sum_{n=-1}^{-\infty} \exp \left[-t\left(j+ \frac{\omega}2 (\Phi_+ -n) + \frac{\omega}2 (N-\Phi_--n) + \frac12\right)^2  \right] \,;\nn
\eea

\item for $\Phi_-<0$, $\Phi_+>0$,
\bea
S^{(+)}_{\rm s} &=& e^{\tau t} \sum_{j=0}^\infty \sum_{n=N+1}^\infty \exp \left[-t\left(j+\frac{\omega}2 (n-\Phi_+) + \frac{\omega}2 (n-N+\Phi_-) + \frac12\right)^2 \right] \,,\nn\\
S^{(0)}_{\rm s} &=& e^{\tau t} \sum_{j=0}^\infty \sum_{n=1}^{N} \exp \left[ -t\left(j+\frac{\omega}2 (n-\Phi_+) + \frac{\omega}2 (N-\Phi_- -n) + \frac12\right)^2 \right] \nn \,,\\
S^{(-)}_{\rm s} &=& e^{\tau t} \sum_{j=0}^\infty \sum_{n=0}^{-\infty} \exp \left[-t\left(j+\frac{\omega}2 (\Phi_+ -n) + \frac{\omega}2 (N-\Phi_--n) + \frac12\right)^2  \right] \,;\nn
\eea

\item for $\Phi_-<0$, $\Phi_+<0$,
\bea
S^{(+)}_{\rm s} &=& e^{\tau t} \sum_{j=0}^\infty \sum_{n=N+1}^\infty \exp \left[-t\left(j+\frac{\omega}2 (n-\Phi_+) + \frac{\omega}2 (n-N+\Phi_-) + \frac12\right)^2 \right] \,,\nn\\
S^{(0)}_{\rm s} &=& e^{\tau t} \sum_{j=0}^\infty \sum_{n=0}^{N} \exp \left[ -t\left(j+\frac{\omega}2 (n-\Phi_+) + \frac{\omega}2 (N-\Phi_- -n) + \frac12\right)^2 \right] \nn \,,\\
S^{(-)}_{\rm s} &=& e^{\tau t} \sum_{j=0}^\infty \sum_{n=-1}^{-\infty} \exp \left[-t\left(j+\frac{\omega}2 (\Phi_+ -n) + \frac{\omega}2 (N-\Phi_--n) + \frac12\right)^2  \right] \,.\nn
\eea
\end{itemize}
If we allow ourselves to extend the notation $\epsilon_\pm$ to
\be
\epsilon_x := \frac{1 + \mathrm{sgn}(x)}2 = \left\{
\begin{array}{ccc}
1 &,\,& \mathrm{sgn}(x) := x/|x| = +1 \\
0 &,\,& \mathrm{sgn}(x) := x/|x| = -1
\end{array}
\right.
\ee
(recall that, before, $\epsilon_b = (1+ b)/2 = 1$ or $0$), these four cases (for which $N>0$) can be succintly written as
\bea
S^{(+)}_{\rm s} &=& e^{\tau t} \sum_{j=0}^\infty \sum_{n=\epsilon_{-\Phi_-}}^\infty \exp \left[-t\left(j+\omega n -\frac{\omega\Delta\Phi}2  + \frac{1+\omega N}2 \right)^2 \right] \nn\\
S^{(0)}_{\rm s} &=& (N-1+\epsilon_{-\Phi_-}+\epsilon_{-\Phi_+}) \times e^{\tau t} \sum_{j=0}^\infty \exp \left[ -t\left(j- \frac{\omega\Phi}2  +\frac{1+\omega N}2 \right)^2\right] \nn\\
S^{(-)}_{\rm s} &=& e^{\tau t} \sum_{j=0}^\infty \sum_{n=-\epsilon_{-\Phi_+}}^{-\infty} \exp \left[-t\left(j-\omega n +\frac{\omega\Delta\Phi}2 + \frac{1+\omega N}2 \right)^2 \right]
\eea
where $\Delta \Phi := \Phi_+ - \Phi_-$.

Performing a similar calculation for the case where $N<0$, we find something slightly different:
\bea
S^{(+)}_{\rm s} &=& e^{\tau t} \sum_{j=0}^\infty \sum_{n=\epsilon_{\Phi_+}}^\infty \exp \left[-t\left(j+\omega n -\frac{\omega\Delta\Phi}2  + \frac{1-\omega N}2 \right)^2 \right] \nn\\
S^{(0)}_{\rm s} &=& (-N-1+\epsilon_{\Phi_-}+\epsilon_{\Phi_+}) \times e^{\tau t} \sum_{j=0}^\infty \exp \left[ -t\left(j+ \frac{\omega\Phi}2  +\frac{1-\omega N}2 \right)^2\right] \nn\\
S^{(-)}_{\rm s} &=& e^{\tau t} \sum_{j=0}^\infty \sum_{n=-\epsilon_{\Phi_-}}^{-\infty} \exp \left[-t\left(j-\omega n +\frac{\omega\Delta\Phi}2 + \frac{1-\omega N}2 \right)^2 \right] \,.
\eea

Writing $\eta:=\mathrm{sgn}(N) = N/|N|$, we find that these expressions -- for arbitrary $N$ -- take the form
\bea
S^{(+)}_{\rm s} &=& e^{\tau t} \sum_{j=0}^\infty \sum_{n=\epsilon_{-\eta\Phi_{-\eta}}}^\infty \exp \left[-t\left(j+\omega n -\frac{\omega\Delta\Phi}2  + \frac{1+\omega |N|}2 \right)^2 \right] \nn\\
S^{(0)}_{\rm s} &=& (|N|-1+\epsilon_{-\eta\Phi_-}+\epsilon_{-\eta\Phi_+}) \times e^{\tau t} \sum_{j=0}^\infty \exp \left[ -t\left(j- \eta\, \frac{\omega\Phi}2  +\frac{1+\omega |N|}2 \right)^2\right] \nn\\
S^{(-)}_{\rm s} &=& e^{\tau t} \sum_{j=0}^\infty \sum_{n=-\epsilon_{-\eta\Phi_\eta}}^{-\infty} \exp \left[-t\left(j-\omega n +\frac{\omega\Delta\Phi}2 + \frac{1+\omega |N|}2 \right)^2 \right] \,.
\eea

Using the final rearrangement
\be
\sum_{n=\epsilon_\mp}^\infty f(n) = \frac12 \left( \sum_{n=-\infty}^\infty f(|n|) \pm f(0) \right)
\ee
and writing $a_\ssN := (1+\omega |N|)/2$, we find that the massage is complete:
\bea
S^{(+)}_{\rm s} &=& \frac12 \Big[ \tilde d(\omega,t,-\omega \Delta\Phi/2 + a_\ssN,\tau) + \eta\,\mathrm{sgn}(\Phi_{-\eta})\,\tilde c(t,-\omega\Delta\Phi/2 + a_\ssN,\tau) \Big] \nn\\
S^{(0)}_{\rm s} &=& \Big[|N|-\eta\big(\mathrm{sgn}(\Phi_+) + \mathrm{sgn}(\Phi_-))/2\Big] \,\tilde c(t,-\eta\,\omega\Phi/2+a_\ssN,\tau) \\
S^{(-)}_{\rm s} &=& \frac12 \Big[ \tilde d(\omega,t,\omega\Delta\Phi/2 + a_\ssN,\tau) +\eta\,\mathrm{sgn}(\Phi_{\eta})\, \tilde c(t,\omega\Delta\Phi/2 + a_\ssN,\tau) \Big] \,.\nn
\eea
This result is, by construction, invariant under $N \to -N$, $\Phi_b \to - \Phi_{-b}$.

Evaluating these expressions, we find, using throughout the definitions
\be \label{Fbdefs}
F_b:=|\Phi_b|\left(1-|\Phi_b|\right)\,,\quad F^{(n)}:= \sum_b F_b^n \,,\quad F^{(1)}:=F \,, \quad G(x):=(1-x)(1-2x)
\ee
that $s^{\rm s}_{-1} = 1/\omega$,
\bea
s^{\rm s}_0(\omega,N,\Phi_b) &=& \frac1\omega \left[ \frac16 + \frac{\omega^2}6(1-3F) \right] \,, \\
s^{\rm s}_1(\omega,N,\Phi_b) &=& \frac1\omega\Bigg[ \frac1{180} - \frac{\cN^2}{24} + \frac{\omega^2}{18}(1 - 3 F) -\frac{\omega^3\cN}{12} \sum_b \Phi_b \, G(|\Phi_b|) + \frac{\omega^4}{180} (1 -15F^{(2)}) \Bigg] ,\qquad \\
s^{\rm s}_2(\omega,N,\Phi_b) &=& \frac1\omega \Bigg[ -\frac1{504} - \frac{11\,\cN^2}{720} + \left(\frac1{90} -\frac{\cN^2}{144} \right) (1-3F) \omega^2 -\frac{\omega^3\cN}{24} \sum_b \Phi_b \, G(|\Phi_b|)  \nn\\
&& \qquad  + \frac{\omega^4(1-\cN^2)}{360}(1 - 15F^{(2)}) - \frac{\omega^5\cN}{120}\sum_b \Phi_b \,G(|\Phi_b|)(1+3F_b)  \nn\\
&&\qquad   + \Bigg(\frac1{1260} -\frac{F^{(2)}}{120} - \frac{F^{(3)}}{60} \Bigg)\omega^6 \Bigg]  \,.
\eea
%

\section*{Spectrum and Mode Sum for the Spin-1/2 Field}
\label{app:SpecNSumSpinors}

This appendix computes the spectrum $\lambda^\mathrm{f}_{jn}$ and the corresponding small-$t$ limit of the mode sum
\be
S_\mathrm{f}(t):= -\sum_\xi\sum_{j,n} e^{-t\lambda^{\mathrm{f}\xi}_{jn}} = \sum_{i=-1}^2 s^\mathrm{f}_i \, t^i
\ee
for a 6D charged spin-1/2 field $\psi$, with mass $m$ and charge $q$, on the rugby ball. Here, the minus sign denotes fermionic degrees of freedom, and the additional sum is over spin states.

\subsection*{Eigenvalues}

The spinor equation of motion is
\be
(\Dsl+m)\psi =0
\ee
where $D_\ssM$ is the covariant derivative and where the metric and background gauge field, respectively, are
\bea
\exd s^2 &=& r^2 ( \exd \theta^2 + \alpha^2 \sin^2\theta \, \exd \varphi^2 ) +\eta_{\mu\nu} \exd x^\mu \exd x^\nu \\
q A_\ssM \,\exd x^\ssM &=& q A_\varphi \,\exd \varphi = \left[-\frac{N-\Phi}2 (\cos\theta -b) + b\Phi_b\right]\,  \exd \varphi \label{gAdefspinor}
\eea
(as before). Here, $q$ is the charge (in units of $\tilde g$) of the field $\psi$ under the gauge field.

We make use of the frame fields ${e_\ssA}^\ssM$ and the spin connection $\Omega_\ssM^{\ssA\ssB}$ to ensure that our Cartesian Clifford algebra (i.e. $\{\Gamma_\ssA,\Gamma_\ssB \} = 2 \eta_{\ssA\ssB}$) conforms to the spherical geometry:
\be
\Dsl \psi = \Gamma^\ssA {e_\ssA}^\ssM \left( \partial_\ssM -\frac{i}2 J_{\ssA\ssB} \,\Omega_\ssM^{\ssA\ssB} - i q A_\ssM \right) \psi \,.
\ee
(Here, $J_{\ssA\ssB} = -i[\Gamma_\ssA,\Gamma_\ssB]/4$ are the Lorentz generators in the spinor representation.) More precisely, these are given by two separate patches (in the same way as the gauge potential):
\bea
{e_a}^m &=&
\frac1r \left(
  \begin{array}{cc}
\cos\varphi & -\frac{b\,\sin\varphi}{\alpha \sin\theta} \\
b\,\sin\varphi & \frac{\cos\varphi}{\alpha \sin\theta}
  \end{array}
\right) \,,\,\, {e_\alpha}^\mu = \delta_\alpha^\mu \,,\,\, {e_\alpha}^m = 0
\quad\textrm{(so that }{e_\ssA}^\ssM {e_\ssB}^\ssN g_{\ssM\ssN} = \eta_{\ssA\ssB})\,,\qquad \\
\Omega_\ssM^{\ssA\ssB} &=& \eta^{\ssA\ssC} ({e_\ssC}^\ssN \Gamma_{\ssM\ssN}^\ssK {e_\ssK}^\ssB - {e_\ssC}^\ssN \partial_\ssM {e_\ssN}^\ssB) \quad \textrm{(where }{e_\ssM}^\ssA := ({e_\ssA}^\ssM)^{-1}) \,.
\eea
(Recall that $b+1$ ($b=-1$) for the north (south) brane.) Given that $\Gamma_{\ssM\ssN}^\ssK = 0$ with the exception of $\Gamma_{\theta\varphi}^\varphi = \cot \theta$ and $\Gamma_{\varphi\varphi}^\theta = -\alpha^2 \sin\theta \cos\theta$, we find that the spin connections mostly vanish as well, with the exception of $\Omega_\varphi^{45} = \alpha \cos\theta -b = -\Omega_\varphi^{54}$.

In order to expand out the above Dirac equation, let's use the representation
\be
\Gamma^\mu = \left(
\begin{array}{cc}
0 & \gamma^\mu \\
\gamma^\mu & 0
\end{array}
\right) \,,\quad
\Gamma^4 = \left(
\begin{array}{cc}
0 & \gamma_5 \\
\gamma_5 & 0
\end{array}
\right) \,,\quad
\Gamma^5 = \left(
\begin{array}{cc}
0 & -i \one_4 \\
i \one_4 & 0
\end{array}
\right)
\ee
where $\gamma^\mu$ are the usual 4D Dirac matrices:
\be
\gamma^\mu = -i \left(
\begin{array}{cc}
0 & \ol \sigma^\mu \\
-\sigma^\mu & 0
\end{array}
\right) \,,\quad
\gamma_5 = -i \gamma^0 \gamma^1 \gamma^2 \gamma^3 = \left(
\begin{array}{cc}
\one_2 & 0 \\
0 & -\one_2
\end{array}
\right)
\ee
and $\sigma_\mu = (\one_2,\sigma_i) = \ol\sigma^\mu$ are the Pauli matrices.

Substituting these definitions, we find that the covariant derivative becomes
\bea
\Dsl &=& \Gamma^\mu \partial_\mu + \Gamma^4\left[\frac{\cos\varphi}r \partial_\theta - \frac{b\,\sin\varphi}{\alpha\, r\sin\theta}(\partial_\varphi - i q A_\varphi) +  \frac{\cos\varphi}{\alpha\, r\sin\theta} \frac{\Omega_\varphi^{45}}2 \right] \nn\\
&&\qquad+ \Gamma^5 \left[\pm\frac{\sin\varphi}r \partial_\theta + \frac{\cos\varphi}{\alpha\, r\sin\theta} (\partial_\varphi - i q A_\varphi)+\frac{b\,\sin\varphi}{\alpha\, r\sin\theta} \frac{\Omega_\varphi^{45}}2 \right]
\eea
or
\be
\centering{\Dsl = \left(
\begin{array}{cc}
0 & \gamma^\mu \partial_\mu +  P_\ssL O_2 -  P_\ssR O_1 \\
\gamma^\mu \partial_\mu +  P_\ssL O_1 - P_\ssR O_2 & 0
\end{array}
\right)}
\ee
where $P_{\ssL(\ssR)} = (1\pm\gamma_5)/2$ and where
\bea
O_1 &=& \frac{e^{ i b\varphi}}r \left(\partial_\theta - \frac{(-i\partial_\varphi - q A_\varphi - \Omega_\varphi^{45}/2)}{\alpha\sin\theta}\right) \\
O_2 &=& \frac{e^{- i b \varphi}}r \left(\partial_\theta + \frac{(-i\partial_\varphi - q A_\varphi + \Omega_\varphi^{45}/2)}{\alpha\sin\theta}\right) \,.
\eea
Since $\psi$ is a 6D Dirac spinor, we can decompose it into two 4D Dirac spinors or, equivalently, four 4D Weyl spinors:
\be
\psi = \left(
\begin{array}{c}
\psi_1 \\ \psi_2
\end{array}
\right) = \left(
\begin{array}{c}
\psi_{1\ssL} + \psi_{1\ssR} \\
\psi_{2\ssL} + \psi_{2\ssR}
\end{array}
\right)
\ee
where the 4D Weyl spinors satisfy $\gamma_5 \psi_{i\ssL} = + \psi_{i\ssL}$ and $\gamma_5 \psi_{i\ssR} = - \psi_{i\ssR}$. In terms of these, the Dirac equation gives
\bea
\gamma^\mu \partial_\mu \psi_{2\ssR} + O_2 \psi_{2\ssL} +m \psi_{1\ssL} = 0 \,,&&\quad\gamma^\mu \partial_\mu \psi_{2\ssL} - O_1 \psi_{2\ssR} +m \psi_{1\ssR} = 0 \label{fcoupled1}\\
\gamma^\mu \partial_\mu \psi_{1\ssR} + O_1 \psi_{1\ssL} +m \psi_{2\ssL} = 0 \,,&&\quad
\gamma^\mu \partial_\mu \psi_{1\ssL} - O_2 \psi_{1\ssR} +m \psi_{2\ssR} = 0 \label{fcoupled2}
\eea
(where we have applied the projection matrix identities $P_\ssL^2 = P_\ssR^2 =1$, $P_\ssL P_\ssR = 0$). From these expressions, it is clear that setting $m=0$ decouples $\psi_1$ from $\psi_2$. In other words, for massless fermions there is a halving of the number of degrees of freedom.

Decoupling \pref{fcoupled1} and \pref{fcoupled2} gives
\bea
(\square_4+O_2 O_1 - m^2)\psi_{1\ssL} = 0 \,,&&\quad
(\square_4+O_1 O_2 - m^2)\psi_{1\ssR} = 0 \\
(\square_4+O_1 O_2 - m^2)\psi_{2\ssL} = 0 \,,&&\quad
(\square_4+O_2 O_1 - m^2)\psi_{2\ssR} = 0
\eea
where $\square_4 = \eta^{\mu\nu}\partial_\mu \partial_\nu$ is the 4D d'Alembertian. In what follows, we will quote results only for $\psi_1$, since a solution for $\psi_{1\ssL}$ ($\psi_{1\ssR}$) will also be a solution for $\psi_{2\ssR}$ ($\psi_{2\ssL}$). To obtain the mode sum for a massive spin--1/2 field, we should remember to multiply the result for $\psi_1$ by 2.
%
%

Using the ans\"atze
\be
\psi_{1\ssL(\ssR)} = e^{i k_4 \cdot x} \Theta_{\ssL(\ssR)}(\theta) \, e^{in_{\ssL(\ssR)}\varphi}
\ee
where $n_\ssL$ and $n_\ssR$ are integers satisfying $n_\ssR-n_\ssL =b$ given eqs.~\pref{fcoupled1}, \pref{fcoupled2} (and because of the $e^{i\varphi}$ factors in $O_1$, $O_2$), we find that the spinor functions $\Theta_{\ssL(\ssR)}$ must satisfy the following equations:
\bea \label{ThetaEqL}
\frac{\exd^2\Theta_\ssL}{\exd\theta^2} &+& \cot\theta \frac{\exd\Theta_\ssL}{\exd\theta} + \Bigg[ \lambda^\mathrm{f}_{jn_{1/2}} -\frac14  \nn\\
\qquad\qquad&-& \left(\frac{\omega^2\left(n_{1/2}-qA_\varphi\right)^2-\omega \left[ (n_{1/2} - q A_\varphi)\cos\theta +  q F_{\theta\varphi}\sin\theta \right]+\frac14}{\sin^2\theta} \right) \Bigg]\Theta_\ssL =0 \qquad\qquad \\
\label{ThetaEqR} \frac{\exd^2\Theta_\ssR}{\exd\theta^2} &+& \cot\theta \frac{\exd\Theta_\ssR}{\exd\theta} + \Bigg[ \lambda^\mathrm{f}_{jn_{1/2}} -\frac14  \nn\\
\qquad\qquad&-& \left(\frac{\omega^2\left(n_{1/2}-qA_\varphi\right)^2+\omega \left[ (n_{1/2} - q A_\varphi)\cos\theta +  q F_{\theta\varphi}\sin\theta \right]+\frac14}{\sin^2\theta} \right) \Bigg]\Theta_\ssR = 0 \qquad\qquad
\eea
where $F_{\theta\varphi} = \exd A_\varphi/\exd\theta$ (in our gauge, where $A_\theta=0$),
\be
n_{1/2}:=n_\ssL+ b/2 = n_\ssR - b/2 \,,
\ee
and where $\lambda_{jn_{1/2}}^\mathrm{f}:= - (k_4^2+m^2) r^2$ are the required eigenvalues (in units of the KK scale). These equations transform into one another when $L \leftrightarrow R$ and $n_{1/2} \leftrightarrow - n_{1/2}$, $A_\varphi \leftrightarrow -A_\varphi$. Since such a transformation is equivalent to a change of coordinates (in particular, the exchange of north and south poles), we expect that the resulting eigenvalues will be the same when solving either equation, as needed. The exception to this logic is the situation where $k_4^2=-m^2$ (i.e. when $\lambda^{\rm f}=0$); in this case, eqs.~\pref{fcoupled1} and \pref{fcoupled2} partially decouple. In this case, we will find that only one of $\psi_{1\ssL}$, $\psi_{1\ssR}$ will have zero mode solutions, given some fixed value $N\neq 0$.

Remarkably, after: 1) performing the coordinate substitution $x:=\cos\theta$; and 2) substituting
\be
\Theta_{\sigma}(x) = (1-x)^{y_{\sigma}}(1+x)^{z_{\sigma}} f_{\sigma}(x) \,
\ee
where $\sigma=+1$ ($\sigma=-1$) denotes the left--handed (right--handed) field, we find that equations \pref{ThetaEqL} and \pref{ThetaEqR} become
\bea \label{fsigmaEq}
(1-x^2)\frac{\exd^2 f_\sigma}{\exd x^2} &-&2[(y_\sigma+z_\sigma)x +y_\sigma-z_\sigma]\frac{\exd f_\sigma}{\exd x} + \Bigg[ \lambda^\mathrm{f}_{jn_{1/2}} -\frac14 + \frac{\sigma\cN}2   \\
&-& \left(\frac{\omega^2\left(n_{1/2}-qA_\varphi\right)^2-\sigma\omega  (n_{1/2} - q A_\varphi)x+k(x,y_\sigma,z_\sigma)+\frac14}{1-x^2} \right) \Bigg] f_\sigma = 0 \nn
\eea
where we have borrowed the previous definition for $k(x,y,z)$ in eq.~\pref{kxyzdef}. (We have substituted for $F_{\theta\phi}$ to obtain the term proportional to $\cN$ in the equation above.) Requiring the numerator in the second line of eq.~\pref{fsigmaEq} to be proportional to $(1-x^2)$ makes eq.~\pref{fsigmaEq} a regular ODE and gives the following result for $y_\sigma$, $z_\sigma$:
\be
y_\sigma = \frac12\left|\omega\Big(n_{1/2}+(N-\Phi)\epsilon_{-b} -b\,\Phi_b \Big)-\frac\sigma2\right|  \,,\quad z_\sigma = \frac12 \left|\omega\Big(n_{1/2}-(N-\Phi)\epsilon_b - b\,\Phi_b \Big)+\frac\sigma2\right| \,.
\ee
The resulting differential equation is
\bea
(1-x^2) \frac{d^2f_\sigma}{dx^2} &-& 2 [(1+y_\sigma+z_\sigma)x+y_\sigma-z_\sigma] \frac{df_\sigma}{dx}  \nn\\
&+& \left[ \lambda^\mathrm{f}_{jn_{1/2}} -\frac14+ \frac{\cN^2}{4} - (y_\sigma+z_\sigma)(y_\sigma+z_\sigma+1) \right] f_\sigma = 0 \,, \label{fODEfermi}
\eea
In passing, it should be mentioned that, if the goal is regularity of eq.~\pref{fsigmaEq}, it is equally valid to take $y_\sigma$ and/or $z_\sigma$ to be definitely negative, rather than positive, in eliminating the singularities from eqs.~\pref{ThetaEqL} and \pref{ThetaEqR}. However, we should discard these solutions since they cause for $\psi$ to diverge at one or both of the poles.

In the case where the spinor does not couple to the background gauge field ($q=0$ or, equivalently, $N=\Phi_b=0$) and the background geometry is that of a sphere ($\alpha=\omega=1$), the above equations simplify as follows. Keeping in mind that $|n_{1/2}|\geq 1/2$, we have that $y_\ssL=z_\ssR = \frac12(|n_{1/2}|-\frac12)$ and $y_\ssR=z_\ssL=\frac12(|n_{1/2}|+\frac12)$ and so eq.~\pref{fsigmaEq} becomes
\bea
(1-x^2) \frac{d^2f_\sigma}{dx^2} - 2 \left[(1+|n_{1/2}|)x-1\right] \frac{df_\sigma}{dx} + \left[ \lambda^\mathrm{f}_{jn_{1/2}}  - \left(|n_{1/2}|+\frac12 \right)^2 \right] f_\sigma &=& 0 \,\,(\mathrm{sphere}). \qquad\quad
\eea
The power series solutions to these equations terminate as long as
\bea
\lambda^\mathrm{f}_{jn_{1/2}} &=& (j+|n_{1/2}|+1/2)^2  \quad(\textrm{sphere}) 
\eea
for some integer $j\geq 0$. If we define $\ell:=j+|n_{1/2}|-1/2$, the above expression for $\lambda^\mathrm{f}$ simplifies to the familiar $(\ell+1)^2$, as found in \cite{CandelasWeinberg}. The degeneracy in this case is $2(\ell+1)$, which accounts for the number of different ways one can choose $(j,n_{1/2})$ to get the same $\ell$. These two formulations are equivalent.

In the general case, the condition for termination of the power series solution to eqs.~\pref{fODEfermi} (at the same order $x^j$) is that
\be
\lambda^\mathrm{f}_{jn_{1/2}}  = \left(j_\sigma+y_\sigma+z_\sigma+\frac12\right)^2 -\frac{\cN^2}4
\ee
where $j_\sigma \in \{0,1,\ldots\}$ (and $j_\ssL \neq j_\ssR$ generally).
\subsection*{Mode Sum}

The spin-1/2 mode sum for $\psi_1$ is
\be
S_\mathrm{f}(\omega,N,\Phi_b,t) = 2 S_\mathrm{f\ssL}(\omega,N,\Phi_b,t) + 2 S_\mathrm{f\ssR}(\omega,N,\Phi_b,t)
\ee
where
\be
S_\mathrm{f\sigma}(\omega,N,\Phi_b,t) = -2 \times e^{ t\cN^2/4} \sum_{j = 0}^\infty \sum_{n = - \infty}^\infty \exp \Bigl[ - t (j + y_\sigma+z_\sigma+1/2)^2 \Bigr] \,.\label{GeneralSpinorS}
\ee
Here, $\sigma=L=+1$ ($\sigma=R=-1$) gives the result for the left-handed (right-handed) spinors. (Note that we have dropped the $\sigma$ subscript on the dummy indices $j$, $n$.) Recalling that $n_{1/2}:=n+b\sigma/2$, we can write $y_\sigma$ and $z_\sigma$ as follows:
\bea
y_\sigma &=& \frac12\left|\omega\left[n+(N-\Phi)\epsilon_{-b} -b\,\left(\Phi_b - \frac\sigma2\right) \right] - \frac\sigma2 \right| = \frac\omega2\left|n+(N_{\rm f\sigma}-\Phi_{\rm f\sigma})\epsilon_{-b} -b\,\Phi^{\rm f\sigma}_b \right|\qquad \\
z_\sigma &=& \frac12\left|\omega\left[n-(N-\Phi)\epsilon_{b} -b\,\left(\Phi_b - \frac\sigma2\right) \right] + \frac\sigma2 \right| = \frac\omega2\left|n-(N_{\rm f\sigma}-\Phi_{\rm f\sigma})\epsilon_{b} -b\,\Phi^{\rm f\sigma}_b \right|\qquad
\eea
where, in the second expressions, we have introduced
\be
N_{\rm f\sigma} := N-\sigma \,,\,\,  \Phi^{\rm f\sigma}_b := \Phi_b - \frac\sigma2\left(1-\frac1\omega\right) \,\left(\textrm{and } \Phi_{\rm f\sigma} := \sum_b \Phi^{\rm f\sigma}_b = \Phi - \sigma \left(1-\frac1\omega\right) \right)\,.
\ee
(Later on, we will also use $\Phi_0^{\rm f} :=(1-\omega^{-1})/2$.) Given these identifications, we can write the fermionic mode sums in terms of the previously--derived scalar result:
\bea
S_{\rm f\sigma}(\omega,N,\Phi_b,t) &=& -2 \times e^{t[\cN^2-(1+\cN_{\rm f\sigma}^2)]/4}  S_{\rm s}(\omega,N_{\rm f\sigma},\Phi^{\rm f\sigma}_b,t)
\eea
where $\cN_{\rm f\sigma}:=\omega(N_{\rm f\sigma} - \Phi_{\rm f\sigma}) = \cN -\sigma$. Expanding this out, we find (using the notation introduced in eq.~\pref{Fbdefs}) that $s^{\rm f\sigma}_{-1} = -2/\omega$,
\bea
s^{\rm f\sigma}_0(\omega,N_{\rm f\sigma},\Phi^{\rm f\sigma}_b) &=& \frac1\omega \left[ \frac16 -\sigma\cN + \left( \frac16 + \Delta F_{\rm f\sigma} \right)\omega^2 \right] \,, \\
s^{\rm f\sigma}_1(\omega,N_{\rm f\sigma},\Phi^{\rm f\sigma}_b) &=& \frac1\omega\Bigg[ \frac7{720} - \frac{\cN^2}{6} + \left(\frac1{72} - \frac{(1-3\sigma\cN)}{6} \Delta F_{\rm f\sigma} \right)\omega^2  \nn\\
&&\qquad +\frac{\omega^3(\cN-\sigma)}{6} \sum_b \Delta\Big[\Phi_b^{\rm f\sigma} \,G(|\Phi_b^{\rm f\sigma}|)\Big] + \Bigg(\frac7{720} +\frac1{6} \Delta[F_{\rm f\sigma}^{(2)}]  \Bigg)\omega^4 \Bigg] , \\
s^{\rm f\sigma}_2(\omega,N_{\rm f\sigma},\Phi^{\rm f\sigma}_b) &=& \frac1\omega \Bigg[ \frac{31}{20160} - \frac{31\,\cN^2}{1440} + \left(\frac7{2880} -\frac{\cN^2}{144} -\frac{(1-5\cN^2)}{60} \Delta F_{\rm f\sigma}  \right) \omega^2   \nn\\
&&  -	\frac{\omega^3\cN(1-\sigma\cN)}{12} \sum_b \Delta[\Phi_b^{\rm f\sigma} \, G(|\Phi_b^{\rm f\sigma}|)] + \Bigg(\frac{7}{2880} -\frac{7\,\cN^2}{1440}  \nn\\
&&  -\frac{(1-3\sigma\cN+\cN^2)}{12} \Delta[F_{\rm f\sigma}^{(2)}]  \Bigg) \omega^4 +\frac{\omega^5(\cN-\sigma)}{60} \sum_b \Delta[\Phi_b^{\rm f\sigma} \, G(|\Phi_b^{\rm f\sigma}|)(1+3F^{\rm f\sigma}_b)]  \nn\\
&&  + \Bigg(\frac{31}{20160} +\frac{\Delta[F_{\rm f\sigma}^{(2)}]}{60} +\frac{\Delta[F_{\rm f\sigma}^{(3)}]}{30} \Bigg)\omega^6 \Bigg]
\eea
where the $F^{(n)}_{\rm f\sigma}$'s are calculated using $\Phi_b^{\rm f\sigma}$, and where we have taken
\be
\Delta [X] := X-X\Big|_{\Phi_b =0}
\ee
to ensure that $\Delta [X] \to 0$ as $\Phi_b \to 0$. For example, $\Delta F^{\rm f\sigma}_b = |\Phi_b^{\rm f\sigma}|(1-|\Phi_b^{\rm f\sigma}|) -(1-\omega^{-2})/4$. (Recall that $\omega> 1$ for the rugby--ball.) It is important to remember that the derivation of these results has assumed $|\Phi_b^{\rm f\sigma}|<1$, which is distinct from the restriction $|\Phi_b|<1$ in the scalar case but of little difference near $\omega= 1$.

Since the 6D the chiral spinor $\psi_1$ is the sum of a left--handed and a right--handed 4D chiral spinor, its $s^{\rm f}_i$ coefficients are given by
\be
s^{\rm f}_i = \sum_{\sigma\in\{\pm 1\}} s_i^{\rm f\sigma} \,.
\ee
%
%
(For massive 6D spinors, we must double this result.) Since we expect some cancelation when considering this sum of coefficients, let's work out the result. We must be careful, though, since there is some subtlety in this. For example, the spin--averaged quantity $\vev { \Delta F^{\rm f}_b}$ depends differently on $\Phi_b$ for $|\Phi_b|>\Phi_0^{\rm f} := (1-\omega^{-1})/2$, as compared to when $|\Phi_b|<\Phi_0^{\rm f}$:
\be
\vev {\Delta F_b^{\rm f}} := \frac12 \sum_\sigma \Delta F^{\rm f\sigma}_b = \left\{
\begin{array}{cc}
 - \Phi_b^2 & ,\quad |\Phi_b|\leq\Phi_0^{\rm f} \\
F_b - \Phi_0^{\rm f} & ,\quad |\Phi_b|\geq\Phi_0^{\rm f}
\end{array}
\right.\,.
\ee
Both of these expressions have useful domains of validity: when considering small fluxes, $\Phi_b$, in the presence of some $\Phi_0^{\rm f}\neq 0$, it is the upper function which is of interest; in the case of the larger fluxes, or in the sphere case where $\omega=1$ and $\Phi_0^{\rm f}=0$, it is the lower function which is relevant. We should therefore treat these two situations differently.

In the case where both $|\Phi_b|\leq \Phi_0^{\rm f}$, we find that
\bea
s^{\rm f}_0(\omega,N,\Phi_b) &=& \frac1\omega \left[ \frac13 + \left( \frac13 -2\sum_b\Phi_b^2 \right)\omega^2 \right] \,, \\
s^{\rm f}_1(\omega,N,\Phi_b) &=& \frac1\omega\Bigg[ \frac7{360} -\frac{\omega\cN\Phi}2 - \frac{\cN^2}{3} + \left(\frac1{36} -\frac{1}6\sum_b\Phi_b^2  \right)\omega^2  \nn\\
&&\qquad -\frac{\omega^3\cN}{6}\sum_b \Phi_b (1-4\Phi_b^2) + \Bigg(\frac7{360} -\frac1{6}\sum_b \Phi_b^2 (1-2\Phi_b^2) \Bigg)\omega^4 \Bigg] , \\
s^{\rm f}_2(\omega,N,\Phi_b) &=& \frac1\omega \Bigg[ \frac{31}{10080} -\frac{\omega\cN\Phi}{16} - \frac{31\,\cN^2}{720}  + \left(\frac7{1440} -\frac{\cN^2}{72}\bigg(1-6\sum_b\Phi_b^2 \bigg) - \frac7{240} \sum_b\Phi_b^2 \right) \omega^2   \nn\\
&& -\frac{\omega^3\cN}{24}\sum_b \Phi_b (1-4\Phi^2_b)  + \Bigg(\frac{7}{1440} -\frac{7\,\cN^2}{720} -\frac{(1-2\cN^2)}{24}\sum_b \Phi_b^2(1-2\Phi_b^2) \Bigg) \omega^4 \nn\\
&&  -\omega^5\cN\left( \sum_b \Phi_b \bigg( \frac7{240} -\frac{\Phi_b^2}6 +\frac{\Phi_b^4}5\bigg) \right)  \nn\\
&&  + \Bigg(\frac{31}{10080} -\sum_b \Phi_b^2\left(\frac7{240} -\frac{\Phi_b^2}{12} + \frac{\Phi_b^4}{15} \right) \Bigg)\omega^6 \Bigg] \,.
\eea

When both $|\Phi_b|\geq \Phi_0^{\rm f}$, the small--$t$ coefficients are given in terms of
\bea
\rho_b := \mathrm{sgn}(\Phi_b) &=& \frac{\Phi_b}{|\Phi_b|} \,,\quad \tilde \Phi_b := \omega(\Phi_b - \rho_b \Phi_0^{\rm f}) \,,\quad \rho:= \sum_b \rho_b \nn\\
 \tilde F_b &:=& |\tilde \Phi_b| (1-|\tilde \Phi_b|) \,,\quad \tilde F:=\sum_b \tilde F_b \quad\textrm{etc.}
\eea
as follows:
\bea
s^{\rm f}_0(\omega,N,\Phi_b) &=& \frac1\omega \left[ -\frac23 +2\tilde F+2\omega  -\frac{2\,\omega^2}3 \right] \,, \\
s^{\rm f}_1(\omega,N,\Phi_b) &=& \frac1\omega\Bigg[ -\frac1{45}  +\frac{\rho\cN}6 - \cN\sum_b \tilde\Phi_b \bigg( \tilde F_b + \frac{\tilde \Phi_b^2}3\bigg) - \frac{\cN^2}{3} + \frac{\tilde F^{(2)}}3 \nn\\
&&\qquad+ \left(\frac19 -\frac{\tilde F}3 +\frac{\cN\tilde\Phi}3 - \frac{\rho\cN}6 \right)\omega^2 - \frac{\omega^4}{45} \Bigg] \\
s^{\rm f}_2(\omega,N,\Phi_b) &=& \frac1\omega \Bigg[ -\frac1{315} +\frac{\tilde F^{(2)}}{30} + \frac{\tilde F^{(3)}}{15} + \frac{\rho\cN}{30} - \frac{\cN}3 \sum_b \tilde \Phi_b \bigg( \tilde F_b^2 + \frac{\tilde\Phi_b^2 \tilde F_b}2 +\frac{\tilde\Phi_b^4}{10} \bigg)  \nn\\
&&\qquad -\cN^2\bigg(\frac1{45}+ \frac{\tilde F^{(2)}}6\bigg) + \Bigg( \frac1{90} - \frac{\tilde F^{(2)}}6  - \frac{\cN}6 \sum_b \tilde \Phi_b \, G(|\tilde \Phi_b|) \nn\\
&&\qquad - \cN^2\bigg(\frac{1}{18} - \frac{\tilde F}6\bigg)\Bigg)\omega^2 + \Bigg( \frac{1}{90} -\frac{\tilde F}{30} -\frac{\rho\cN}{30} + \frac{\cN\tilde\Phi}{30} +\frac{\cN^2}{90} \Bigg)\omega^4 -\frac{\omega^6}{315} \Bigg]   \,\,\qquad
\eea
(Note that we should take $\tilde\Phi_b =0$ when comparing these sets of equations when they overlap at $|\Phi_b|=\Phi_0^{\rm f}$.) In the case of an uncharged, massive fermion on a sphere (i.e.~when $\omega=1$, $\cN=\Phi_b=0$), these two sets of expressions both give
\be
s^\mathrm{f}_{-1}(1,0,0) = -8 \,,\quad s^\mathrm{f}_0(1,0,0) = \frac43 \,,\quad s^\mathrm{f}_1(1,0,0) = \frac2{15} \,,\quad s^\mathrm{f}_2(1,0,0) = \frac2{63}
\ee
in agreement with the result in \cite{KandM}.


\section*{Spectrum and Mode Sum for the Massless Spin-1 Field}

This appendix computes the spectrum $\lambda^\mathrm{gf}_{jn}$ and the corresponding small-$t$ limit of the mode sum
\be
S_\mathrm{gf}(t):= \sum_{j,n} e^{-t\lambda^\mathrm{gf}_{jn}} = \sum_{i=-1}^2 s^\mathrm{gf}_i \, t^i
\ee
for a 6D massless spin-1 field $A_\ssM$ on the rugby ball.

\subsection*{Eigenvalues}

The spin--1 equation of motion is
\be
g^{\ssM\ssN}D_\ssM F_{\ssN\ssP} =0
\ee
where $F_{\ssM\ssN} = \pd_\ssM \cA_\ssN - \pd_\ssN \cA_\ssM$, $D_\ssM$ is the covariant derivative,
\be
{(D_\ssM)^\rho}_\sigma F^\sigma_{\ssN\ssP} = \pd_\ssM F^\rho_{\ssN\ssP} - \Gamma^\ssQ_{\ssM\ssN} F^\rho_{\ssQ\ssP} - \Gamma^\ssQ_{\ssM\ssP} F^\rho_{\ssN\ssQ} - i {(t^\ssA)^\rho}_\sigma A_\ssM F^\sigma_{\ssN\ssP} \,,
\ee
and the metric is
\bea
\exd s^2 &=& r^2 ( \exd \theta^2 + \alpha^2 \sin^2\theta \, \exd \varphi^2 ) +\eta_{\mu\nu} \exd x^\mu \exd x^\nu \nn\\
&=& r^2 ( \exd \theta^2 + \alpha^2 \sin^2\theta \, \exd \varphi^2 ) + \delta_{ij} \exd x^i \exd x^j + 2 \exd x^- \exd x^+ \,.\label{gAdefvector}
\eea
In the last line above, we introduce light--cone coordinates $x^\pm := (x^3 \pm x^0)/\sqrt{2}$ which are convenient for what follows. (Also, $i$, $j$ run only over $1$, $2$.) More formally, we can introduce the (sub--)metrics $g_{\alpha\beta}$ and $g_{mn}$ over the light--cone coordinates $\alpha,\beta \in \{+,-\}$ and extra--dimensional coordinates $m,n \in \{\theta,\varphi\}$, respectively:
\be
g_{\alpha\beta} = \left(
\begin{array}{cc}
0&1\\
1&0
\end{array}
\right) \,,\quad g_{mn} = \left(
\begin{array}{cc}
r^2 & 0 \\
0 & \alpha^2 r^2 \sin^2\theta
\end{array}
\right)\,.
\ee
The full metric $g_{\ssM\ssN}$ is then composed as follows:
\be
g_{\ssM\ssN} = g_{mn} \oplus \delta_{ij} \oplus g_{\alpha\beta} \,.
\ee

In what follows, we assume that the generator for the background gauge field $A_\ssM$ is diagonal in the adjoint representation: ${(t^\ssA)_\rho}^\sigma = q \, \delta_\rho^\sigma$. Given this choice, it is convenient to suppress the group indices in what follows. We also choose the light--cone gauge, in which
\be
\cA_- := \frac{\cA_3 - \cA_0}{\sqrt{2}} = 0 \,.
\ee
As before, the background gauge field is taken to be
\be
q\,A_\ssM \exd x^\ssM = q\,A_\varphi \exd\varphi = \left[-\frac{N-\Phi}2 \left( \cos\theta -b \right) +b\Phi_b\right] \,\exd\varphi \,,
\ee
where $N$ is an integer. Expanding out the equation of motion for $P=-$, we find that
\bea
P=-:\qquad 0 &=& - \delta^{ij} \partial_i \partial_- \cA_j - g^{-+} \partial_-^2 \cA_+ - g^{mn} D_m \partial_- \cA_n \nn\\
&=& -\partial_- \left( \partial_i \cA^i + \partial_- \cA_+ + g^{mn} D_m \cA_n \right) \,.
\eea
This can be integrated to give the constraint
\be
F_{-+} = \partial_- \cA_+ = -\partial_i \cA^i - g^{mn} D_m \cA_n \,.
\ee
This constraint tells us that $\cA_+$ is completely determined (up to some initial conditions) once we have solved for $\cA_k$ and $\cA_p$. As we shall see, this constraint also simplifies the equations of motion for $\cA_k$ and $\cA_p$.

Evaluating the equations of motion using the above constraint for $P=k \in \{1,2\}$ and $P=p \in \{\theta,\varphi\}$, respectively, we find that
\bea
P=k:\qquad 0 &=& \partial^i F_{ik} + g^{-+} \partial_- F_{+k} + g^{+-} \partial_+ F_{-k} + g^{mn} D_m F_{nk} \nn\\
&=& \left(\square_4 + g^{mn} D_m D_n \right) \cA_k \label{eomAk}\\
P=p:\qquad 0 &=& \partial^i F_{ip} + g^{-+} \partial_- F_{+p} + g^{+-} \partial_+ F_{-p} + g^{mn} D_m F_{np} \nn\\
&=& \left(\square_4 + g^{mn} D_m D_n \right) \cA_p + D_p (g^{mn} D_m \cA_n) - g^{mn} D_m D_p \cA_n \label{eomAp}
\eea
where $\square_4 := \eta^{\mu\nu} \partial_\mu \partial_\nu =\partial^i \partial_i + 2\,\partial_- \partial_+$. Eqs.~\pref{eomAk} are exactly the same as the one encountered in the case of a scalar field, so the corresponding spectrum for $\cA_1$ and $\cA_2$ will be that of a scalar. Eqs.~\pref{eomAp}, however, are coupled and our task is to decouple them.

Given that the Christoffel symbols mostly vanish, with the exception of $\Gamma_{\theta\phi}^\phi = \Gamma_{\phi\theta}^\phi = \cot \theta$ and $\Gamma_{\phi\phi}^\theta = -\alpha^2 \sin\theta \cos\theta$, we find that, using the ansatz
\be
\cA_m(x,\theta,\varphi) = e^{i k_4\cdot x} \, \Theta_m(\theta) e^{i n \varphi} \,,
\ee
Eqs.~\pref{eomAp} become
\bea
\frac{\exd^2 \Theta_\theta}{\exd \theta^2} + \cot\theta \frac{\exd \Theta_\theta}{\exd \theta}+ \lambda \Theta_\theta - \frac{K_\theta}{\alpha^2 \sin^2\theta} &=& 0 \\
\frac{\exd^2 \Theta_\varphi}{\exd \theta^2} - \cot\theta \frac{\exd \Theta_\varphi}{\exd \theta}+ \lambda \Theta_\varphi - \frac{K_\varphi}{\alpha^2 \sin^2\theta} &=& 0
\eea
where $\lambda := - k_4^2 r^2$ are the required eigenvalues and where
\bea
K_\theta &=& \left[ (n-q A_\varphi)^2 + \alpha^2 \right] \Theta_\theta + 2i(n-q A_\varphi)\cot\theta\, \Theta_\varphi \\
K_\varphi &=& (n-q A_\varphi)^2 \Theta_\varphi - 2 i (n-q A_\varphi) \alpha^2 \cos\theta\sin\theta \,\Theta_\theta \,.
\eea
In order to algebraically decouple the $K_m$ terms, we should first perform a field redefinition such that the differential operators are the same for both equations. (Note that they are not currently, due to the sign difference in the $\cot\theta \,\exd \Theta_m/\exd\theta$ terms.)

With this goal in mind, we should (for convenience) switch to $x:=\cos\theta$, and substitute
\be
\Theta_m (x) = (1-x)^{y_m} (1+x)^{z_m} f_m(x) \,,
\ee
where $y_m$, $z_m$ are to be related in some specific way. Doing so gives
\bea
(1-x^2) \frac{\exd^2 f_\theta}{\exd x^2} -2\left[ (1+z_\theta+y_\theta)x +y_\theta-z_\theta \right] \frac{\exd f_\theta}{\exd x} + \lambda f_\theta - \frac{\hat K_\theta}{1-x^2} &=& 0 \\
(1-x^2) \frac{\exd^2 f_\varphi}{\exd x^2} -2\left[ (z_\varphi+y_\varphi)x +y_\varphi-z_\varphi \right] \frac{\exd f_\varphi}{\exd x} + \lambda f_\varphi - \frac{\hat K_\varphi}{1-x^2} &=& 0
\eea
for some $\hat K_m$, to be specified shortly. Equating the coefficients of $\exd f_m/\exd x$, we find that the required relations are
\be
y_\theta = y_\varphi -1/2 := y \,,\quad z_\theta = z_\varphi -1/2 := z \,.
\ee
Given these identifications, we find that the $\hat K_m$ are given by
\bea
\hat K_\theta &=& \left(\omega^2 (n-q\cA_\varphi)^2 + 1+  k(x,y,z)\right) f_\theta + 2i\omega^2(n-q\cA_\varphi) x \, f_\varphi \\
\hat K_\varphi &=& \left( \omega^2 (n-q\cA_\varphi)^2 +1 + k(x,y,z)\right) f_\varphi - 2i(n-q\cA_\varphi) x \, f_\theta
\eea
where $k(x,y,z)$ is defined as in eq.~\pref{kxyzdef}.
%

From here, we can see that the complex choice
\be
\cF(x) := f_\theta(x) + i \omega f_\varphi(x)
\ee
yields decoupled equations of motion for $\cF$ (and $\cF^*$):
\be
(1-x^2) \frac{\exd^2 \cF}{\exd x^2} - 2 \left[ (1+y+z)x + y-z\right] \frac{\exd \cF}{\exd x} + \left( \lambda - \frac{K_\cF}{1-x^2} \right) \cF = 0
\ee
where
\be
K_\cF \, \cF := \hat K_\theta + i\omega \hat K_\varphi = \left[ \omega^2 (n-q\cA_\varphi)^2 +1+k(x,y,z) + 2\omega (n-q A_\varphi) x \right] \cF \,.
\ee
Now that we have decoupled the equations of motion, we can eliminate the singular behaviour as $x \to \pm 1$ by requiring that $K_\cF$ be proportional to $(1-x^2)$. Doing so gives
\be
y=\frac{\omega}2 \left| n + (N-\Phi) \epsilon_{-b} -b\Phi_b +\frac1\omega \right| \,,\quad z = \frac{\omega}2 \left| n  - (N-\Phi) \epsilon_b - b\Phi_b - \frac1\omega \right| \,,
\ee
yielding the spectrum
\be
\lambda^{\rm gf}_{jn}(\cF) = \left(j+y+z+\frac12\right)^2 - \frac14 - \frac{\omega^2 N^2}4 \,.
\ee
The spectrum for $\cF^*$ will be slightly different, since complex conjugation is equivalent to taking $n \to -n$ and $q \to -q$ (or $N \to -N$, $\Phi_b \to - \Phi_b$). In this case,
\be
y_* = \frac{\omega}2 \left| n + (N-\Phi) \epsilon_{-b}-b\Phi_b -\frac1\omega \right| \,,\quad z_* = \frac{\omega}2 \left| n  - (N-\Phi) \epsilon_b - b\Phi_b + \frac1\omega \right| \,.
\ee
As is done previously, we have taken $y$, $y_*$, $z$, and $z_*$ to be positive-definite, so that the corresponding fields do not diverge near the poles at $x=\pm1$.

Before computing the mode sum, it is helpful to introduce a notion of helicity similar to that used for the spin--1/2 field. If we define a quantity $\upsilon$ such that $\xi=+1$ ($\xi=-1$) corresponds to the case of $\cF$ ($\cF^*$), then we can efficiently write
\be
\lambda^{\rm gf\xi}_{jn}(\omega,N,\Phi_b) = \left(j+y_\xi+z_\xi+\frac12\right)^2 - \frac14 - \frac{\omega^2 N^2}4
\ee
where
\be
y_\xi = \frac{\omega}2 \left| n + (N-\Phi) \epsilon_{-b} - b\Phi_b + \frac\xi\omega \right| \,,\quad z_\xi = \frac{\omega}2 \left| n  - (N-\Phi) \epsilon_b - b\Phi_b - \frac\xi\omega \right| \,.
\ee
The gauge field is then understood to be comprised of two spin states with $\xi=0$, and two spin states with each of $\xi =\pm 1$, respectively.

\subsection*{Mode Sum}

We wish to compute the small--$t$ limit of the sum
\be
S_{\rm gf}(\omega,N,\Phi_b,t) := \sum_\xi S_{\rm gf}^\xi (\omega,N,\Phi_b,t) = \sum_\xi \sum_{j,\,n} e^{-t \lambda^{\rm gf\xi}_{jn}} \,,
\ee
Here, the additonal sum is over spin states, $\xi \in \{0,0,1,-1\}$. Given the spectrum derived previously, we see that two of the spectra will be identical to that of a scalar, and two will be  slightly different. Similar what was done for the spin--1/2 field, we can make the identifications
\be
N_{\rm gf\xi}=N \,,\quad\Phi^{\rm gf\xi}_b := \Phi_b -\frac\xi\omega
\ee
(or $\Phi^{\rm gf\xi}_b := \Phi_b -\xi\Phi_0^{\rm gf}$ where $\Phi_0^{\rm gf}=1/\omega$) so that
\be
\lambda_{jn}^{\rm gf\xi}(\omega,N,\Phi_b) + \frac{\cN^2}4 = \lambda_{jn}^{\rm s}(\omega,N^{\rm gf\xi},\Phi_b^{\rm gf\xi}) + \frac{\cN_{\rm gf\xi}^2}4 \,,
\ee
where $\cN_{\rm gf\xi} := \omega(N_{\rm gf\xi} -\Phi_{\rm gf\xi}) = \cN +2\xi$. In this case, $S_{\rm gf}$ can be written largely in terms of the scalar result:
\bea
S_{\rm gf}(\omega,N,\Phi_b,t) &=& 2\times S_{\rm s}(\omega,N,\Phi_b,t) + \sum_{\xi\in\{1,-1\}} e^{t(\cN^2-\cN^2_{\rm gf\xi})/4}\,S_{\rm s}(\omega,N,\Phi_b-\xi/\omega,t) \nn\\
&=& 4\times S_{\rm s}(\omega,N,\Phi_b,t) + \Delta S_{\rm gf} (\omega,N,\Phi_b,t) \,.
\eea
In the last line above, we introduce the quantity $\Delta S_{\rm gf} = \sum_\xi (S^\xi_{\rm gf} - S_{\rm s})$, which is a convenient one since the small--$t$ limit of $S_{\rm gf}^\xi$ is very close to that of $S_{\rm s}$. (In what follows, we will dispense with the level of detail as was given for the spin--1/2 field and skip to the complete result, summed over spins.) Expanding out the argument of the sum above, we find a result that behaves similarly to the one found for the spin--1/2 field, in that the result will depend on whether the fluxes $\Phi_b$ are either greater or smaller in magnitude when compared to $\Phi^{\rm gf}_0=1/\omega$.

When both $|\Phi_b|\leq\Phi_0^{\rm gf}$, we find that the differences, $\Delta s^{\rm gf}_i$, are given by $\Delta s^{\rm gf}_{-1} = 0$,
\bea
\Delta s^{\rm gf}_0(\omega,N,\Phi_b) &=& \frac1\omega\left[-2\omega+ \omega^2 \sum_b |\Phi_b| \right] \,, \\
\Delta s^{\rm gf}_1(\omega,N,\Phi_b) &=& \frac1\omega\Bigg[ \cN^2 + \omega \cN \Phi +\frac{\omega^2}3 \sum_b |\Phi_b| -\frac{\omega^3 \cN}2 \sum_b \Phi_b \, |\Phi_b| - \frac{\omega^4}3 \sum_b |\Phi_b|^3 \Bigg] \,,\qquad \\
\Delta s^{\rm gf}_2(\omega,N,\Phi_b) &=& \frac1\omega \Bigg[ -\frac{\omega\cN^2}4 + \omega^2 \bigg( \frac1{15} -\frac{\cN^2}{24}\bigg) \sum_b |\Phi_b|  - \frac{\omega^3\cN}4 \sum_b \Phi_b \, |\Phi_b|  \nn\\
&& \qquad  -\frac{\omega^4 (1-\cN^2)}6 \sum_b |\Phi_b|^3 + \frac{\omega^5\cN}4 \sum_b \Phi_b \, |\Phi_b|^3 +\frac{\omega^6}{10} \sum_b |\Phi_b|^5 \Bigg]  \,.
\eea
These expressions serve to cancel all the terms in $s^s_i$ which are odd under $|\Phi_b| \to - |\Phi_b|$, for those spin states with $\xi = \pm 1$.

Since the condition $|\Phi_b|\geq \Phi_0^{\rm gf}=1/\omega$ does not include very much parameter space for small $\omega$ (since we are already implicitly assuming $|\Phi_b| \leq 1$), we will not consider this case in detail here.


\subsection*{Spectrum and mode sum for massive gauge fields}

Let us now turn to the computation of
the spectrum $\lambda^\mathrm{mgf}_{jn}$ and the corresponding small-$t$ limit of the mode sum
\be
S_\mathrm{mgf}(t):= \sum_{j,n} e^{-t\lambda^\mathrm{mgf}_{jn}}
\ee
for a massive 6D spin--1 field $A_\ssM$ on the rugby ball.

The simplest field content for a massive gauge field includes,
in addition to the metric and the gauge field, the scalar field $\Phi$ whose vev breaks the gauge group down
 to some subgroup $H$. The relevant part of the action is
\be \int \exd^6x \, \sqrt{-g} \; \left( -\frac{1}{4} F_{\ssM\ssN} F^{\ssM\ssN}
-\left(D_{\ssM}\Phi\right)^{\dagger}D^{\ssM}\Phi-U(\Phi)\right) \, ,\ee
where the potential  $U(\Phi)$ is assumed to have a minimum for $\Phi\neq 0$.

We would like to compute the linearized theory around a certain background solution
with $\Phi\neq 0$ (to give a bulk mass to some of the gauge fields) and the metric and gauge field set to an unwarped
configuration with 4D Poincar\'e invariance.
So we substitute $$A_\ssM \rightarrow A_\ssM+ V_\ssM\quad \mbox{and} \quad \Phi\rightarrow\Phi+\eta\, ,$$ where $V_\ssM$ and $\eta$ are
small perturbations and now $g_{\ssM \ssN}$, $A_\ssM$ and $\Phi$ represent
the given background solution. In order to interpret $\Phi\neq 0$ as a 6D spontaneous symmetry breaking, we require $\Phi$ to
be constant and to lie at the minimum of $U$. Then, to solve the background scalar equation, $D_\ssM D^\ssM \Phi=0$,
we also
demand that $\Phi$ does not break the $U(1)$ where the background gauge field lies: otherwise
it would not be  possible to set $\Phi$ to a constant value, at least in the sphere compactification of interest
in this paper \cite{RandjbarDaemi:2006gf}. Such requirement is equivalent to  demanding that the background
gauge field lives in the Lie algebra of $H$.

The linearized theory for the perturbations is rather complicated with mixing between $V_\ssM$, $\eta$
and the metric fluctuations. However,
choosing the light-cone gauge defined in the previous subsection, the bilinear action for $V_\ssM$ (such that Tr$(A_\ssM V_\ssM)=0$) and
$\eta$ has a relatively simple form \cite{RandjbarDaemi:2002pq}, that is
\bea \int \exd^6 x \, \sqrt{-g} \;  \left[ -\frac{1}{2} D_\ssM V_i D^\ssM V^i
-\frac{1}{2} D_\ssM V_m D^\ssM V^m  -\frac{1}{2} R^{mn}V_m V_n \right. \nn \\
 \left. -\frac{e^2}{2} \left(V^a_i V^{i b}+V^a_m V^{m b}\right)\Phi^\dagger \{T^a,T^b\}\Phi-
 \tilde{g} F^{mn} V_m \times V_n  \right. \label{bil-action-lcg}\\ \nn \left. -\left(D_\ssM \eta\right)^\dagger D^\ssM \eta
 -\frac12\eta \frac{\partial^2 U}{\partial \Phi^2}\eta-\frac12\eta^* \frac{\partial^2 U}{\partial \Phi^{*2}}\eta^*
 -\eta^* \frac{\partial^2 U}{\partial \Phi \partial \Phi^*}  \eta  \right. \nn  \\ \left.
 +\frac{e^2}{2} \left(\eta^\dagger T^a \Phi- \Phi^\dagger T^a \eta \right)
 \left(\eta^\dagger T^a \Phi- \Phi^\dagger T^a \eta \right)\right] \, ,\nn\eea 
where the indices are raised and lowered by the background metric and $e$ is a collective name for the gauge couplings of the generators broken  by $\Phi$.

The simplest model of this sort having
a non-trivial background flux is a $U(1)_1 \times U(1)_2$ gauge theory with $\Phi$ charged under the $U(1)_2$ only,
the background flux embedded in the $U(1)_1$ and the potential having a simple mexican hat form,
\be U(\Phi)= - \mu^2|\Phi|^2 + \lambda_4 |\Phi|^4\, ,\label{mexican-hat}\ee
with
$\mu^2 > 0$ (and so tachyonic) and $\lambda_4$ a positive constant. Let us first consider
this simple option. The bilinear action for $V^2_\ssM$ and
$\eta:=(\eta_1+i\eta_2)/\sqrt2$ is
\bea \int \exd^6 x \, \sqrt{-g} \;\left[-\frac{1}{2} D_\ssM V^2_i D^\ssM V^{i2}
-\frac{1}{2} D_\ssM V^2_m D^\ssM V^{m2}  -e^2 \Phi^2 \left(V^2_i V^{i 2}+V^2_m V^{m 2}\right)
 \right. \nn \\ \left. -\frac{1}{2} R^{mn}V^2_m V^2_n -\frac{1}{2}D_\ssM\eta_1 D^\ssM \eta_1
-\frac{1}{2}D_\ssM\eta_2 D^\ssM \eta_2 +m^2_{\rm tach} \eta_1^2-e^2 \Phi^2\eta_2^2
 \right] \, , \label{bilinear-mgf}
 \eea
where we have assumed $\Phi$ real and positive without loss of generality.
The fields $V^2_M$ and $\eta_2$ form together
a massive 6D vector field with bulk mass $m=2|e|\Phi$, while $\eta_1$ is a genuine scalar with bulk mass squared
$2 \mu^2$.

In the example above the massive gauge field is not charged under the background $U(1)$, that is $N=0$. A slightly more
complicated non-abelian model can be used to illustrate the charged case. Let us consider for example
an electroweak-like $SU(2)\times U(1)_b$
gauge theory, where $\Phi$ is in the ${\bf 2}_{1/2}$ representation. Like in the Standard Model we trigger
$SU(2)\times U(1)_b \rightarrow U(1)_q$ through a non vanishing VEV, $\Phi^T= (0,v/\sqrt2)$,
where $U(1)_q$ is the analogue of the
electromagnetic $U(1)$, and we assume $v$ real and positive without loss of generality. Also we embed the background gauge field in the $U(1)_q$ so that we can solve the scalar bulk
equations for $v$ constant and equal to the point of minimum of $U$. In this case
\be \eta =  \left(\begin{array} {c} \frac{1}{\sqrt2}(\eta_1+i \eta_2) \\
\frac{1}{\sqrt2}(v+\sigma+i \eta_0) \end{array}
            \right)\, , \quad V_M=W_M^a \frac{\sigma^a}{2} + B_M b\, , \ee
where $\sigma^a$ are the Pauli matrices, the $\eta_i$ and $\sigma$ are real scalars and $W_M$ and $B_M$ are the gauge field perturbations corresponding to
the $SU(2)$ and $U(1)_b$ group factors. By using the bilinear action in the light-cone gauge, eq.~(\ref{bil-action-lcg}), we find that
$W^\pm=(W^1\pm iW^2)/\sqrt2$ and $\eta^{\pm}=(\eta_1\pm i \eta_2)/\sqrt2$ have the same bulk mass,
$m=\frac{1}{2}v | g_2|$, where $g_2$ is the gauge coupling of $SU(2)$,  and represent altogether a massive gauge field
with $N=\pm1$. Apart from this bulk mass term the light-cone-gauge bilinear action for $W$ and $\eta^{\pm}$ coincide
with that of a massless bulk gauge field and a massless scalar respectively.

The bottom line of the examples above is that a massive gauge field leads to the 4D spectrum
of a massless gauge field, given in eq. (\ref{4D-gaugescalar-spectrum}),
plus that of a scalar, which we provided before. It follows that  the $s_i$ coefficient of a massive gauge
field are
 \bea  s_{-1}^{\rm mgf}&=& 5s_{-1}^{\rm s}\, , \quad  s_{0}^{\rm mgf}=5s_{0}^{\rm s} +\Delta s_0^{\rm gf}\, , \nonumber \\
\quad s_{1}^{\rm mgf}&=&5s_{1}^{\rm s}+\Delta s_1^{\rm gf}\, , \quad s_{2}^{\rm mgf}=5s_{2}^{\rm s} + \Delta s_2^{\rm gf} \, ,\eea
where the $s_i^s$ are the corresponding quantities for a 6D scalar, those given in eqs.~(\ref{eq:simplescalars1})--(\ref{eq:simplescalars2}), and  we
used $|N|\leq 1$, to ensure the stability (see the appendix on the massless gauge fields).

\end{document}